\documentclass{aa}
\usepackage{hyperref}
\usepackage{xcolor}
\hypersetup{
    colorlinks=true,
    linkcolor=red,
    filecolor=magenta,
    urlcolor=orange,
    citecolor=blue
}
\usepackage{graphicx}
\usepackage{mathtools}

\usepackage{mathptmx}
\usepackage[varg]{txfonts}
\usepackage{txfonts}
\usepackage{subfigure}
\usepackage{lscape}
\usepackage{pifont}
\usepackage{comment}
\usepackage{natbib}
\bibpunct{(}{)}{;}{a}{}{,} 

\newcommand*\CHECK{\ding{51}}

\begin{document}
\title{Ubiquitous cold and massive filaments in cool core clusters}

\author{V.~Olivares\inst{1}, P.~Salome\inst{1}, F.~Combes\inst{1,2}, S.~Hamer\inst{3}, P.~Guillard\inst{4,5}, M.~D.~Lehnert\inst{4}, F.~Polles\inst{1}, R.~S.~Beckmann\inst{4}, Y.~Dubois\inst{4}, M.~Donahue\inst{7}, A.~Edge\inst{11},
A.~C.~Fabian\inst{3}, B.~McNamara\inst{10}, T.~Rose\inst{11}, H.~Russell\inst{3}, G.~Tremblay\inst{8}, A.~Vantyghem\inst{10}, R.~E.~A.~Canning\inst{9}, G.~Ferland\inst{12}, B.~Godard\inst{6}, M.~Hogan\inst{10}, S.~Peirani\inst{4,13}, G.~Pineau des Forets\inst{1}}

\institute{LERMA, Observatoire de Paris, PSL Research Univ., CNRS, Sorbonne Univ., 75014 Paris, France.        
\and
Coll\`ege de France, 11 Place Marcelin Berthelot, 75005 Paris, France. 
\and
Institute of Astronomy, University of Cambridge, Madingley Road, Cambridge CB1 0HA, UK.                                     
\and
Sorbonne Universit\'e, CNRS, UMR7095, Institut d'Astrophysique de Paris, 98 bis Bd Arago, 75014 Paris, France.              
\and
Institut Universitaire de France, Minist\'ere de l'Education Nationale, de l'Enseignement Sup\'erieur et de la Recherche, 1 rue Descartes, 75231 Paris Cedex 05, France. 
\and 
Ecole Normale Sup\'erieure, PSL Research Univ., CNRS, 75005 Paris, France. 
\and
Physics \& Astronomy Department, Michigan State University, East Lansing, MI 48824-2320, USA. 
\and
Harvard-Smithsonian Center for Astrophysics, 60 Garden St., Cambridge, MA 02138, USA. 
\and
Kavli Institute for Particle Astrophysics and Cosmology (KIPAC), Stanford University, 452 Lomita Mall, Stanford, CA 94305-4085, USA. 
\and
Physics \& Astronomy Department, Waterloo University, 200 University Ave. W., Waterloo, ON, N2L, 2GL, Canada. 
\and
Centre for Extragalactic Astronomy, Durham University, DH1 3LE, UK. 
\and 
Department of Physics and Astronomy, University of Kentucky, Lexington, Kentucky 40506-0055, USA. 
\and 
Universit\'e Cote d'Azur, Observatoire de la Cote d'Azur, CNRS, Laboratoire Lagrange, Bd de l'Observatoire, CS 34229, F-06304 Nice Cedex 4, France. 
}

\date{Received XX, ; accepted XX,}

\abstract{Multi-phase filamentary structures around Brightest Cluster Galaxies (BCG) are likely a key step of AGN-feedback. We observed molecular gas in 3 cool cluster cores: Centaurus, Abell\,S1101, and RXJ1539.5 and gathered ALMA (Atacama Large Millimeter/submillimeter Array) and MUSE (Multi Unit Spectroscopic Explorer) data for 12 other clusters. Those observations show clumpy, massive and long, 3--25~kpc, molecular filaments, preferentially located around the radio bubbles inflated by the AGN (Active Galactic Nucleus). Two objects show nuclear molecular disks. The optical nebula is certainly tracing the warm envelopes of cold molecular filaments. Surprisingly, the radial profile of the H$\alpha$/CO flux ratio is roughly constant for most of the objects, suggesting that (i) between 1.2 to 7 times more cold gas could be present and (ii) local processes must be responsible for the excitation.

Projected velocities are between \mbox{100--400~km~s$^{-1}$}, with disturbed kinematics and sometimes coherent gradients. This is likely due to the mixing in projection of several thin (as yet) unresolved filaments. The velocity fields may be stirred by turbulence induced by bubbles, jets or merger-induced sloshing. Velocity and dispersions are low, below the escape velocity. Cold clouds should eventually fall back and fuel the AGN.

We compare the filament's radial extent, r$_{\rm fil}$, with the region where the X-ray gas can become thermally unstable. The filaments are always inside the low-entropy and short cooling time region, where t$_{\rm cool}$/t$_{\rm ff}$<20 (9 of 13 sources). The range t$_{\rm cool}$/t$_{\rm ff}$, 8-23 at r$_{\rm fil}$, is likely due to (i) a more complex gravitational potential affecting the free-fall time t$_{\rm ff}$ (sloshing, mergers…); (ii) the presence of inhomogeneities or uplifted gas in the ICM, affecting the cooling time t$_{\rm cool}$. For some of the sources, r$_{\rm fil}$ lies where the ratio of the cooling time to the eddy-turnover time, t$_{\rm cool}$/t$_{\rm eddy}$, is approximately unity.

}


\authorrunning{V. Olivares et al.}
\titlerunning{Ubiquitous cold and massive filaments}

\maketitle

\section{Introduction}
\label{sec:introduction}

The cores of galaxy clusters are very dynamic and host some of the most energetic active galactic nuclei (AGN) known in the local universe. Brightest cluster galaxies, BCGs, located at the centers of galaxy clusters, are the largest, most luminous and massive elliptical galaxies in the Universe. These BCGs have extended, diffuse stellar envelopes, and predominantly old, ``red and dead stellar'' populations.

Deep X-ray observations of the central regions of clusters show intense X-ray emission suggesting that the Intra Cluster Medium (ICM) is cooling radiatively. This rapidly cooling gas should condense into cold gas clouds and/or form stars on short timescales relative to the age of the cluster \citep[e.g.,][]{fabian94,fabian02b}. However, the cold gas mass and star formation rates observed are too low to be consistent with the mass of gas that is predicted to be cooling from the ICM, suggesting that some form of heating balances the radiative cooling. Mechanical and radiative heating from AGN is the most widely accepted scenario for preventing rapid cooling of the ICM and this notion is supported by observations of obvious interaction between the radio--jets and the high-filling factor gas within the ICM. For example, X-ray images clearly show that super massive black holes (SMBH) strongly interact with their cluster environments, generating outflows/jets, bubbles, turbulence, and shocks that heat thermally and dynamically the surrounding hot atmosphere \citep[e.g.,][for a review]{mcnamaraandnulsen07, fabian12}.

Multi-wavelength observations of central regions of galaxy clusters have determined that the ICM is multiphase. The existence of a multiphase ICM implies
that the gas is cooling down to form atomic and molecular gas and that this warm/cold gas may provide the fuel necessary to regulate the triggering of the AGN activity. BCGs contain large amounts of cold molecular gas, with masses of 10$^{9}$ to 10$^{11}$ M$_{\odot}$ \citep{edge01, salome03, salome06, salome08, salome11}. BCGs have luminous filamentary emission-line nebulae \citep{heckman89, crawford99} and spatial distribution of these nebulae coincides with soft X-ray emission features \citep{werner14}. They also have dust lanes \citep{mittal12} and sometimes UV emission \citep{mcdonald09}, suggesting that some filaments are forming stars. Multiphase intra-cluster media is observed preferentially in systems where the central cooling time is below $\rm \sim5\times10^{8}$~yr or, similarly, where the entropy is less than 30~keV~cm$^{-2}$ \citep{cavagnolo08,rafferty08}. This threshold is interpreted as the onset of thermal instabilities (TI) in hot atmospheres \citep{cavagnolo08, rafferty08}.

Recent theoretical analyses and simulations have suggested that the hot atmospheres can become thermally unstable locally when the ratio of the cooling to free-fall timescales, t$_\mathrm{cool}$/t$_\mathrm{ff}$ falls below $\sim$ 10--20 \citep{mccourt12, gaspari12, prasad15, voitdonahue15, li15, mcnamara16, voit15, voit17}. In this model, the condensed cold gas rains down on the central supermassive black hole, fuelling it. The SMBH, then being  actively accreting AGN, injects back the energy in the form of massive outflows, often entrained by relativistic jets, establishing a tight self-regulated loop \citep[a feedback loop driven by an AGN or ``AGN feedback''; ][]{Gaspari_2017}. Molecular gas observations in the central region of clusters have shown  that cold filaments extend along the trajectories of the radio bubbles or around the X-ray cavities \citep{salome04,salome06, david14, mcnamara14, russell14, russell16, russell17, tremblay16, vantyghem16, vantyghem18}. The molecular clouds have likely cooled in-situ from (i) either low-entropy gas that have been uplifted by the bubbles at an altitude where it becomes thermally unstable (ii) or by direct thermal instability of small perturbation in the hot halo. This paper discusses these two scenarios. 



In this paper we present Atacama Large Millimeter Array (ALMA) Cycle 3 observations that maps the kinematics and morphology of the cold molecular gas phase of three well-studied BCGs: Centaurus (also known as NGC4696), Abell\,S1101 (so-called Sersic 159-03) and RXJ1539.5-8335, traced by CO(1-0) emission. In addition, we include a revised analysis of the ALMA observations of 12 BCGs: PKS\,0745-191, Abell\,1664, Abell\,1835, Abell\,2597, M87, Abell\,1795, Phoenix-A, RXJ0820.9+0752, 2A0335+096, Abell\,3581, Abell\,262 and Hydra-A. We compare ALMA observations with new Multi-Unit Spectroscopic Explorer (MUSE) IFU data that maps the kinematics and morphology of the warm ionized gas. The ALMA sample is described in section~\ref{sec:sample}, while the MUSE and ALMA data are described in section~\ref{sec:observations}, results are presented in section~\ref{sec:results}, discussed in section~\ref{sec:discussion} and summarized in section~\ref{sec:conclusions}. We assume a standard cosmology throughout with H$\rm_{0}$=70 km s$^{-1}$ Mpc$^{-1}$ and $\Omega\rm_{m}$=0.3.

\section{Sample}
\label{sec:sample}

\begin{table*}
\caption{Sample: ALMA and MUSE data used in our analysis.}
\label{tab:sample}
\centering
\begin{tabular}{llccccccc}
\noalign{\smallskip}\hline \hline \noalign{\smallskip}
\rm Cluster ID & BCG ID&  ALMA ID project  & $z$  &  CO(1-0) & CO(2-1) & CO(3-2) & MUSE  \\
\rm 	(1)    	  	& (2)		& (3) 			&  (4)		& 	&(5)    &		& (6)\\
\noalign{\smallskip} \hline \noalign{\smallskip}
\textit{New ALMA observations}\\
\noalign{\smallskip}
Centaurus		&\small NGC4696 				&2015.1.01198.S	&  0.01016      &  \CHECK & ...         & ...       &  \CHECK   \\
RXJ1539.5-8335	&\small 2MASXJ15393387-8335215  &2015.1.01198.S	&  0.07576      &  \CHECK & ...         & ...       &  \CHECK   \\
Abell\,S1101	&\small ESO 291-G009 		    &2015.1.01198.S	&  0.05639      &  \CHECK & ...         & ...       &  \CHECK   \\
\noalign{\smallskip} \hline \noalign{\smallskip}
\textit{ALMA archive}\\
\noalign{\smallskip}
PKS\,0745-191 	&\small PKS\,0745-191 			&2012.1.00837.S  	& 0.10280  	&  \CHECK  & ...        &  \CHECK   &  \CHECK   \\
Abell\,1664		&\small 2MASXJ13034252-2414428  &2011.0.00374.S  	& 0.12797  	&  \CHECK  & ...        &  \CHECK   &  ...     \\
Abell\,1835		&\small 2MASXJ14010204+0252423  &2011.0.00374.S 	& 0.25200  	&  \CHECK  & ...        &  \CHECK   &  ...      \\
Abell\,2597		&\small PKS\,2322-12		    &2012.1.00988.S	    & 0.08210 	& ...      &  \CHECK    & ...       &  \CHECK   \\
M87				&\small M87				    	&2013.1.00862.S 	& 0.00428 	& ...      &  \CHECK    & ...       &  ...      \\
Abell\,1795		&\small Abell\,1795				&2015.1.00623.S	    & 0.06326	& ...      &  \CHECK    & ...       &  \CHECK   \\
Phoenix-A		&\small SPT-CLJ2344-4243		&2013.1.01302.S	    & 0.59600	& ...      & ...        &  \CHECK   & ...       \\
RXJ0820.9+0752	&\small 2MASXJ08210226+0751479  &2016.1.01269.S    	& 0.10900 	&  \CHECK  & ...        &  \CHECK   &  \CHECK   \\
2A0335+096		&\small 2MASXJ03384056+0958119	&2012.1.00837.S	    & 0.03634  	&  \CHECK  & ...        &  \CHECK   &  \CHECK   \\ 
Abell\,3581 	&\small IC4374				    &2015.1.00644.S 	& 0.02180 	& ...      &  \CHECK    & ...       &  \CHECK   \\
Abell\,262  	&\small NGC 708				    &2015.1.00598.S 	& 0.01619 	& ...      &  \CHECK    & ...       & ...       \\ 
Hydra-A			&\small Hydra-A				    &2016.1.01214.S 	& 0.05435 	& ...      &  \CHECK    & ...       &  \CHECK   \\
\noalign{\smallskip} \hline \noalign{\smallskip}
\end{tabular}
\small
\raggedright (1) Source name, the actual target studied in this paper is the central brightest cluster galaxy of the cluster.\\
\raggedright (2) Central brightest cluster galaxy name. \\
\raggedright (3) ALMA project ID of the observations analyzed.\\
\raggedright (4) The redshifts $z$ of the BCGs which were derived from optical emission lines/features. \\
\raggedright (5) The lines observed with ALMA used in our study. \\
\raggedright (6) The availability of MUSE data for each source.
\tablebib{
Centaurus from \citet{postman&lauer95}; RXJ1539.5-8335 from \citet{huchra12}; Abell\,S1101 from \citet{nicolacidacosta91};
PKS\,0745-191 from \citet{russell16}; Abell\,1664 from \citet{pimbblet06}; Abell\,1835 from \citet{crawford99}; Abell\,2597 from \citet{tremblay16}; Abell\,1795 from \citet{russell17b}; Phoenix-A from \citet{russell16}; RXJ0820.9+0752 from \citet{crawford95}; 2A0335+096 from \citet{mcnamara90}; Abell\,3581 from \citet{Canning_2013}; Hydra from \citet{rose19}.}
\end{table*}

The 15 cool-core BCGs, with z$<$0.6, that comprise our sample are listed in Table~\ref{tab:sample}, and split into two groups: (1) those with new ALMA observations and (2) those for which we used ALMA archive data. All are well-studied and have nearly complete spectral data coverage from the radio to the X-ray.
Our new ALMA observations sample presented here include three well-known BCGs: Centaurus, RXJ1539.5-8335, and Abell\,S1101. All three are known to have extended optical emission nebulae. Furthermore, we describe in depth the three galaxies for which we have obtained CO(1-0) data. We adopted distances of 49.5, 261, and 331.1~Mpc for Centaurus, Abell\,S1101, and RXJ1539 which
then yield physical scales of 0.208, 1.437 and 1.094~kpc~arcsec$^{-1}$, respectively.

In order to have a more complete picture of cool core clusters, we have included a complementary sample of twelve BCGs from the ALMA archive that are known to have detections in CO(1-0), CO(2-1) and/or CO(3-2) (as indicated specifically in Table~\ref{tab:sample}). Each of these clusters has a central cooling time less than 1$\times$10$^{9}$~yr and all have been observed in H$\alpha$. Throughout the rest of this paper we refer to the central galaxies by the names of the clusters in which they lie (Table~\ref{tab:sample}).

Our sample spans a wide range of X-ray mass deposition rates and SFRs, Balmer and forbidden line luminosities, and in the
power of their AGN, radio sources, and X-ray cavities. Our sample of sources span from high to low SFRs and strong to weak signatures of AGN feedback. A summary of these various properties can be found in Table~\ref{tab:sourceprop}. In the following paragraphs, we describe in detail the three ALMA sources for which we obtained new ALMA observations.

\paragraph{\bf Centaurus} NGC 4696, at $z$=0.0114 and the brightest galaxy of the Centaurus cluster of galaxies (Abell\,3526), is one of the X-ray brightest and nearest galaxy clusters in the sky. \textit{Chandra} observations of the cluster revealed a plume-like structure of soft X-ray emission extending from the nucleus \citep{sandersandfabian02,sanders16}.
Optical observations reveal bright line-emitting dusty filaments in H$\alpha$ and [NII] \citep{fabia82,fabian16}. These filaments extend over the central arcminute \citep[$\sim$14~kpc; ][]{sparks89, crawford05} and have approximately the same morphology as the spiral structure observed in the X-ray. They also show a remarkable spatial correlation with the dust features, in particular the dust sweeping around the core of the galaxy to the south and west, looping around the core of the NGC 4696 \citep{sparks89, laine03}.
NGC 4696 hosts a compact steep-spectrum FR I-type radio source, PKS\,1246-410 \citep[spectral index of 1.75;][]{werner11}, with a total radio luminosity of 9.43$\times$10$^{40}$~erg~s$^{-1}$ \citep{mittal09}. VLA observations indicate that the extended component of the central radio source appears to be interacting with and displacing the ICM and it has excavated small cavities in the X-ray gas \citep{taylor02, taylor06}. 
\citet{mittal11} measured two of the strongest cooling lines of the interstellar medium, [CII]157.74~$\mu$m and [O I]63.18~$\mu$m. The [CII] emission has a similar morphology and velocity structure compared to the H$\alpha$ emission. 
In addition, they found 1.6$\times$10$^{6}$~M$_{\odot}$ of dust at 19~K, although the low FIR luminosity suggests a star formation rate of only 0.13 M$_{\odot}$ yr$^{-1}$. \citet{fabian16} detected similar CO(2-1) emission that our ALMA CO(1-0) observations. In this paper, we present an analysis of the molecular gas maps from CO(1-0) Cycle 3 ALMA observations of the Centaurus Cluster.

\paragraph{\bf RXJ1539.5-8335} The cluster RXJ1539.5-8335 (BCG: 2MASX J15393387-8335215) is the most distant of our three targets, $z$=0.0728, but is the most luminous in H$\alpha$. It has bright clumpy filaments in H$\alpha$ that extend to the east and west of the BCG over the central arcminute ($\sim$ 83.7 kpc) and has disturbed morphology and velocity field \citep{hamer16}. \citet{hogan15} showed that the source has a flat spectrum, with a spectral index of 0.45$\pm$0.07, and is core-dominated. The X-ray morphology is extended and bar-like in the core of the cluster which is qualitatively similar to the observed H$\alpha$ morphology. We note that no radio jets have been observed in this source.

\paragraph{\bf Abell\,S1101} Abell\,S1101 (Sersic 159-03) is a relatively low-mass cool-core moderately distant cluster, $z$=0.058, that has been the subject of many studies. The central dominant galaxy of the cluster reveals a complex structure at all wavelengths and shows dust lanes, molecular gas, excess UV emission, and a bright, 44~kpc long H$\alpha$+[NII] filamentary nebula, associated with low entropy, high metallicity, cooling X-ray emitting gas \citep{oonk10, werner11, farage12, mcdonald12a}. The mass of cold dust in the center of the cluster is  $\sim$10$^{7}$ M$_{\odot}$ \citep{mcdonald15}. This cluster shows signs of strong AGN feedback. The radio source has a very steep power-law spectrum, $\alpha$ $\sim$ $-$2.2, and is extended to the east, northeast, and northwest of the core. The radio emission from the core is flatter than the more extended radio emission \citep[$\rm \alpha \sim -1$ to $-$1.5; ][]{birzan04, werner11}. The radio lobes seen at 5 GHz and 8.5 GHz are extended over $\sim$10 kpc and are aligned with the filaments and southern cavity \citep{birzan04, birzan08, werner11}.
In a multi-wavelength study of the BCG core, \citet{werner10} found evidence for strong interaction between the jets and hot gas of the cluster. They suggested that the jets and radio bubbles have removed most of the cooling X-ray gas from the core (within $\sim$7.5~kpc). They also proposed that the filamentary emission line nebula has been uplifted from the central galaxy by the jets and lobes, but that gas in the filaments is not currently being impacted by the radio jets and is cooling and forming stars.
\citet{farage12}, based on X-ray and radio data, also claimed that Abell\,S1101 is dominated by feedback from the central AGN. However, the kinematics of the optical emission are consistent with either infall or outflow of material along the bright filaments. 

\begin{table*}
\caption{ALMA Observation properties.}
\centering
\begin{tabular}{lcccccc}
\noalign{\smallskip}\hline \hline \noalign{\smallskip}
 \rm Source		& Band  & Date & Obs. time 	& RMS			& Angular Res. & References \\
  \rm  	   		&		& 	   & (hrs) 		& (mJy bm$^{-1}$)	& ($\arcsec$) 			&\\
\noalign{\smallskip} \hline \noalign{\smallskip}
Centaurus		&3& 2016-01-27	& 1.310	& 0.43 	&2.90$\arcsec\times$2.14$\arcsec$  (0.6 kpc $\times$ 0.4 kpc)   	&... \\
 \noalign{\smallskip} \hline \noalign{\smallskip}
RXJ1539.5-8335	&3& 2016-03-06	& 0.890	& 0.41 	& 3.10$\arcsec\times$1.85$\arcsec$ (4.5 kpc $\times$ 2.6 kpc)	    &	 ...\\
 \noalign{\smallskip} \hline \noalign{\smallskip}
Abell\,S1101		 &3& 2016-01-26	& 1.596	& 0.44	&2.20$\arcsec\times$1.80$\arcsec$  (2.3 kpc $\times$ 1.9 kpc)		& ... \\
 \noalign{\smallskip} \hline \noalign{\smallskip}
PKS\,0745-191	 &3& 2014-04-26	& 0.907	& 0.47	&1.80$\arcsec\times$1.34$\arcsec$  (3.4 kpc $\times$ 2.5 kpc)	    & (1) \\
				 &7& 2014-08-19	& 0.403	& 0.82	&0.27$\arcsec\times$0.19$\arcsec$  (0.5 kpc $\times$ 0.4 kpc)		& (1) \\
 \noalign{\smallskip} \hline \noalign{\smallskip}
Abell\,1664		 &3& 2012-03-27	& 0.840	& 0.69	&1.57$\arcsec\times$1.27$\arcsec$  (3.6 kpc $\times$ 2.9 kpc)		& (2) \\
				 &7& 2012-03-28	& 1.176	& 1.60 	&0.70$\arcsec\times$0.42$\arcsec$  (1.6 kpc $\times$ 0.9 kpc)		& (2)\\
 \noalign{\smallskip} \hline \noalign{\smallskip}
Abell\,1835		&3& 2013-06-04	& 1.008	& 0.48	&1.73$\arcsec\times$1.52$\arcsec$  (2.7 kpc $\times$ 2.4 kpc)		& (3)\\
				&7& 2013-06-04	& 1.008	& 1.00 	&0.65$\arcsec\times$0.52$\arcsec$  (1.0 kpc $\times$ 0.8 kpc)		& (3)\\
\noalign{\smallskip} \hline \noalign{\smallskip}
Abell\,2597		 &6& 2013-11-17	& 3.024	& 0.12	&0.96$\arcsec\times$0.76$\arcsec$  (1.5 kpc $\times$ 1.2  kpc)	    & (4)\\
\noalign{\smallskip} \hline \noalign{\smallskip}
M87				&6& 2013-08-24	& 2.957	& 0.27	&1.90$\arcsec\times$0.20$\arcsec$  (0.15 kpc $\times$0.01  kpc) 	& (5)\\
\noalign{\smallskip} \hline \noalign{\smallskip}
Abell\,1795		&6& 2016-06-11	& 1.193	& 0.46	&0.81$\arcsec\times$0.62$\arcsec$  (0.9 kpc $\times$ 0.7 kpc)		& (6)\\
\noalign{\smallskip} \hline \noalign{\smallskip}
Phoenix-A		&6& 2015-09-02	& 0.932	& 0.32 	&0.74$\arcsec\times$0.72$\arcsec$  (4.9 kpc $\times$ 4.8 kpc)   	& (7)\\
\noalign{\smallskip} \hline \noalign{\smallskip}
RXJ0820.9+0752	&3& 2016-10-01	& 1.445	& 0.17	&0.71$\arcsec\times$0.69$\arcsec$  (1.4 kpc $\times$ 1.3  kpc)	    & (8)\\
				&7& 2016-10-01	& 0.378	&  0.44	&0.16$\arcsec\times$0.13$\arcsec$  (0.3 kpc $\times$ 0.2 kpc)		& (8)\\
\noalign{\smallskip} \hline \noalign{\smallskip}
2A0335+096		&3& 2014-07-04	& 1.142	& 0.44 	&1.30$\arcsec\times$1.00$\arcsec$  (0.9 kpc $\times$ 0.7  kpc)	    & (9)\\
				&7& 2014-08-12	& 0.546	& 0.91	&0.61$\arcsec\times$0.52$\arcsec$  (0.4 kpc $\times$ 0.3 kpc)		& (9)\\
\noalign{\smallskip} \hline \noalign{\smallskip}
Abell\,3581 	&6& 2017-10-19	& 1.462	& 0.18	&0.99$\arcsec\times$0.77$\arcsec$  (0.4 kpc $\times$ 0.3 kpc)		& (10) \\
\noalign{\smallskip} \hline \noalign{\smallskip}
Abell\,262  	&6& 2017-08-23	& 0.185	& 0.62	&0.95$\arcsec\times$0.61$\arcsec$  (0.3 kpc $\times$ 0.2  kpc)	    & ... \\
\noalign{\smallskip} \hline \noalign{\smallskip}
Hydra-A			&6& 2018-10-23	& 0.756	& 0.49	&0.29$\arcsec\times$0.21$\arcsec$  (0.28 kpc $\times$ 0.2 kpc)		& (11)\\
\noalign{\smallskip} \hline \noalign{\smallskip}
\label{tab:obslog}
\end{tabular}
\tablebib{
(1) \citet{russell16};  (2) \citet{russell14}; (3) \citet{mcnamara14}; (4) \citet{tremblay16};
(5) \citet{simionescu18};  (6) \citet{russell17b}; (7) \citet{russell17} ; (8) \citet{vantyghem17}; (9) \citet{vantyghem16}; (10) (PI: Y. Fujita); (11) \citet{rose19}.
}
\end{table*}


\section{Observations and Data Reduction}
\label{sec:observations}
\subsection{ALMA data reduction}
The BCGs Centaurus, RXJ1539.5-8335 and Abell\,S1101 were observed in band 3 with ALMA as a Cycle 3 program (ID = 2015.1.01198.S; PI: S. Hamer).
The Centaurus data were obtained in two sets of observations of 40 min on the 27th January 2016; RXJ1539.5-8335 was observed during two sessions of 26 min each on 6th March 2016 and Abell\,S1101 data was obtained in two runs of 48 min observations on the 26th January 2016, see Table~\ref{tab:obslog}. On average forty-four 12m antennas were used in the compact configuration with baseline 15--312~m. The observations used a single pointing, centered on the BCGs in each cluster.

The frequency division correlator mode was used with a 8 GHz bandwidth and frequency resolution of 3.9 kHz ($\sim$5 km s$^{-1}$) for Centaurus and 1.9 kHz ($\sim$5 km s$^{-1}$) for both RXJ1539.5-8335 and Abell\,S1101. Channels were binned together to improve the signal-to-noise ratio.
The precipitable water vapor (PWV) was 2.46, 3.0, 2.36~mm for the observations taken in January, 26th and 27th, and 3rd March 2016, respectively. The flux calibrators were J1107-4449 and J1427-4206 for Centaurus, whereas Titan and Neptune were used for the observations of RXJ1539.5-8335 and Abell\,S1101, respectively.
The observations were calibrated using the Common Astronomy Software Application (CASA) software \citep[version 4.5.1,][]{mcmullin07} following the processing scripts provided by the ALMA science support team. 

To make channel maps suitable for our own needs, first we performed the continuum subtraction where the level of the continuum emission was determined from line-free channels and subtracted from visibilities using the task UVCONTSUB. The calibrated continuum-subtracted visibilities were then imaged and deconvolved with the CLEAN algorithm. We found that self-calibration did not produce a significant reduction in the rms noise of the final image. To find the optimal signal-to-noise, we made several channel maps with different velocity resolution and with natural weighting using the Hogbom method.

The final data for Centaurus has a 2.90$\arcsec\times$ 2.14$\arcsec$ synthesized beam with a P.A.= 81.5~deg and the rms noise level is 0.43~mJy beam$^{-1}$ for a velocity resolution of 30~km~s$^{-1}$. For RXJ1539.5-8335, the rms noise level was 0.41~mJy~beam$^{-1}$ for a velocity resolution of 30~km~s$^{-1}$ and a synthesized beam of 2.20$\arcsec\times$1.85$\arcsec$ with a P.A.= 11.7 deg. Finally, the Abell\,S1101 data cube has an rms noise level of 0.44~mJy~beam$^{-1}$ for a velocity width of 30~km~s$^{-1}$ and a synthesized beam of 2.20$\arcsec\times$1.80$\arcsec$ with a P.A.= 66.2~deg.
Images of the continuum emission were also produced using CLEAN task of CASA by averaging channels free of any line emission. Continuum images of the flux calibrators were performed in order to test the veracity of the flux and measure the uncertainty. To do this, we compared the flux measurements with the flux computed at the same frequency in a similar date of the observations using the SMA calibrator catalogue.

ALMA archived data were performed following the processing script provided by the ALMA science support team and using the CASA version that they suggest. In particular, we followed the CASA guides to modify the packaged calibration script in order to apply the proper amplitude calibration scale (which has changed since Cycle 0). The imaging and the continuum subtraction procedure was the same as for new ALMA observations described before. In Table~\ref{tab:obslog} lists the RMS sensitivities and the angular resolutions for the ALMA archive data. As each observation is described in details in the references listed in table~\ref{tab:obslog} and table~\ref{tab:sample}, we only summarize the main characteristics of these observations.

\subsection{MUSE Optical Integral Field Spectroscopy}

We also present new optical nebular emission line kinematics and morphologies for 11 sources in our sample from an analysis of data obtained with the Multi-Unit Spectroscopic Explorer (MUSE). These sources are, Centaurus, RXJ1539.5-8335, Abell\,S1101, PKS\,0745-191, Abell\,2597, Abell\,1795, RXJ0820.9+0752, 2A0335+096, Abell\,3581, and Hydra-A. MUSE is an
optical image slicing integral field unit (IFU) with a wide field-of-view (FOV) which is mounted on UT4 of the European Southern Observation Very Large Telescope (ESO-VLT). The data were obtained as a part of the ESO programme 094.A-0859(A) (PI: S. Hamer). The observations where carried out in MUSE\'s seeing limited WFM-NOAO-N configuration with a FOV of 1$\arcmin\times$1$\arcmin$ covering the entire BCG in a single pointing and with an on-source integration time of 900~s.

The data were reduced using version 1.6.4 of the MUSE pipeline \citet{weilbacher14}, including a bias subtraction, wavelength and flux calibration, illumination correction, flat-fielding, and correction for differential atmospheric diffraction. In addition to the sky subtraction in the pipeline, we have performed an additional sky subtraction using ZAP \citep[Zurich Atmosphere Purge; ][]{soto16}. The final MUSE datacube maps the entire galaxy between 4750 $\AA <\lambda<$ 9300~$\AA$ with a spectral resolution of 1.5~$\AA$.

We have created higher level MUSE data products by modeling the stellar and nebular components of each galaxy. To separate the nebular emission from underlying stellar continuum, we did full-spectral fitting using the \textsc{PLATEFIT} code \citep{tremonti04}. First, we determined the redshift of the stellar component using the \textsc{AUTOZ} code, which determines redshifts using cross-correlations with template spectra \citep{Baldry_2014}. The stellar velocity dispersion is determined using \textsc{VDISPFIT}. This code uses a set of eigenspectra, convolved for different velocity dispersions. From this, the best fit velocity dispersion is determined. This value is the observed value and is not corrected for the contribution from the instrumental velocity dispersion.

For the spectral fitting we use the \textsc{PLATEFIT} spectral-fitting routine. \textsc{PLATEFIT}, which was developed for the SDSS, fits the stellar continuum and emission lines separately. In this continuum fitting stage, regions including possible emission lines are masked out. The stellar continuum is fitted with a collection of the stellar population synthesis model templates of simple stellar populations \citep{bruzual03}. The template fit is performed using the previously derived redshift and velocity dispersion. The second \textsc{PLATEFIT} emission-line fitting stage is now performed on the residual spectrum, that is, after continuum subtraction. Each of the emission lines are modelled as a single Gaussian profile. All emission lines share a common velocity offset and a common velocity dispersion which are treated as a free parameters in the fit. For this paper, we make a requirement that all our emission-line flux measurements have signal-to-noise ratio, S/N$>$7. We have also corrected the datacubes for Galactic foreground extinction estimated from the \citet{schlafly11} recalibration of the \citet{schlegel98} IRAS+COBE Milky Way dust map assuming R$_{V} =$~3.1. Finally, we shifted the velocity field of the nebular emission to redshifts of the BCGs in order to match the CO emission velocities from the ALMA data.
 

\section{Results}\label{sec:results}

\subsection{ALMA Continuum}

\begin{figure*}
\centering
\subfigure{\includegraphics[width=0.3\textwidth]{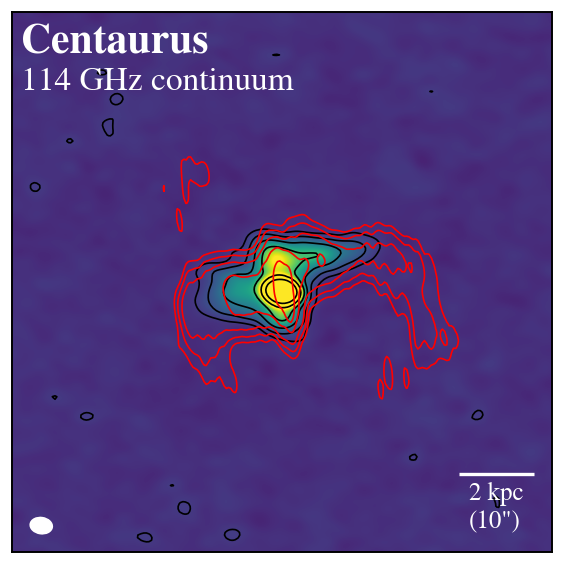}}
\subfigure{\includegraphics[width=0.3\textwidth]{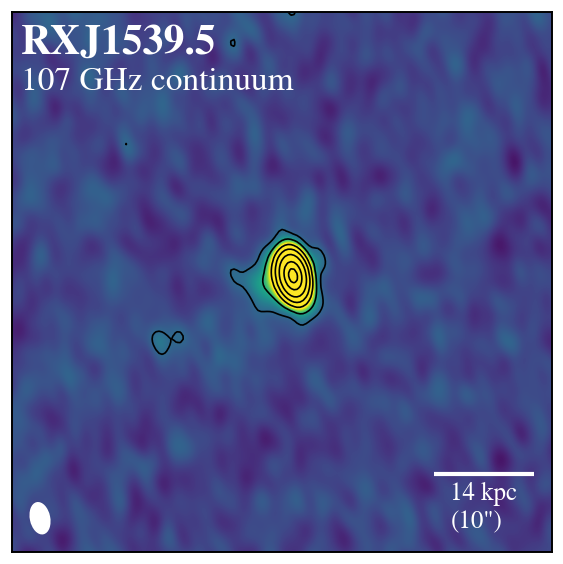}}
\subfigure{\includegraphics[width=0.3\textwidth]{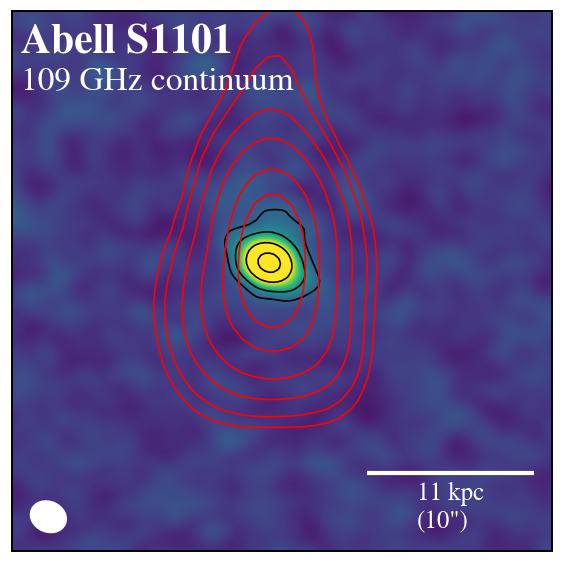}}
\caption{ALMA continuum observations of Centaurus (114 GHz), RXJ1539.5-8335 (107 GHz) and Abell\,S1101 (109 GHz). The black contours are relative to RMS noise in each of the data sets and were set to:  3$\sigma$, 10$\sigma$, 25$\sigma$, 50$\sigma$, 100$\sigma$, 200$\sigma$, 300$\sigma$. The beam is plotted in white at the bottom left side of each panel. The red contours indicate the radio emission from the VLA observations at 1.45 GHz for Centaurus and 1.46 GHz for Abell\,S1101. The mm continuum maps are dominated by the synchrotron emission from the AGN at the galaxy center.}
\label{fig:continuum}
\end{figure*}

A continuum source is detected in band 3 in RXJ1539.5, Abell\,S1101 and Centaurus. The continuum is partially resolved for all sources but dominated by a bright point source (Fig.~\ref{fig:continuum}). The continuum map extents are $\sim$9--10$\arcsec$ (13~kpc and 10~kpc) for RXJ1539.5 and Abell\,S1101, and $\sim$26$\arcsec$ (5.5~kpc) for Centaurus. The location of the continuum source of Abell\,S1101 is consistent with the 1.46 GHz Very Large Array (VLA) radio position. The mm-continuum source position and morphology of Centaurus also coincides with the radio emission at 8.33 GHz. The total continuum emission in Centaurus has a flux density of 40.9~mJy. No radio image is available for RXJ1539.5.

The continuum maps are dominated by a mm synchrotron continuum point source associated with the AGN at the galaxy center, with flux density at 3~mm of 11.4~mJy for RXJ1539.5, 2.2~mJy for S1101, and 37.8~mJy for Centaurus (see Table~\ref{tab:continuum}). We fitted a power-law to the radio spectral energy distribution (SED) using measurements from VLA and ATCA \citep{hogan15} plus our ALMA observations. The form of the power-law was $\rm f_{\upsilon} \propto \upsilon^{-\alpha}$ and we derived $\rm \alpha\sim-$1 for Centaurus and Abell\,S1101, and a flat-spectrum for RXJ1539.5 with $\rm \alpha\sim-$0.3. We do not detect any CO absorption features in bins from 4 to 50~km~s$^{-1}$ \citep[e.g.,][]{david14, tremblay16,rose19} against the weak nuclear continuum in those three systems.

\tabcolsep=0.05cm
\begin{table}
\caption{ALMA continuum properties.}
\label{tab:continuum}
\begin{tabular}{lrrrrc}
    \noalign{\smallskip}
    \hline
    \hline
    \noalign{\smallskip}
    \rm  Source			& RMS			    & Flux      	& Peak 		   &Size	 & \rm$\alpha$\\
    \rm     			& \small(mJy/bm)    &\small (mJy)	&\small (Jy/bm)   & (kpc)	&\\
  \noalign{\smallskip}
  \hline
  \noalign{\smallskip}
  Centaurus		    	&0.051	        &40.9$\pm$0.1	    	&0.035	        	&5.5    &$-$1.02$\pm$0.01\\
  RXJ1539.5 			&0.028	        &11.4$\pm$0.1		 	&0.009	        	&13     &$-$0.32$\pm$0.03\\
  Abell\,S1101			&0.038	        &2.2$\pm$0.1			&0.002		        &10 	&$-$1.13$\pm$0.09\\
     \noalign{\smallskip} \hline \noalign{\smallskip}
\end{tabular}
\end{table}

%
\subsection{Molecular and ionized Warm Gas in RXCJ1539.5-8335, Abell\,S1101 and Centaurus}
\label{sec:hatoco}

\begin{figure*}
\centering
\subfigure{\includegraphics[width=0.3\textwidth]{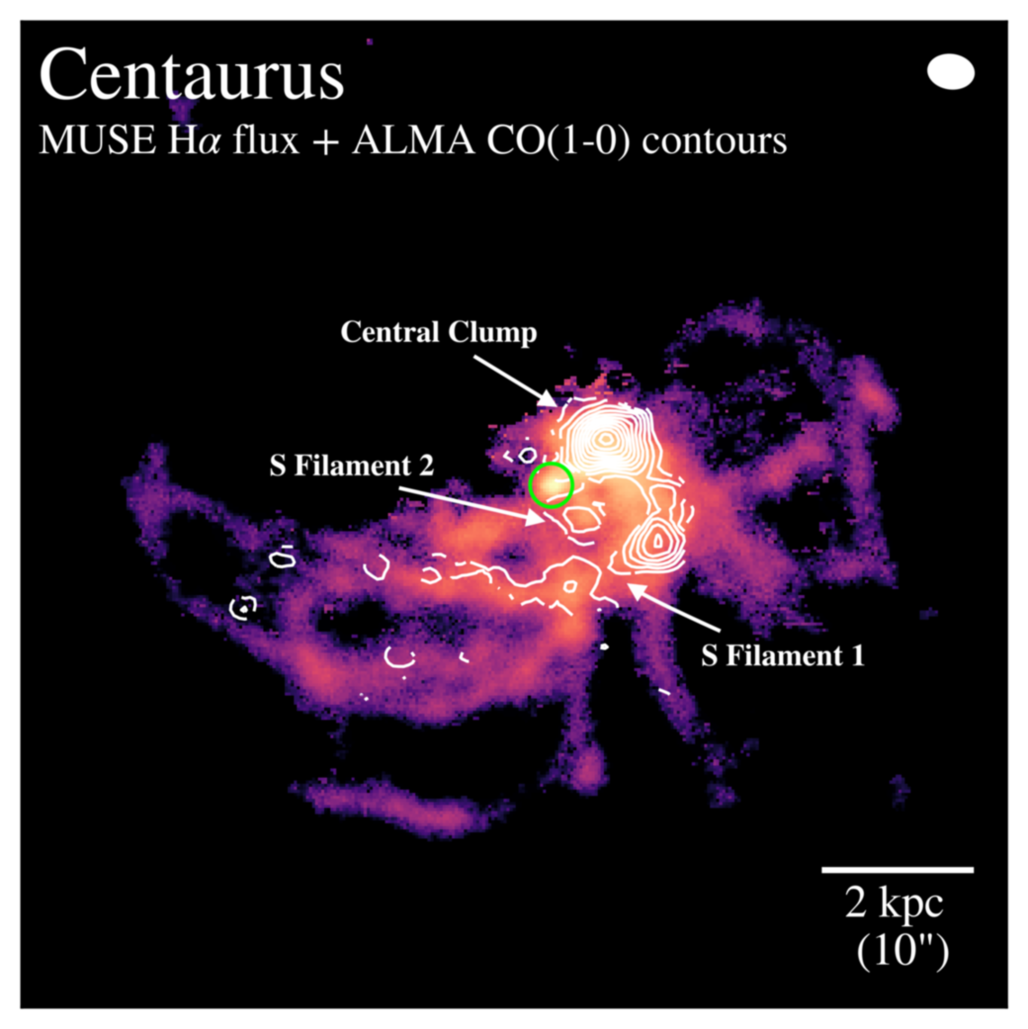}\label{fig:centaurus}}
\subfigure{\includegraphics[width=0.3\textwidth]{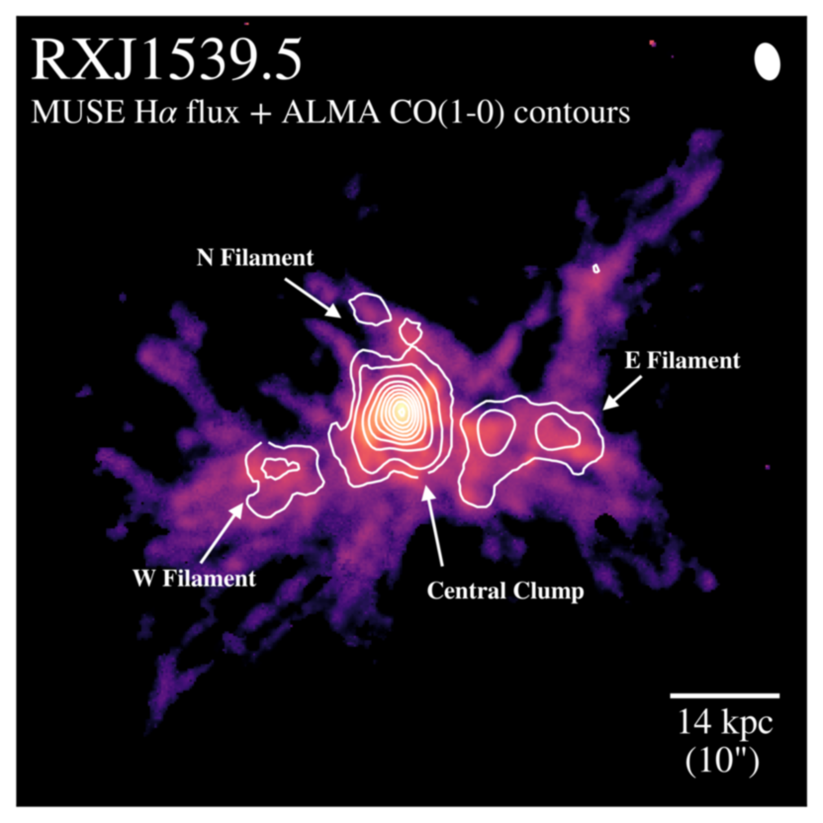}\label{fig:rxj1539}}
\subfigure{\includegraphics[width=0.3\textwidth]{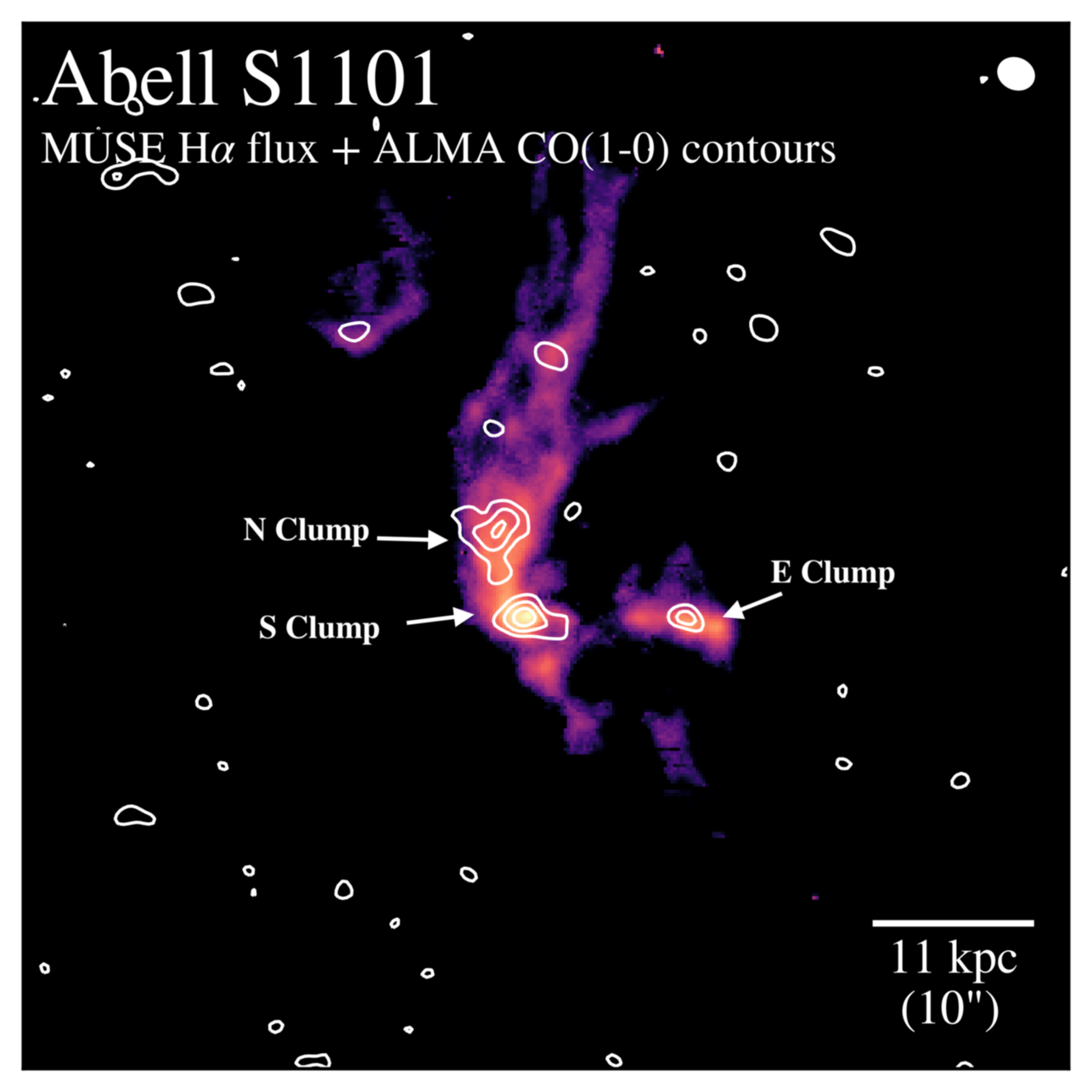}\label{fig:s1101}}
\caption{Logarithmically-scaled H$\alpha$ flux maps from MUSE observations overlaid with contours from the CO(1-0) integrated intensity maps for three the new ALMA sources: Centaurus (left panel), RXCJ1539.5-8335 (middle panel) and Abell\,S1101 (right panel). The co-spatial and morphological correlation between the warm ionized and cold molecular nebulae is clear in these maps. The CO(1-0) emission has been continuum-subtracted and binned from $-300$~km~s$^{-1}$ to $+300$~kms$^{-1}$. The ALMA beam is shown upper right of each panel. The CO(1-0) contours for Centaurus are: 3$\sigma$, 5$\sigma$, 7$\sigma$, for RXCJ1539.5-8335 are: 3$\sigma$, 10$\sigma$, 30$\sigma$, 50$\sigma$, 70$\sigma$..., and for Abell\,S1101 are: 3$\sigma$, 5$\sigma$, 7$\sigma$. In the panel showing the data for Centaurus, the small green circle indicates the location of the highest surface brightness H$\alpha$ emission without any CO emission (see text for details).}
\label{fig:haoverco}
\end{figure*}

Figure~\ref{fig:haoverco} shows the continuum-subtracted integrated CO(1-0) line and H$\alpha$ flux distributions of Centaurus, RXCJ1539.5-8335, and Abell\,S1101. The cold molecular emission is located where the H$\alpha$ emission is the strongest. The molecular filaments are co-spatial in projection with the warm ionized gas, similar to what has been found in other cool core BCGs \citep{salome06, russell14, mcnamara14, mcnamara16, vantyghem16, tremblay16, tremblay18,hatch05,lim12}. In these three sources, the molecular gas is not as extended as the warm ionized gas. This is likely due to the limit in sensitivity of the ALMA observations to low surface brightness gas rather than due to a true absence of cold molecular gas. Note that the ALMA and MUSE data cover similar areas in each source; e.g., for Centaurus the FOV (MUSE) is $\sim$12~kpc (1$\arcmin$, while the HPBW of ALMA at 115.3~GHz is $\sim$9.5~kpc (46$\arcsec$). The CO, like the H$\alpha$ emission distribution, shows very bright central emission with the addition of several more diffuse and filamentary structures. We note that secondary bright CO and H$\alpha$ spots also appear inside the filaments, particularly in Centaurus and Abell\,S1101.

\noindent{\bf Centaurus -} The continuum-subtracted CO(1-0) map distribution of Centaurus (Fig.~\ref{fig:haoverco}), reveals a filamentary emission that spans over $\sim$5.6~kpc (28$\arcsec$) and that is co-spatial with almost all the brightest emission from the warm ionized nebula. However, the correspondence is not perfect. The brightest nuclear region in Centaurus (marked with a green circle in the Fig.~\ref{fig:haoverco}) seen in the warm ionized gas shows no detection of molecular emission. This could be due to several factors: the molecular gas in the central regions could have been highly excited, and not particularly bright in the low J transitions of CO or, destroyed or expelled by the AGN, leaving no molecular gas to emit. Perhaps, the molecular gas was consumed by star formation as a consequence of positive feedback. Without information about other transitions or evidence of recent star formation, we can only speculate.

\noindent{\bf RXCJ1539.5-8335 -} In RXCJ1539.5-8335 the CO(1-0) emission distribution appears to be complex and filamentary, spanning along 48.8~kpc (34$\arcsec$) and matching morphologically with the brightest warm ionized nebula from the H$\alpha$ MUSE observations (see central panel of figure~\ref{fig:haoverco}).

\noindent{\bf Abell\,S1101 -} The CO(1-0) integrated emission map of Abell\,S1101 (Fig.~\ref{fig:haoverco}), shows a filament that extends up to 14.4~kpc (13$\arcsec$) in the south-north direction, which is also co-spatial with the strongest line-emitting regions of the warm ionized nebula.

\subsection{Total Mass and Mass Distribution of the Molecular Gas}
\label{sec:molgas}

\begin{table*}
\caption{Fit parameters of the CO emission lines for different regions.}
\label{tab:prop_molecular}\centering
\begin{tabular}{lccccccc}
\noalign{\smallskip} \hline \hline \noalign{\smallskip}
\rm  Source							& $\rm S_{CO}\Delta\nu$	 	    & v$_{\rm offset}$	& FWHM 			& Peak			& L$^\prime_{\rm CO}$	                       & M$_{\rm gas}^{a}$      &$\rm \Sigma_{gas}$\\
\rm     							&  (Jy~km~s$^{-1}$) & (km~s$^{-1}$)		&(km~s$^{-1}$)	&(mJy)			& (10$^{8}$~K~km~s$^{-1}$~pc$^{2}$)  	&(10$^{8}$~M$_{\odot}$) &   (M$_{\odot}$~pc$^{-2}$)\\
    \noalign{\smallskip} \hline \noalign{\smallskip}
    
 Centaurus 						        &	4.1$\pm$0.1     &   -2$\pm$1    &  72$\pm$2	    &  15.1$\pm$0.2     &  0.191$\pm$0.005	&  0.88$\pm$0.02	   & 0.49$\pm$0.01\\
   \hspace{2.0mm}Central component	    &   2.3$\pm$0.1     &   11$\pm$6    &   160$\pm$14 	&  13.9$\pm$0.2  	&  0.107$\pm$0.005  &  0.49$\pm$0.02  	 & 0.89$\pm$0.03\\
   \hspace{2.0mm}Southern Filament 1 	&   1.4$\pm$0.1   	&  242$\pm$7    &   99$\pm$16 	&  12.6$\pm$0.2 	&  0.065$\pm$0.005 	&  0.30$\pm$0.02 	& 0.38$\pm$0.02\\
   \hspace{2.0mm}Southern Filament 2	&  	0.7$\pm$0.1     &  107$\pm$9    &   98$\pm$28	&  8.6$\pm$0.2	    &  0.032$\pm$0.005 	&  0.15$\pm$0.02 	& 0.50$\pm$0.07\\
   
    \noalign{\smallskip} \hline \noalign{\smallskip}
    
  RXJ1539.5     						&  10.5$\pm$0.4	&  27$\pm$3 		&  195$\pm$2		&  41.2$\pm$0.2  	&  26.0$\pm$1.1	    	&  119.7$\pm$4.9    & 18.8$\pm$1.0\\
     									&  0.4$\pm$0.2	&  -250$\pm$3 		&  94$\pm$5		&  5.2$\pm$0.2  	&  1.0$\pm$0.5	    	&  4.9$\pm$2.5      & 0.8$\pm$0.4\\
  \hspace{2.0mm}Central Component		&  7.8$\pm$0.3 	&  40$\pm$2 		&  231$\pm$2	&  31.2$\pm$0.2  	&  21.1$\pm$0.8	    	&  97.2$\pm$3.7     & 39.9$\pm$1.5\\
  \hspace{2.0mm}Western Filament 		&  0.4$\pm$0.1	&  112$\pm$4 	  	&  131$\pm$9		& 3.6$\pm$0.2	    &  10.8$\pm$0.3 	    &  4.9$\pm$1.2 		& 4.3$\pm$1.1\\
   		 		 		 		 		&  0.3$\pm$0.1	& -250$\pm$42 	  	&  98$\pm$9		&  2.8$\pm$0.2	    &  0.8$\pm$0.3 	    	&  3.7$\pm$1.2 		& 3.3$\pm$1.1\\
  \hspace{2.0mm}Eastern Filament 		&  1.9$\pm$0.2	&  33$\pm$1   		& 124$\pm$21		&  14.8$\pm$0.2	 	&  23.7$\pm$0.5 		&  6.2$\pm$1.3      & 10.8$\pm$1.1\\
  \hspace{2.0mm}Northern Filament 		&  0.2$\pm$0.1	& -276$\pm$9   		& 144$\pm$21 	&  1.6$\pm$0.2		&  0.5$\pm$0.3 	    	&  2.5$\pm$1.2 	    & 3.93$\pm$1.9\\

    \noalign{\smallskip} \hline \noalign{\smallskip}
    
Abell\,S1101 							&      1.3$\pm$0.1   &   29$\pm$5			&  112$\pm$9	&  4.8$\pm$0.2		&  1.9$\pm$0.2	        &  8.9$\pm$0.7     & 14.2$\pm$1.0\\
							            &      0.3$\pm$0.1   & -146$\pm$28			&  209$\pm$54	&  0.2$\pm$0.1		&  0.4$\pm$0.1	        &  2.0$\pm$0.7     & 3.2$\pm$1.0\\
  \hspace{2.0mm}Northern Clump			&      0.8$\pm$0.1   &   35$\pm$ 9			&   98$\pm$14 	&  2.5$\pm$0.2 	    &  1.2$\pm$0.2		    &  5.4$\pm$1.4		& 20.1$\pm$5.2\\
  \hspace{2.0mm}Southern Clump      	&      0.6$\pm$0.2   &   -35 $\pm$10 	    &   98$\pm$13 	&  2.5$\pm$0.5 	    &  0.9$\pm$0.3	        &  4.1$\pm$1.4	   	& 11.5$\pm$3.8\\
                                    	&      0.2$\pm$0.1   &   -209 $\pm$79 	    &   210$\pm$54 	&  0.9$\pm$0.2 	    &  0.3$\pm$0.1	        &  1.4$\pm$0.7	   	& 9.5$\pm$1.9\\
  \hspace{2.0mm}Eastern Clump			&      0.04$\pm$0.01 &   -174 $\pm$50 	    &   244$\pm$94 	&  0.01$\pm$0.2 	&  0.06$\pm$0.02	  	&  0.28$\pm$0.06	& 1.7$\pm$0.4\\

    \noalign{\smallskip} \hline \noalign{\smallskip}\\
\end{tabular}\\
\small
Notes: All spectra have been corrected for the response of the primary beam.
\raggedright $^{a}$ The molecular gas mass was calculated from the CO(1-0) integrated intensity as described in section~\ref{sec:molgas}.
\end{table*}

We fitted each CO line with one or two Gaussian profile extracted from a rectangular aperture encompassing all emission with a significance $\geq$3$\sigma$ from the primary beam corrected datacube. Values of (i) the integrated emission, $\rm S_{CO}\Delta\nu$; (ii) the line FWHM; (iii) the velocity offset relative to the systemic velocity of the BCG, v$_{\rm offset}$; (iv) the peak emission flux density; (v) the CO(1-0) line luminosity, L$_{\rm CO}$; (vi) molecular masses, M$_{\rm gas}$; and (vii) molecular surface densities, $\Sigma_{\rm gas}$ are all listed in Tab.~\ref{tab:prop_molecular}. Those values were compared with the \textsc{specflux} task in CASA, recovering similar values within about 10--30\%. The molecular gas masses for the three ALMA sources, Centaurus, RXJ1539.5 and Abell\,S1101, were calculated from the integrated CO(1-0) intensity under the assumption of a Galactic CO-to-H$_{2}$ conversion factor, $X_{\rm CO}$, as,
\begin{equation}
M_{\rm mol} = 1.05 \times 10^{4}~X_{\mathrm{CO}}~\left(\frac{1}{1+z}\right) \left(\frac{S_{\mathrm{CO}} \Delta \nu}{\mathrm{Jy~km s^{-1}}}\right) \left(\frac{D_{\mathrm{L}}}{\mathrm{Mpc}} \right)^{2} M_{\odot},
\label{eq:molmass}
\end{equation}

\noindent
where $X_{\rm CO}$ = 2$\times$10$^{20}$~cm$^{-2}$~(K~km~s$^{-1}$)$^{-1}$ as is found for the Milky Way disk \citep{solomon87}, $D_{\rm L}$ is the luminosity distance, and $z$ is the redshift of the BCG. The major uncertainty in the molecular mass comes from the conversion factor, $X_{\rm CO}$, since that the Galactic conversion factor is not expected or observed to be universal \citep[see, ][for reviews]{bolatto13, vantyghem18}. The value of the conversion factor depends on the metallicity and whether or not the CO emission is optically thick. Similar CO-to-H$_{2}$ conversion factors were used for the molecular mass calculation in previous studies \citep[e.g.,][]{mcnamara14, russell14, russell17, tremblay18}. We compared the total fraction of molecular gas lying inside filaments, i.e., the gas within coherent structures outside of the main nuclear peak of emission. The integrated CO(1-0) intensities values used are those listed in Tab.~\ref{tab:prop_molecular}.

In Centaurus, the main central gas peak accounts for $\sim$50\% of the total molecular mass and has a molecular mass of (4.9$\pm$0.2)$\times$10$^{7}$~M$_{\odot}$. The southern filament 1 and contain (3.0$\pm$0.2)$\times$10$^{7}$~M$_{\odot}$ and (1.5$\pm$0.2)$\times$10$^{7}$~M$_{\odot}$, respectively, resulting in a total molecular mass of $\sim$9.8$\times$10$^{7}$~M$_{\odot}$ in the filaments.
This is consistent with the value of $\sim$10$^{8}$ M$_{\odot}$ found by \citep{fabian16}, which was based on a 3.4$\sigma$ detection of CO(2-1) emission and by assuming a CO(2-1)/CO(1-0) ratio of 0.7.

In RXJ1539.5, the main central peak, with a total mass of  (97.2$\pm$3.7)$\times$10$^{8}$ M$_{\odot}$, accounts for $\sim$75\% of the total molecular mass. The western filament contains (6.2$\pm$1.3)$\times$10$^{8}$~M$_{\odot}$. The lump of the eastern filament contains (23.6$\pm$2.5)$\times$10$^{8}$ M$_{\odot}$, and the smaller filament to the N of the central components has a molecular mass of (2.5$\pm$1.2)$\times$10$^{8}$~M$_{\odot}$, resulting in a total detected molecular gas mass of $\sim$1.3$\times$10$^{10}$~M$_{\odot}$.

For Abell\,S1101 the northern and southern clumps contain similar amounts of molecular gas. For the northern and southern clumps, we derived a molecular mass of (5.4$\pm$1.4)$\times$10$^{8}$~M$_{\odot}$ and (5.5$\pm$1.4)$\times$10$^{8}$~M$_{\odot}$, respectively. The mass of the the eastern clump is 0.28$\pm$0.06$\times$10$^{8}$~M$_{\odot}$. Summing up all the mass estimates, the total detected molecular mass is $\sim$10.8$\times$10$^{8}~$M$_{\odot}$ for this source.

We computed the molecular mass estimates and filamentary mass fractions in the same way for all sources studied, that is for the ones we observed and those from the archival data. We identified filaments in each of these sources. The molecular masses of each filament are listed in the Table~\ref{tab:sourceprop}. When necessary, we assumed a flux density line ratio CO(2-1)/CO(1-0)$=$3.2 \citep{braine92} and CO(3-2)/CO(1-0)~$\sim$~0.8 \citep{edge01,russell16} in Eq.~\ref{eq:molmass}. The molecular masses of the filaments in the sample lie in the range of 3$\times$10$^{8}$ to 5$\times$10$^{10}$~M$_{\odot}$, with an median value of $\sim$2$\times$10$^{9}$~M$_{\odot}$. The filaments typically account for 20--50\% of the total molecular mass and can be as high as 75 -- 100\% for those objects where the CO line emission is offset from the central BCG.
We emphasize that the cold molecular gas masses are likely to be lower limits, since it is highly probable that the CO emission from the filaments is more extended and may follow the distribution of the optical filaments. High sensitivity ALMA observations are needed to confirm that (see Section~\ref{sec:discussion}).

\subsection{ALMA molecular filaments in a sample of 15 BCGs}
\label{sec:morphology}

Figures~\ref{fig:moments}, \ref{fig:moments1}, \ref{fig:moments2} and \ref{fig:moments3} display the CO moment maps together with the X-ray and optical images from \textit{Chandra}, HST and MUSE observations. The moment maps of each object, i.e., the integrated intensity, line-of-sight velocity and velocity dispersion maps, were constructed by using the masked moment technique \citep{dame11}\footnote{https://github.com/TimothyADavis/makeplots}. This technique creates a Hanning smoothed three-dimensional mask that takes coherence into account in position-velocity space to exclude pixels that do not have significant signal but still captures weak emission that simply clipping by a constant value of $\sigma$ might miss. Zero, first, and second moment maps were created using this mask on the un-smoothed cube which recovers as much flux as possible while optimizing signal-to-noise ratio and keeping only the regions where the CO lines were detected with a significance greater than 3$\sigma$. The detailed description of the cold molecular gas distribution for each individual galaxy cluster\'s core can be found in Appendix~\ref{appendix:morph}.

As can be seen from Figure~\ref{fig:moments}, the CO morphology varies from system to system. We classify these morphologies into 2 distinct categories based on the distribution of their molecular emission: (1) those with extended filamentary molecular gas emission; and (2) with nuclear molecular emission only. The classification and properties of each system enumerated in Table~\ref{tab:sourceprop}.

\begin{figure*}[htbp!]
\centering
{\bf Category extended (1):} Extended molecular gas emission\\
\subfigure{\includegraphics[width=1.0\textwidth]{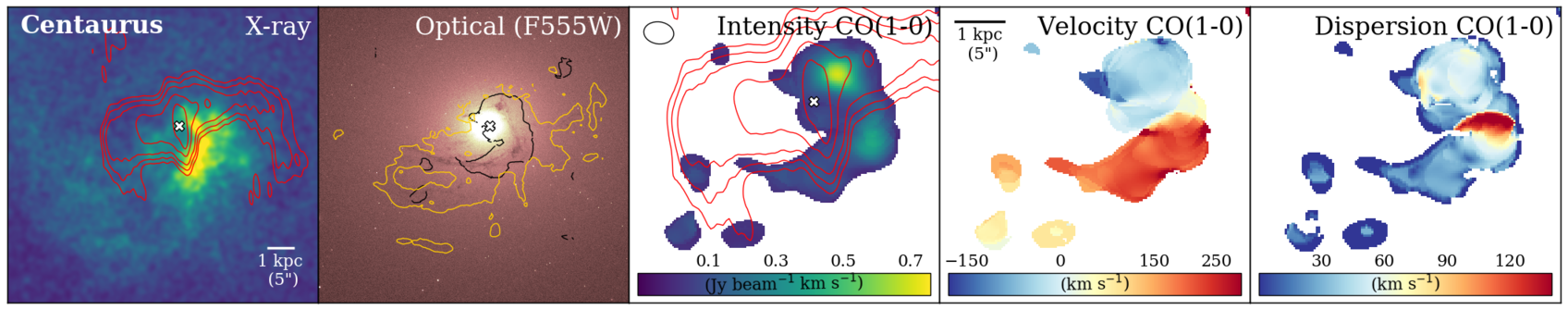}}\\ 
\vspace{-4mm}
\subfigure{\includegraphics[width=1.0\textwidth]{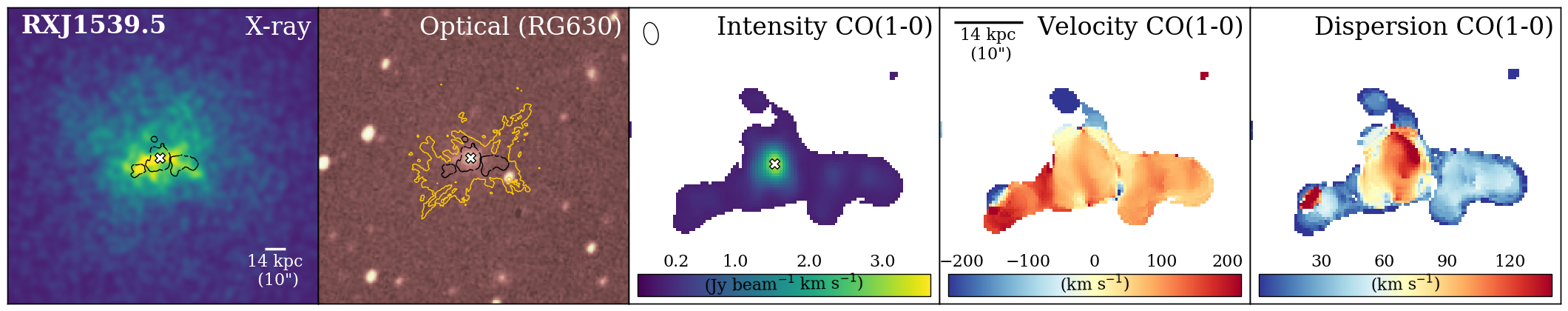}}\\ 
\vspace{-4mm}
\subfigure{\includegraphics[width=1.0\textwidth]{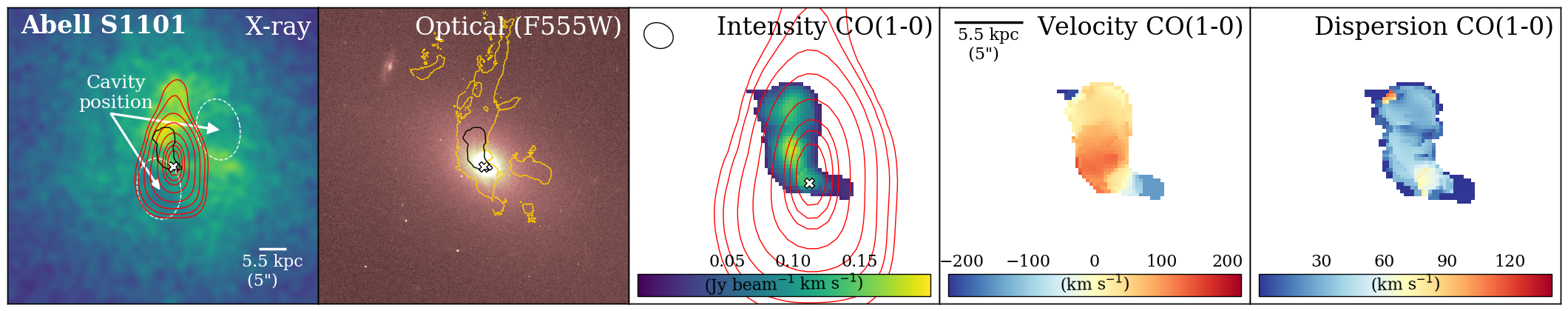}}\\ 
\vspace{4mm}

{\bf Category compact (2):} Nuclear molecular gas emission\\
\subfigure{\includegraphics[width=1.0\textwidth]{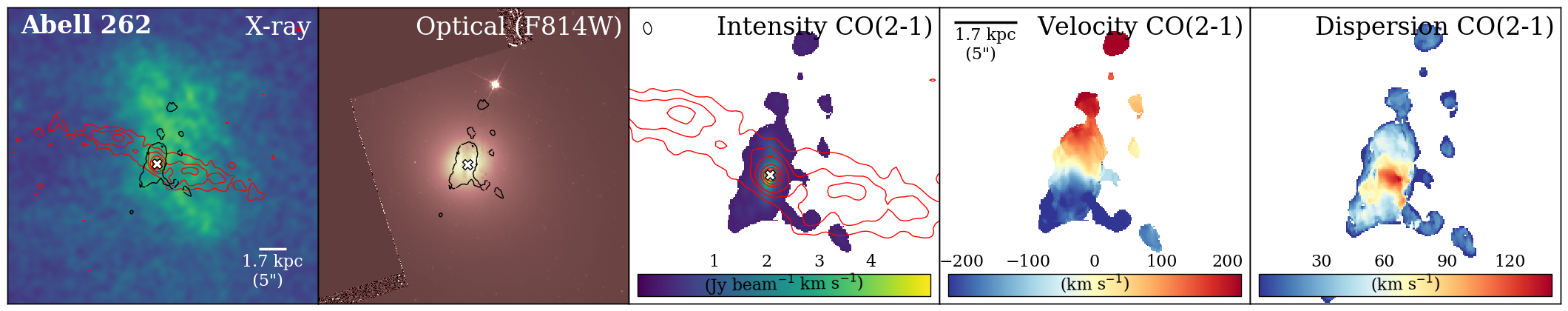}}\\ 
\vspace{-4mm}
\subfigure{\includegraphics[width=1.0\textwidth]{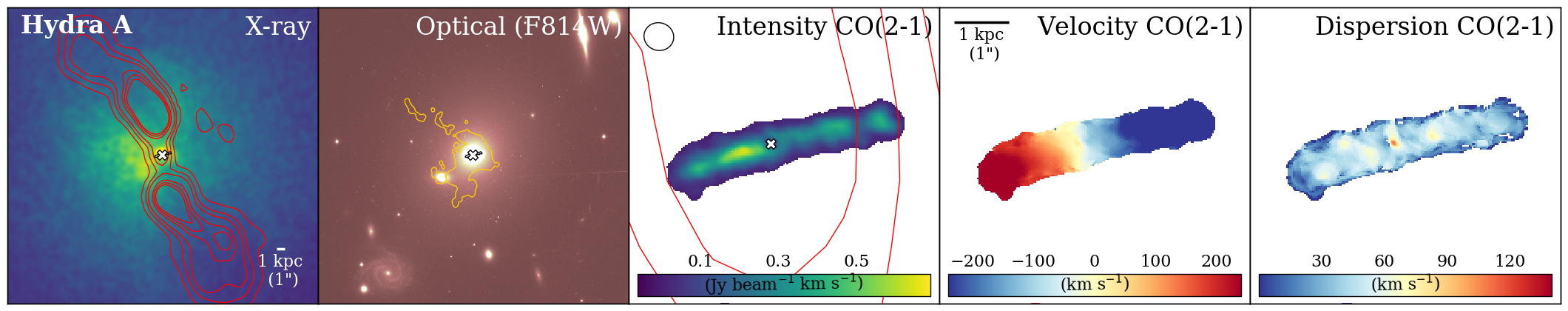}}\\ 
\caption{\textit{First panel of each row}: \textit{Chandra} X-ray image with VLA radio contours ovelaid in red. Dashed white lines indicate previously identified X-ray cavity positions. \textit{Second panel of each row}: Optical image from HST overlaid with black contours representing the distribution of the cold molecular gas.  We show H$\alpha$ emission distribution as orange contours. The first two panels cover the same area. \textit{Panels 3-5 from the left along each row}: The 3 panels are, in order from left to right, the first, second and third moment maps of the CO emission (i.e., intensity map, integrated intensity velocity map, and velocity dispersion map). In decending order, the data shown are for: Centaurus, RXJ1539.5 and Abell\,S1101, Abell\,3581, Abell\,3581, Abell\,1975, 2A~0335+096, RXJ0821+0752, PKS\,0745-191, Abell\,1835, Abell\,1664, M87, Abell\,2597, Phoenix-A from Category~extended (1), Abell\,262 and Hydra A from Category compact (2). All source names are indicated at the top right in the first panel of each row and what each image is, is given in the upper right of each panel. In the moment zero map, the red contours represent the radio emission from VLA observations from 1.45 GHz for Centaurus, 1.46 GHz for Abell\,S1101, 1.46 GHz for PKS\,0745-191, 1.45 GHz for Abell\,262 and 4.89 GHz for Hydra-A, 4.86 GHz for Abell\,2597, 4.91 GHz for M87, 4.19 GHz for Abell\,1795, 4.89 GHz for 2A~0335+096, 1.46 GHz for PKS\,0745-191. The white cross indicates the position of the mm-continuum source. The velocity dispersion sometimes appears high because of the presence of multiple velocity components; each component have small velocity dispersion.}
\vspace{4mm}
\label{fig:moments}
\end{figure*}

\begin{figure*}[htbp!]
\centering
\subfigure{\includegraphics[width=1.0\textwidth]{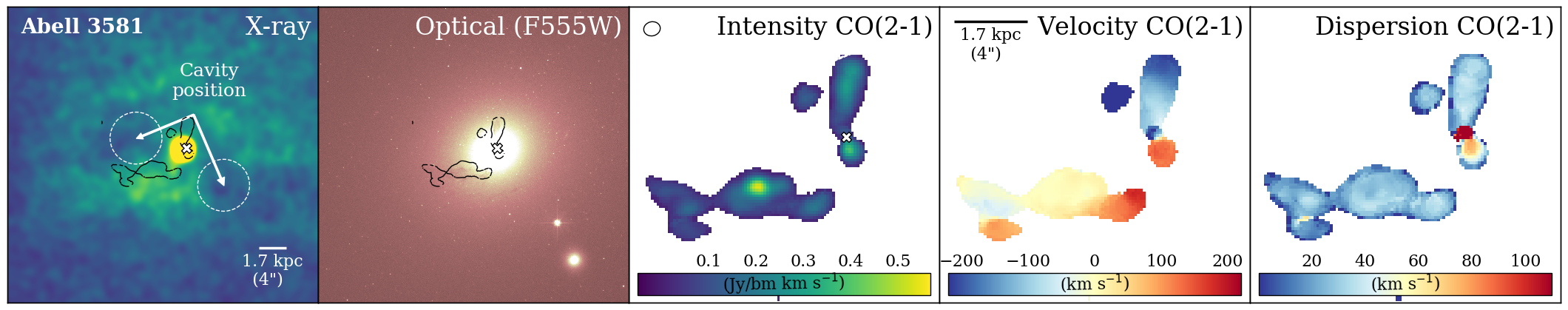}}\\ 
\vspace{-4mm}
\subfigure{\includegraphics[width=1.0\textwidth]{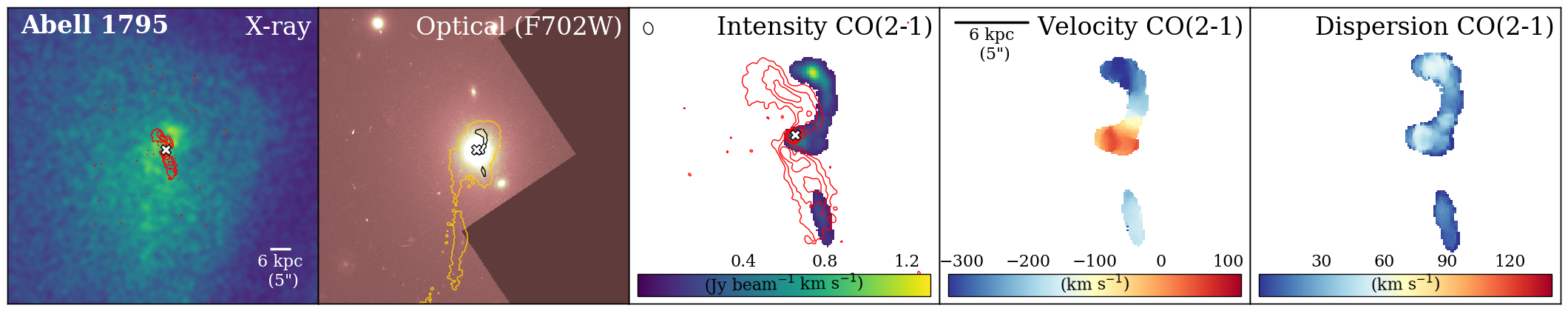}}\\ 
\vspace{-4mm}
\subfigure{\includegraphics[width=1.0\textwidth]{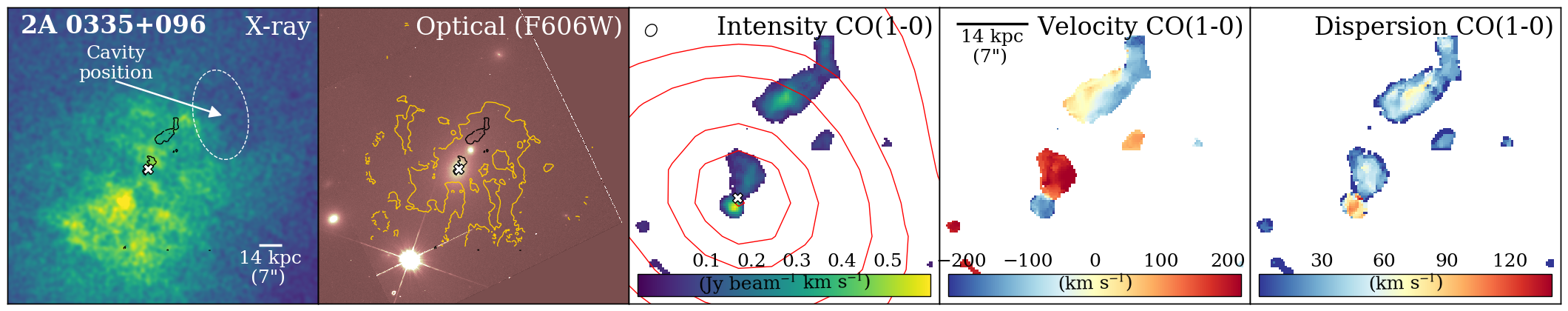}}\\
\vspace{-4mm}
\subfigure{\includegraphics[width=1.0\textwidth]{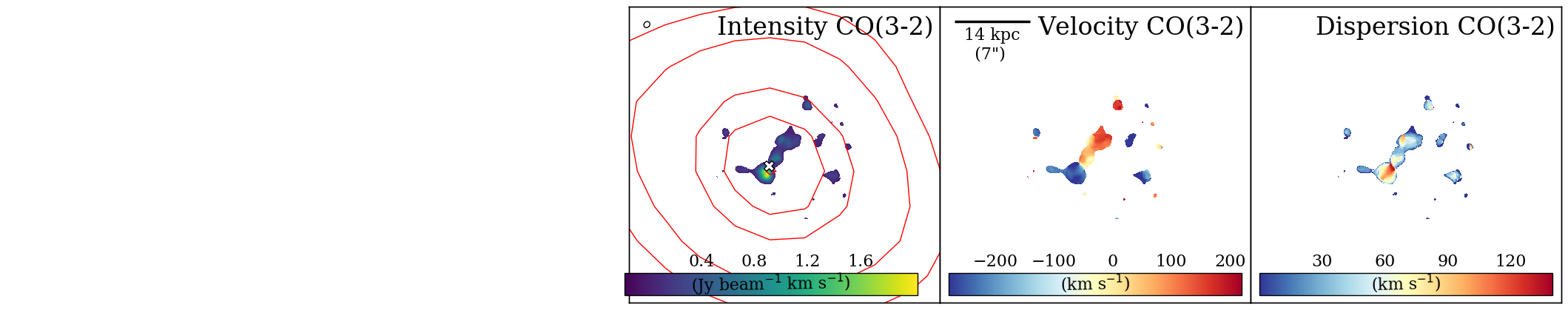}}\\
\vspace{-4mm}
\subfigure{\includegraphics[width=1.0\textwidth]{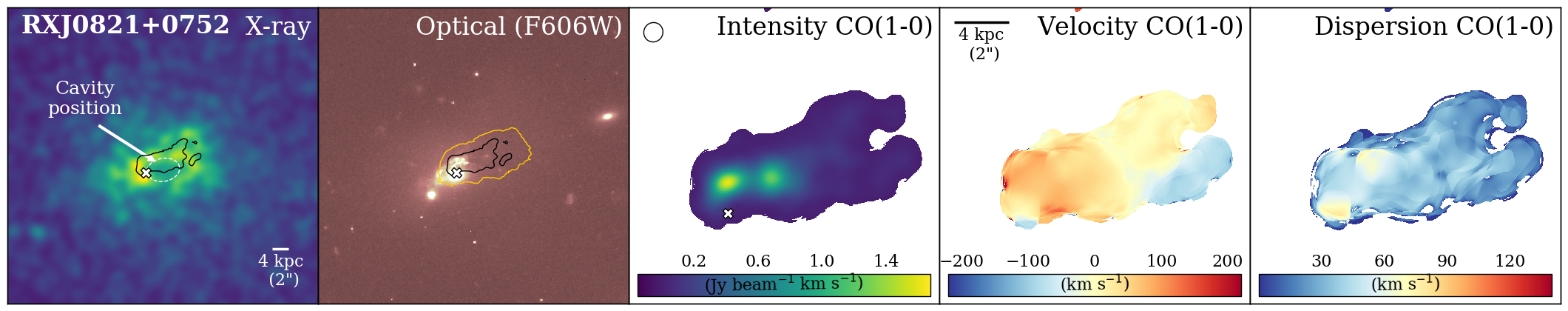}}\\ 
\vspace{-4mm}
\subfigure{\includegraphics[width=1.0\textwidth]{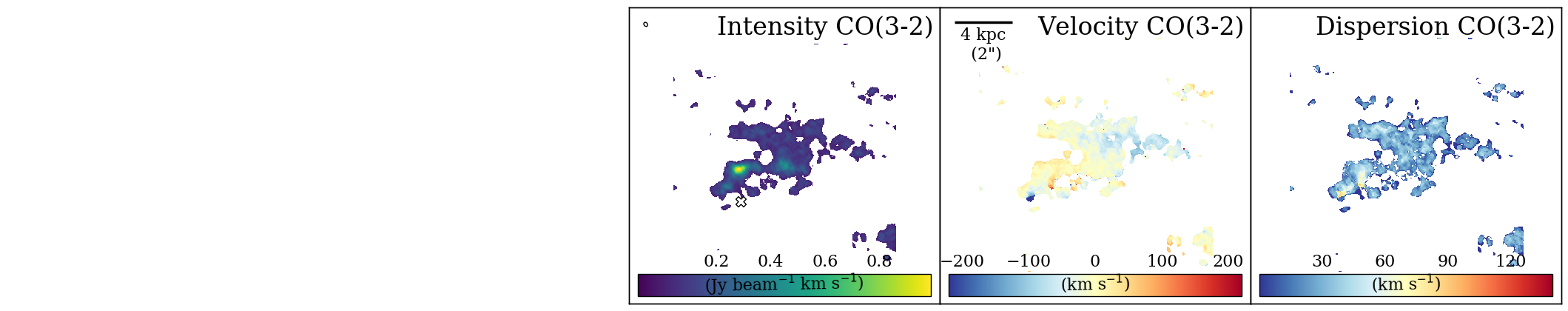}}\\ 
\caption{Continuation of Fig.~3 of systems classified as category~extended (1) (see text for details).}
\vspace{15mm}
\label{fig:moments1}
\end{figure*}

\begin{figure*}[htbp!]
\centering
\subfigure{\includegraphics[width=1.0\textwidth]{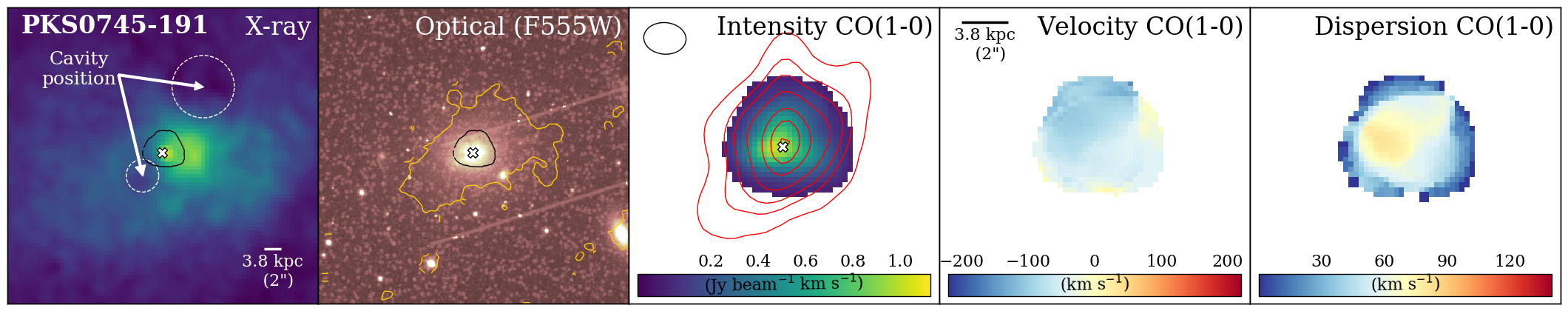}}\\ 
\vspace{-4mm}
\subfigure{\includegraphics[width=1.0\textwidth]{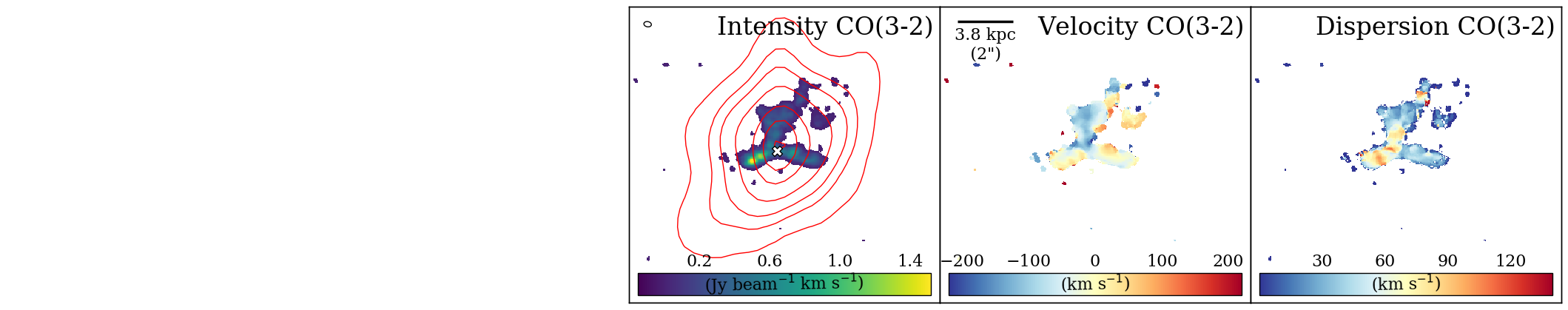}}\\ 
\vspace{-4mm}
\subfigure{\includegraphics[width=1.0\textwidth]{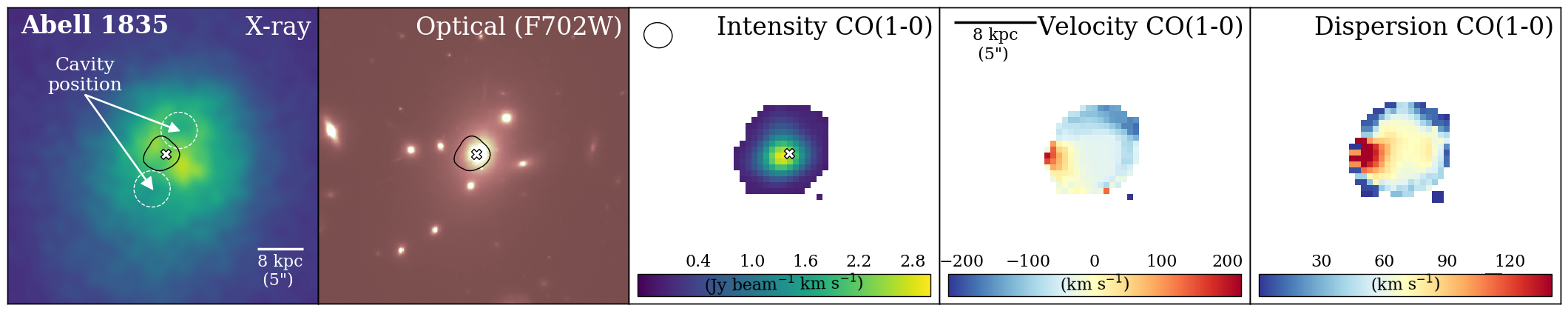}}\\ 
\vspace{-4mm}
\subfigure{\includegraphics[width=1.0\textwidth]{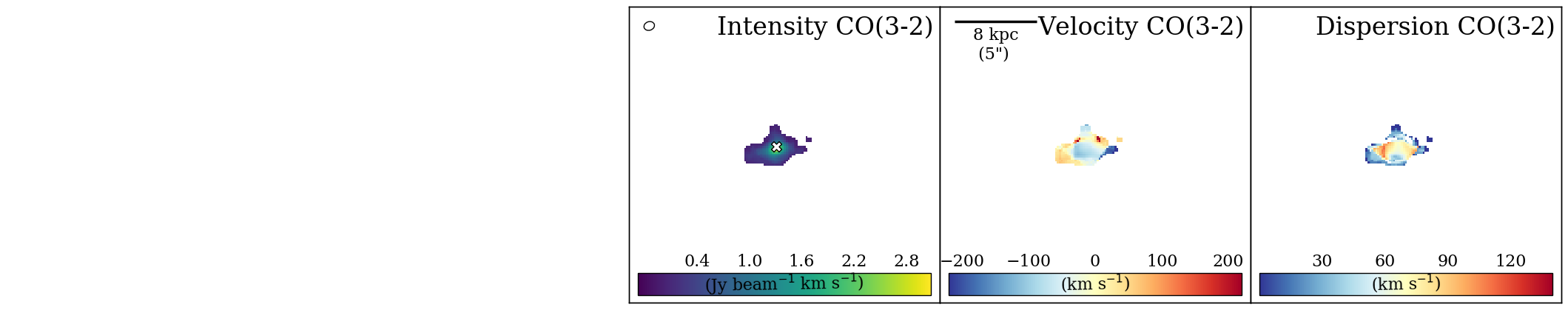}}\\ 
\vspace{-4mm}
\subfigure{\includegraphics[width=1.0\textwidth]{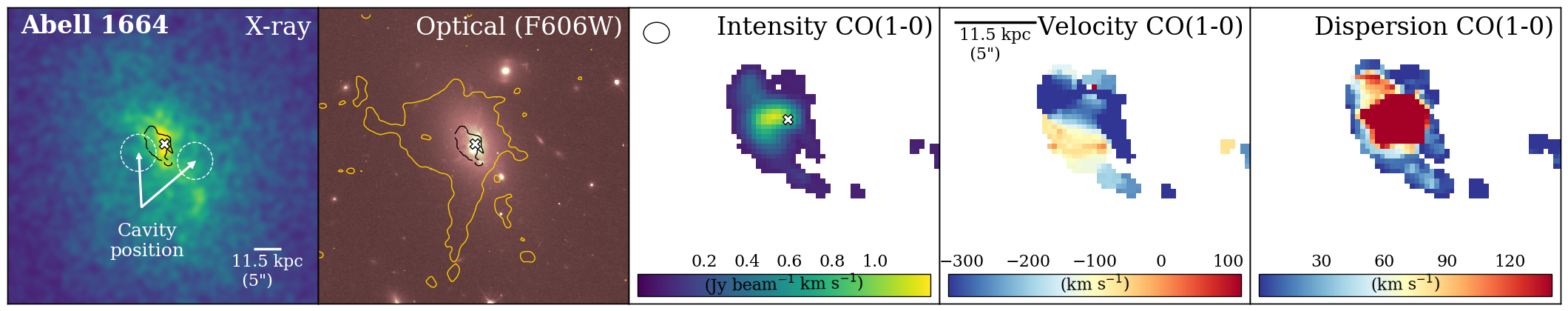}}\\ 
\vspace{-4mm}
\subfigure{\includegraphics[width=1.0\textwidth]{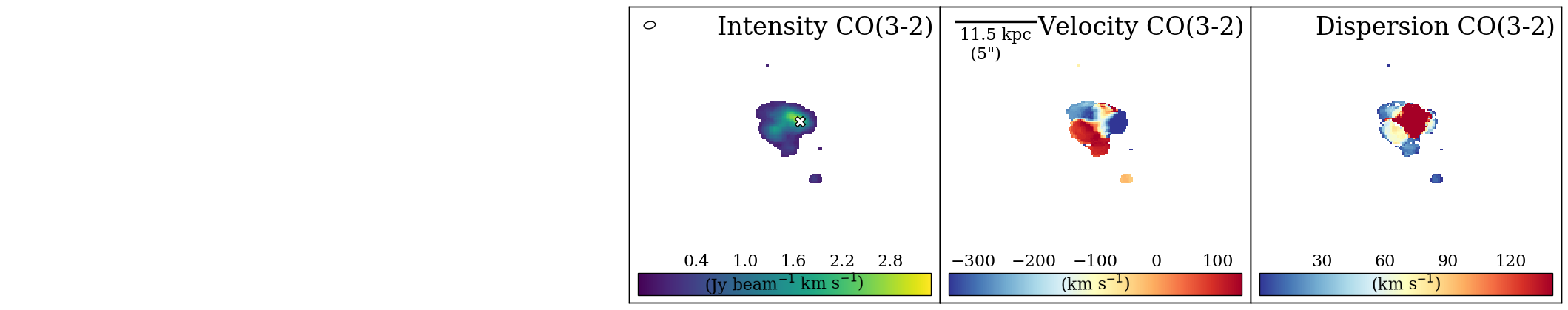}}\\ 
\caption{Continuation of Fig.~3 of systems classified as category~extended (1) (see text for details).}
\vspace{15mm}
\label{fig:moments2}
\end{figure*}

\begin{figure*}[htbp!]
\centering
\subfigure{\includegraphics[width=1.0\textwidth]{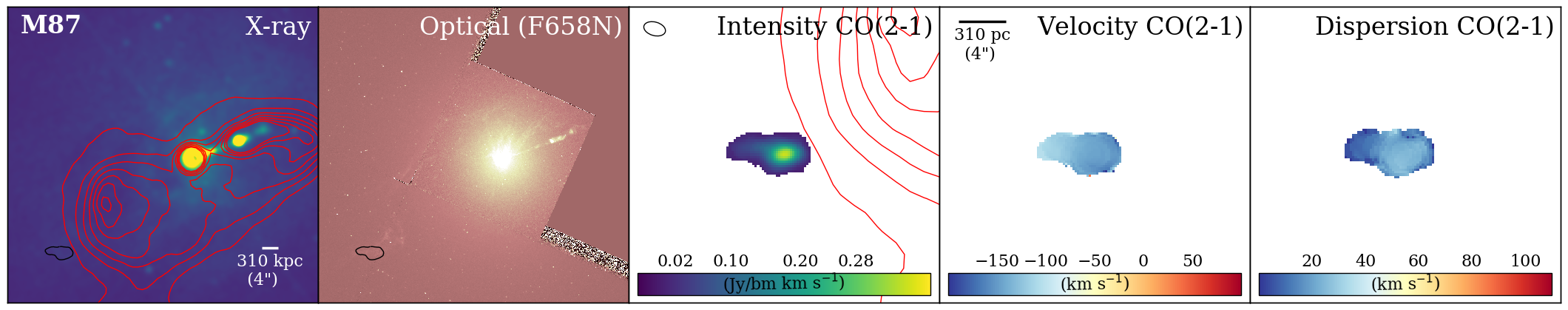}}\\ 
\vspace{-4mm}
\subfigure{\includegraphics[width=1.0\textwidth]{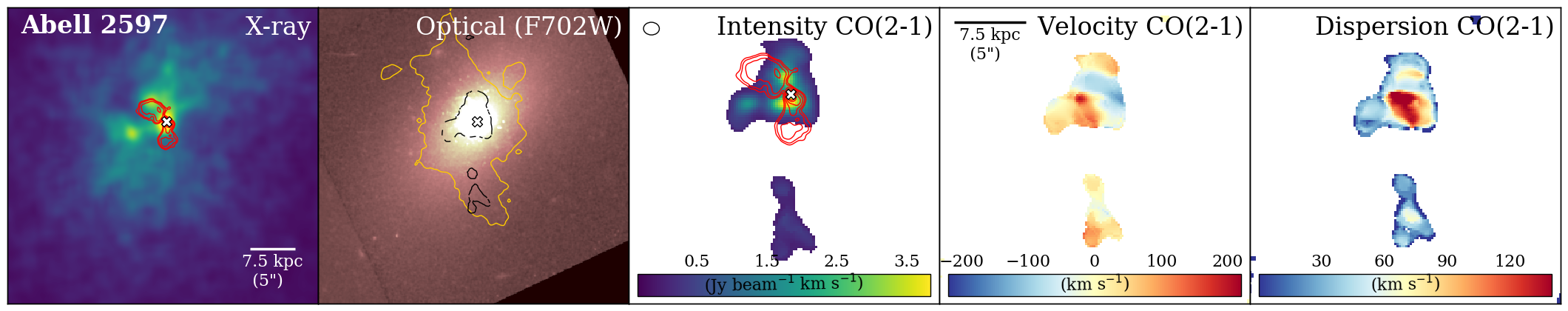}}\\ 
\vspace{-4mm}
\subfigure{\includegraphics[width=1.0\textwidth]{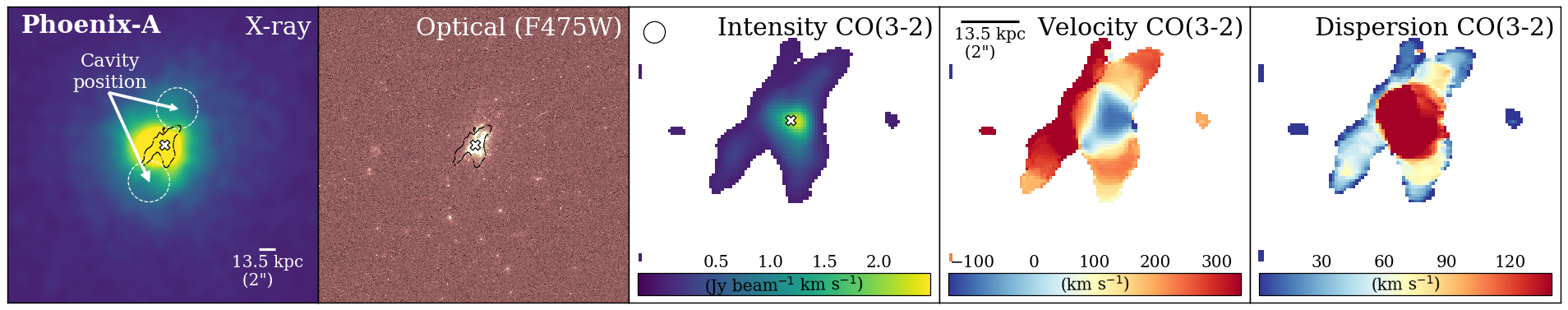}}\\
\caption{Continuation of Fig.~3 of systems classified as category~extended (1) (see text for details).}
\label{fig:moments3}
\end{figure*}


\noindent{\bf Extended molecular gas emission: Category (1) Extended Sources} \label{sec:extended}
Thirteen sources out of 15 were found with complex and extended cold molecular filamentary morphologies (see Fig.~\ref{fig:moments},\ref{fig:moments1},\ref{fig:moments2},\ref{fig:moments3}). A position-velocity diagram was extracted from each data-cube along coherent physical structures to measure the filament length using the Python package \textsc{PV Extractor}. The filaments extend over 2--25~kpc, with a mean length of $\sim$11--12~kpc and, in general, are located preferentially around the X-ray cavities seen with \textit{Chandra}, or at radii less than the extended radio lobes. The cold molecular gas sometimes peaks in the nuclear region of the central galaxy, e.g., Abell\,2597 and RXJ1539.5.
Some other extended systems show an offset between the CO peak position and the location of BCG. In these sources, most of the cold gas lies outside the position of the BCG, e.g., RXJ0820.9+0752 and M87. Some others have two CO peaks, with the second CO peak outside the BCG which contains a majority of the gas (>50\%), e.g., Abell\,3581 and Abell\,1795; and sometimes a similar fraction of the central CO peak, e.g., Centaurus, Abell\,3581 and Abell\,S1101. 

One striking feature is that the filaments are very massive, M$_{\rm gas}$=few$\times$10$^{8}$--10$^{10}$~M$_{\odot}$ or 0.25-50$\times$ the molecular gas mass of the Milky-Way. The filaments encompass at least 20--50\% of the total molecular mass, or even 75--100\% for those objects that are highly offset relative to the central BCG or lack a central emission component. The clumpy molecular filaments detected have a mean length of 11--12~kpc taking into account all the CO transitions and a width smaller than 0.5--3.5~kpc (the filaments are not resolved). The spatial resolution of the ALMA observations is insufficient to resolve the filamentary structures at 60~pc scales observed in H$\alpha$ from high resolution HST imaging of Centaurus \citep{fabian16} or even the lower resolution H$\alpha$ data, 100-500~pc, of Perseus \citep{conselice01}.

For four objects of our sample the molecular gas is projected onto the nuclear region of the host galaxy, e.g., Abell\,1664, Abell\,1835, PKS\,0745-191 and Phoenix-A. However, the low spatial resolution of our data clearly prevents us from resolving more structures in those objects. Higher angular resolution interferometry of these objects would likely show the presence of filaments. PKS\,0745-0191 is a nice example of an unresolved source in CO(1-0) that in fact host 3 clear filaments when observed in CO(3-2) with a smaller synthesized beam.

The low-redshift systems, those with $z<0.1$, e.g., Centaurus, Abell\,3581, and Abell\,S1101, all have low SFR and AGN power, 1--37~M$_{\odot}$~yr$^{-1}$, P$_{\rm cav}$<160$\times$~10$^{42}$~erg~s$^{-1}$ and cold molecular masses below $\sim$10$^{10}$~M$_{\odot}$. Objects which lie at higher redshifts, $z>$0.1, have, in general, higher cold molecular masses, M$_{\rm gas}$=1--5$\times$10$^{10}$~M$_{\odot}$, higher SFR, 13--600~M$_{\odot}$~yr$^{-1}$, and more powerful AGNs with P$\rm_{cav}\sim$100--4700$\times$10$^{42}$~erg~s$^{-1}$  (Table~\ref{tab:sourceprop}).
This is a clear bias due to the small size of the current sample and the lack of sensitivity of the current observations necessary to detect more compact and less powerful clusters at large distances.

\noindent{\bf Nuclear molecular emission: The Category (2) Compact Sources}\label{sec:disks}
Two systems, Abell\,262 (its BCG is NGC~708) and Hydra-A \citep{rose19}, show a compact cold molecular kpc-scale disks with isophotal sizes of $\sim$1.2~kpc and 2.4--5.2~kpc respectively. The molecular gas masses in these disks are 3.4$\times$10$^8$ M$_{\odot}$ and 5.4$\times$10$^9$ M$_{\odot}$, which are 3--10 times the mass of molecular gas in the filaments found in BCGs at roughly the same redshift. Both systems show a double peaked line profile, a clear gradient in their velocity maps. This is characteristic of rotating disks and they have maxima velocity differences across their line emission between 400--600~km~$s^{-1}$ (Fig.~\ref{fig:moments}). We note that the rotation is in a plane perpendicular to the projected orientation of the radio jets from the VLA observations (Fig.~\ref{fig:moments}). 


\subsection{Velocity Structure of the Molecular Gas}
\label{sec:velmoments}

Figure~\ref{fig:moments} shows the maps of the mean line-of-sight velocity (moment 1) and velocity dispersion of each system (moment 2). The detailed descriptions of the velocity structures for each object are found in the Appendix~\ref{appendix:velo}.

\noindent{\bf Slow and disturbed motions}
There is a lack of relaxed structures such as rotating disks in the central core of the galaxy cluster apart for two systems: Abell\,262 and Hydra-A. The cold molecular gas in some of the galaxy cluster cores show disturbed velocities, e.g., RXJ1539, while others show smooth velocity gradients in sub-structures indicating inflow or outflow of gas. The possibility of superposition of several filaments in projection along the line of sight makes the interpretation of velocity fields challenging. Generally, the cold molecular gas has complex velocity fields, with gas found at projected line-of-sight velocities that span over 100--400~km~s$^{-1}$ (without taking into account the high-velocity components). These values are relatively large but still much lower than expected if the gas falls ballistically to the cluster center from a few to a few tens of kpc, e.g., $\sim$300 to $\sim$1000~km~s$^{-1}$. The velocity differences of several 100~km~s$^{-1}$ are likely due to the presence of distinct velocity components and presumably different physical structures. Indeed, velocity gradients inside coherent sub-structures are rather small (\mbox{$\sim$20~km~s$^{-1}$~kpc$^{-1}$} (Tab.~\ref{tab:filament_properties})). 



\noindent{\bf High velocity components} In a few objects we detected high-velocity components (which we term high velocity systems or HVS): in both Abell\,1664 \citep{russell14} and Abell\,3581 these are at $-$570~km~s$^{-1}$ relative to the systemic velocity, and in Phoenix-A at $+$600~km~s$^{-1}$. In Abell\,1664, the velocity of the cold gas in the HVS increases towards the center of the BCG and may be a flow close to the nucleus (inflow or outflow) \citep{russell16}. While, more interestingly, the velocity of the HVS in Abell\,3581 increases from west to east at a PA$\sim-$50~$\degr$ relative to the filament along the south to north direction where a young star cluster has been detected \citep{canning13}. This star cluster could be accelerating the cold gas (inflowing/outflowing) in the opposite projected direction to the radio bubbles from the AGN.


\noindent{\bf Velocity dispersion}
The velocity dispersion maps of most objects in the sample show a profile which peaks towards the centre of the system, $\sigma=$100--150~km~s$^{-1}$, and falls well below 50~km~$s^{-1}$ at larger radii where the intensity of the molecular gas declines. This is not surprising given that in the central region, the AGN jets and the BCG bulk motions may be able to inject more kinetic energy globally, disturbing the cold molecular nebula.
Interestingly some systems show additional peaks away from central galaxy where the gas is more kinematically disturbed, creating regions of high velocity dispersion, e.g. Abell\,1795 (SN curved filament) 2A 0335+096 (NE filament) and Centaurus (Plume filament), etc. In some objects in Figures~\ref{fig:moments1} to \ref{fig:moments3} the velocity dispersion may appear high only because there exist multiple velocity components.
Furthermore the velocity dispersions of molecular nebulae, in general, tend to be well below the velocity dispersions of the stars, where $\sigma_{\rm *}$ for a BCG is typically $\sim$300~km~s$^{-1}$ \citep{vonderlinden07} and up to 480~km~s$^{-1}$ in our sample.

\section{Discussion}
\label{sec:discussion}

\subsection{General Properties of the Cold Molecular Gas}
\label{sec:generalprop}
From the sample studied with ALMA we identify two distinct distribution categories. The nuclear molecular emission systems (Category compact) described earlier account for the smallest fraction with 2/15 objects, while, the filamentary distribution (Category extended) account for 13/15 objects. Category extended systems show, in general, unrelaxed-state structures with disturbed motions along the filaments. On the other hand, the two systems in Category compact are well described by relaxed structures showing ordered motions within a compact ($\sim$2--5~kpc length) thin ($\sim$~kpc) rotating disk located at the very center of the BCG. This cold gas disk may be an important step in driving the gas into the vicinity of the AGN. In an AGN-regulated scenario, one indeed expects that a fraction of the cooled gas will eventually fuel the SMBH to maintain the powerful jets which are injecting mechanical energy into the ICM. Note that both systems show multiphase structures and have similar SFRs, $\sim$0.8~M$_{\odot}$~yr$^{-1}$. Simulations found a stable rotating disk (or torus) at the center of the galaxy clusters \citep{gaspari12,li15,prasad18,prasad15,Gaspari_2018}, as a result of a large burst of cold gas formation.

For the systems in Category extended, the molecular distribution usually consists of a nuclear emission component closely related to the BCG\'s core, and a set of extended clumpy filaments. Those filaments are massive, with cold molecular masses of a few$\times$10$^{8}$--10$^{10}$~M$_{\odot}$, comprising 20--50\% of the total molecular mass, or even 75--100\% for those objects that are offset relative to the central BCG and with lack of central component. Something must explain that large amounts of cold gas (often the dominant part), are found offset from the BCG position. It may be that the gas has cooled in the IGM, far away from the BCG. Motions induced via BCG in the cluster potential wells (cooling-wake) or AGN-driven motions may contribute to displace the cooling gas out from the core of the central galaxy.\\

\noindent{\bf What is slowing down the cold molecular velocities?} As mentioned earlier, on average, the velocities of the cold clouds are slow through the molecular nebulae, lying in a range of 100--400~km~s$^{-1}$. 
Those molecular velocities are inconsistent with the scenario of simple freely in-falling gas, where higher velocities are expected, around $\sim$300--1000~km~s$^{-1}$. As infalling material is likely dynamically coupled to feedback outflows and subject to ram pressure, the gas infall time, t$_{\rm I}$, is generally larger than the free fall time, t$_{\rm ff}$ \citep{mcnamara16}. The infall time could be easily 2-3 times longer than t$_{\rm ff}$, so that the gas in its slow fall has time to entrain and cool more warm gas, and the corresponding ratio t$_{\rm cool}$/t$_{\rm I}$ might be more representative than t$_{\rm cool}$/t$_{\rm ff}$ of the physics involved.

The ram pressure is likely to have a significant influence, and we can compute an order of magnitude of the terminal velocity $V_{\rm t}$ acquired by a molecular cloud, of size $d_{\rm cl}$, and density $\rho_{\rm cl}$ falling in the ICM of density $\rho_{\rm ICM}$, at a distance $R$ from the center of the cluster/BCG. The ram-pressure force exerted on the cloud of surface S is $\rho_{\rm ICM}$ S $V_{\rm t}^{2}$, and by definition of the terminal velocity should be equal to the gravitational force exerted on the cloud of mass $m_{\rm cl} \sim \rho_{\rm cl} d_{\rm cl}$ S. The latter is equal to $g~m_{\rm cl}$, and $g$ = $V_{\rm r}{^2}/R$, where $V_{\rm r}$ is the circular velocity of the potential of the BCG at that radius R. Equating the two forces yields the equation:
\begin{equation}
\frac{V_{\rm t}{^2}}{V_{\rm r}{^2}} = \frac{\rho_{\rm cl}~d_{\rm cl}}{\rho_{\rm ICM}~R}
\end{equation}

For a typical GMC (Giant Molecular Cloud), or GMA (Giant Molecular Association) where N(H$_{2}$)$\sim$10$^{21}$~cm$^{-2}$, with a density $\rho_{\rm ICM}\sim$0.1~cm$^{-3}$ at R=10~kpc \citep[e.g.][]{hogan17b}, the right hand side is equal to unity, meaning that the terminal velocity is comparable to the equivalent rotational velocity in the potential of the BCG at 10~kpc, i.e. $\sim$300~km~s$^{-1}$. This means that the order of magnitude of the observed velocities appear perfectly compatible with a scenario where clouds reach the terminal velocity, and are almost floating freely (without acceleration)  in the ICM medium, slowly infalling onto the center of the BCG, and fueling the central AGN. Of course, this is true on average, with a lot of factors that can be varied, both on the density of the ICM, and column densities of the effective clouds, since there is a size and mass spectrum of molecular clouds (with however an average column density varying slowly with size, \citep[e.g.][]{solomon87, elmegreen96}. Numerical simulations of a multiphase gas have shown that the physical state and fragmentation structure  of the molecular clouds might be quite different in the environment of the hot atmosphere \citep{sparre19}: the clouds could be surrounded by a co-moving gas layer protecting them from destruction, and they could shield themselves by behaving like a single stream \citep{forbes-lin18}. 

The AGN may create an active wind, able to provide an additional ram-pressure against the in-falling molecular clouds in the cluster center (r$<$10 kpc). The efficiency of ram-pressure in slowing down the molecular velocities at small radii requires that the molecular filaments are ''thready``, i.e., that the cold filaments are moving within a skin of warmer gas \citep{li18,jaffe05}. Magnetic field of the ICM can also increase the drag force on the cold clouds decreasing its velocities \citep{mccourt15}. Chaotic motions driven by turbulence, may also cancel some of the extreme velocities, plus the superposition of inflowing and outflowing filaments along the line of sight can mix and cancel velocities structures.

\subsection{Evidence supporting that cold and ionized emission arise from the same ensemble of clouds.}
\label{sec:halpha_co}

\begin{figure*}[htbp!]
\centering
    \subfigure{\includegraphics[width=0.9\textwidth]{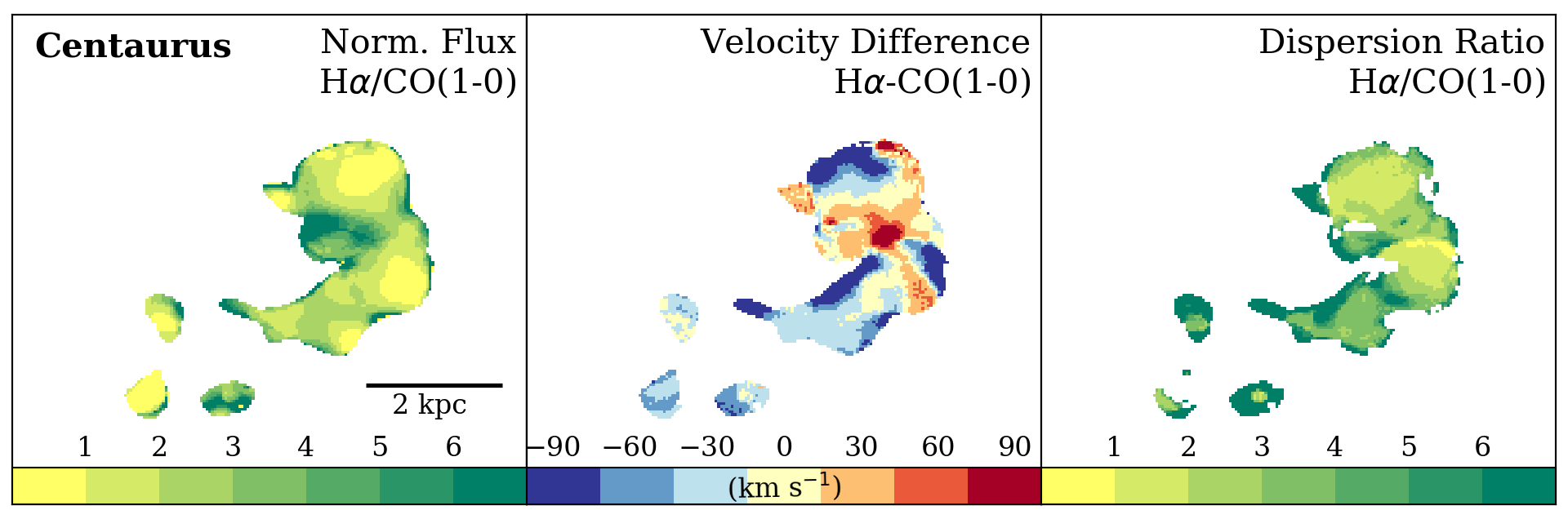}}\\
    \vspace{-2mm}
    \subfigure{\includegraphics[width=0.9\textwidth]{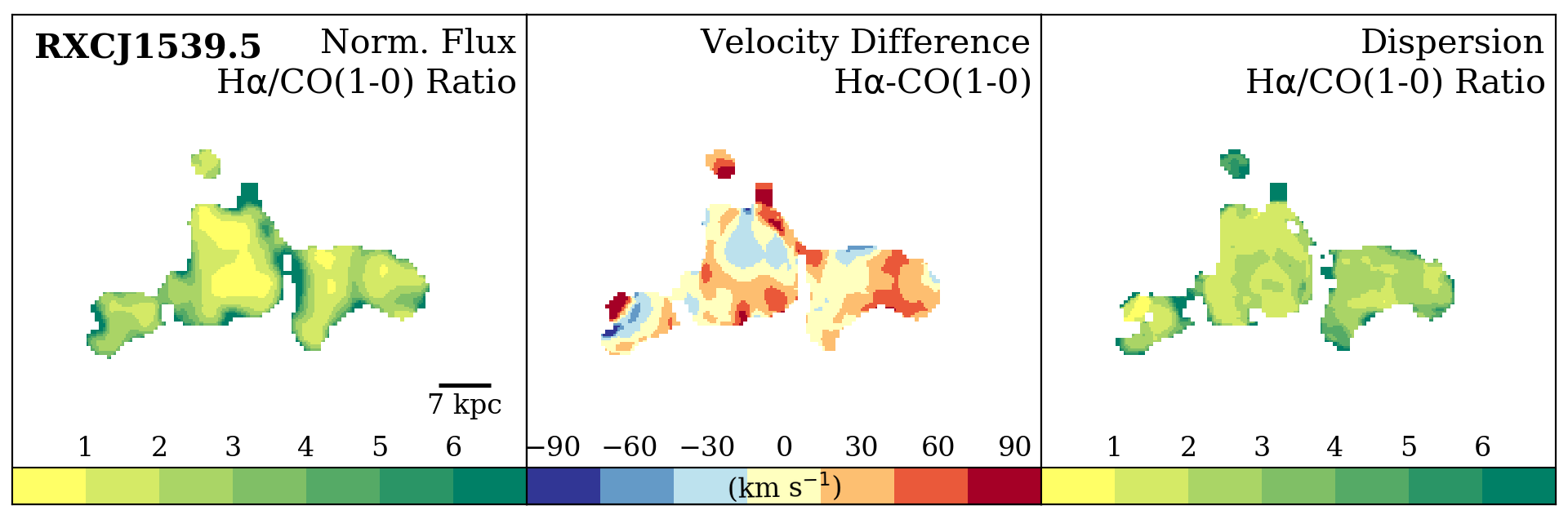}}\\
    \vspace{-2mm}
    \subfigure{\includegraphics[width=0.9\textwidth]{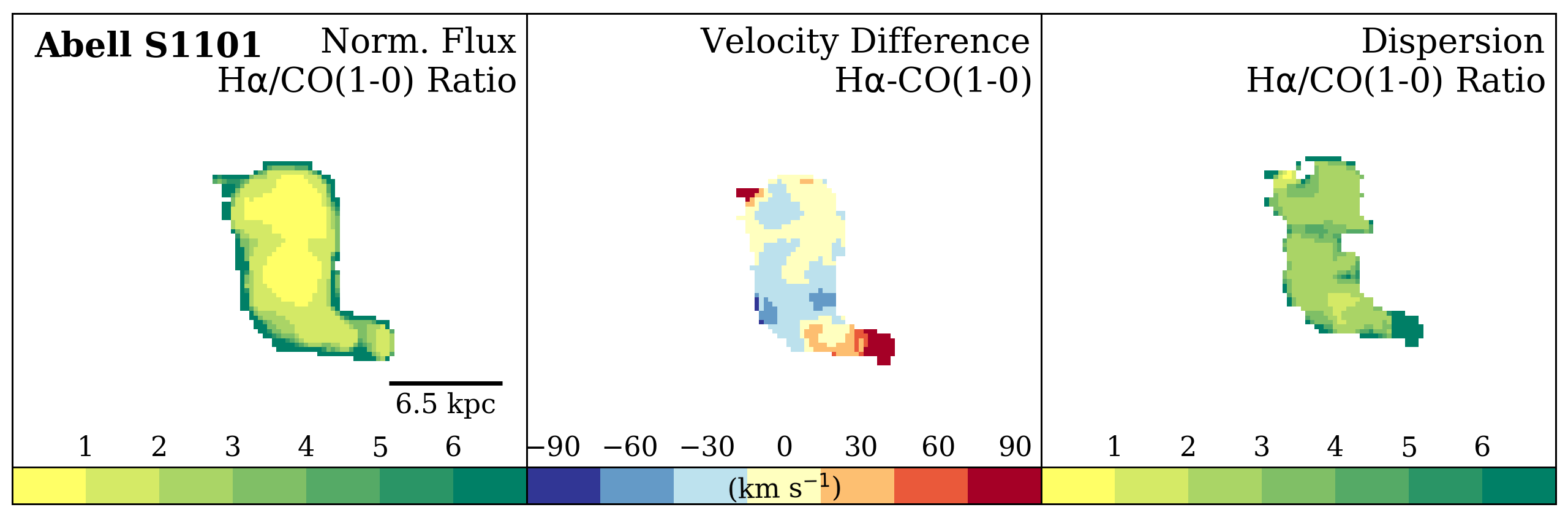}}\\
    \vspace{-2mm}
    \subfigure{\includegraphics[width=0.9\textwidth]{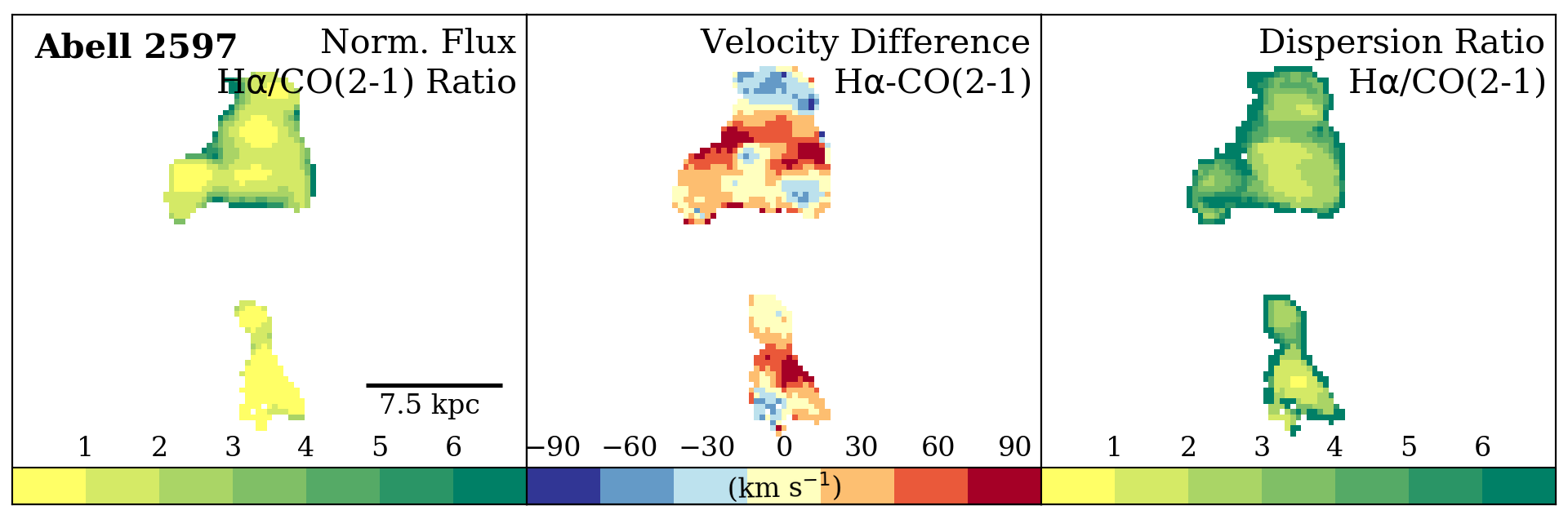}}\\
    \caption{Maps of the ratio between the H$\alpha$ and CO linear normalized flux ratio, line-of-sight velocity difference and velocity dispersion ratio for Centaurus, RXJ1539.5, Abell\,S1101, Abell\,2597, Abell\,3581, PKS\,0745-191, Abell\,1795, RXJ0820.9+0752, 2A0335+096, Hydra-A from ALMA and MUSE observations. Those maps were made by dividing the corresponding MUSE and ALMA moment maps, respectively. We have smooth the datacube to account for different spatial resolutions (see text for details).
    Furthermore, the velocity dispersion ratio map shows that, on average, the H$\alpha$ velocity dispersion is a factor of 2--3 times broader than that for CO.}
    \label{fig:haco}
\end{figure*}

\begin{figure*}[htbp!]
\centering
    \subfigure{\includegraphics[width=0.9\textwidth]{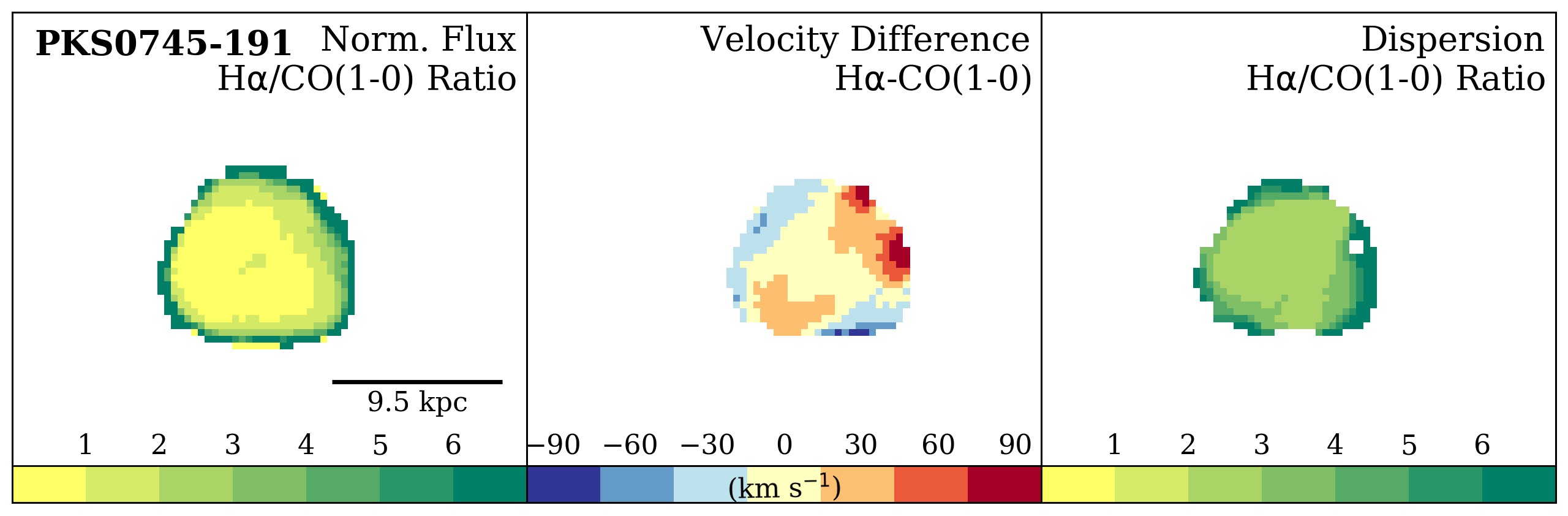}}\\
    \vspace{-2mm}
    \subfigure{\includegraphics[width=0.9\textwidth]{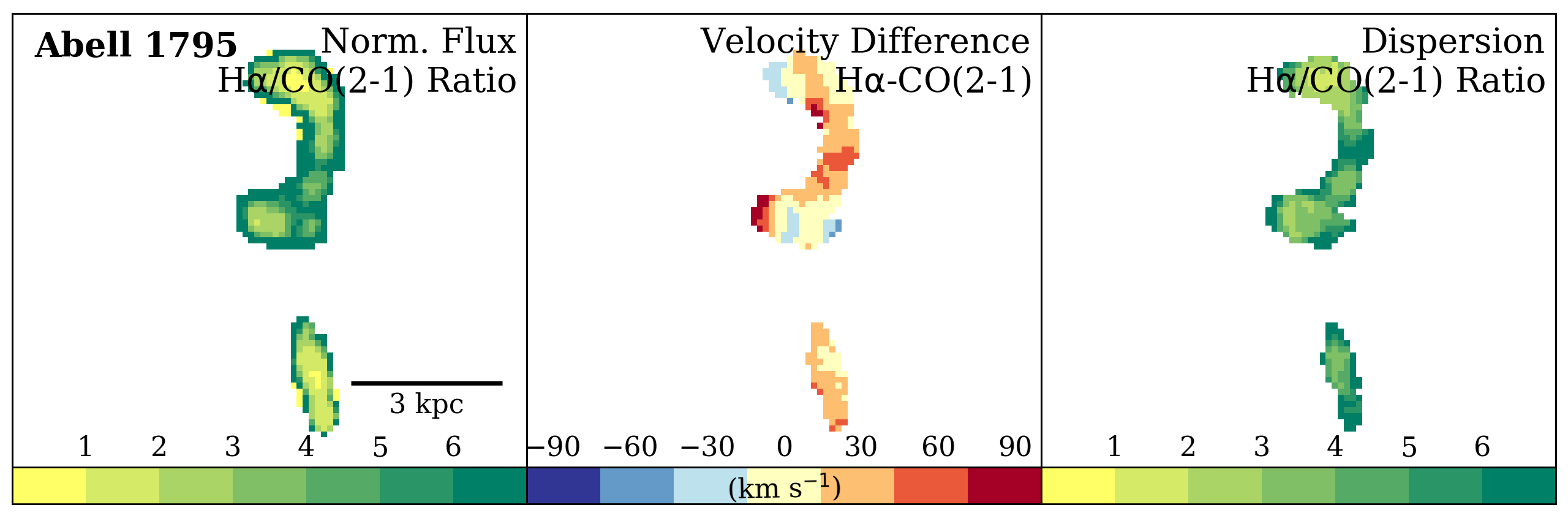}}\\
    \vspace{-2mm}
    \subfigure{\includegraphics[width=0.9\textwidth]{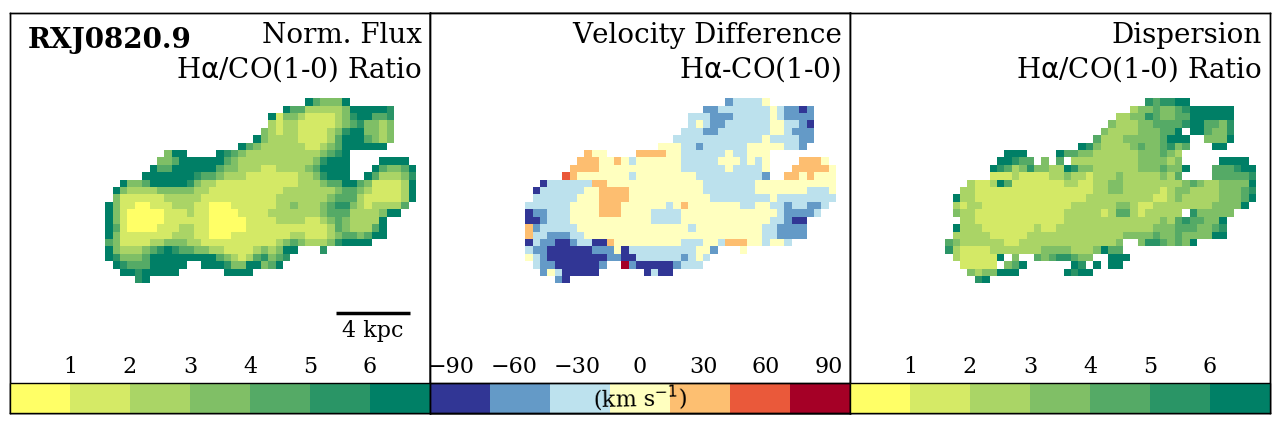}}\\
    \vspace{-2mm}
    \subfigure{\includegraphics[width=0.9\textwidth]{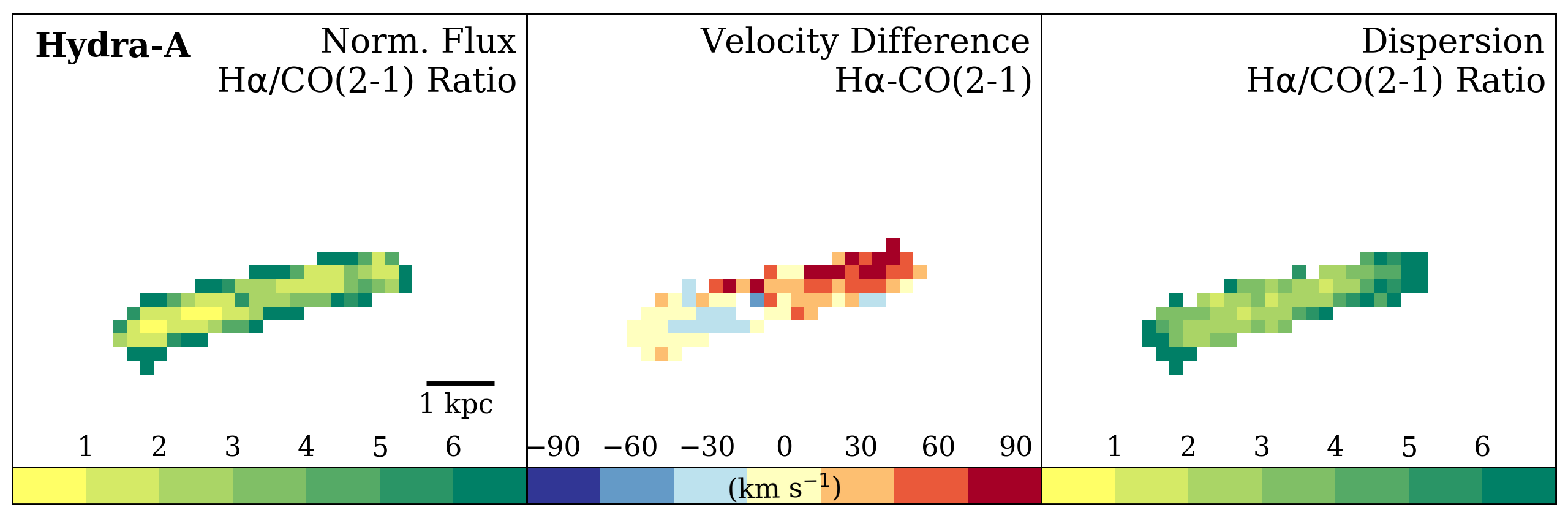}}\\
    \caption{Continuation of Fig.~7, with maps of the ratio between the H$\alpha$ and CO linear normalized flux ratio, line-of-sight velocity difference and velocity dispersion ratio.}
    \vspace{5mm}
    \label{fig:haco1}
\end{figure*}

Figures~\ref{fig:haco} and \ref{fig:haco1} show the relative characteristics of the H$\alpha$ and CO in three different ways: (i) normalized flux ratio (normalized by each emission peak); (ii) line-of-sight velocity difference; and (iii) velocity dispersion ratio for the sources that have both ALMA and MUSE observations (Table~\ref{tab:sample}).


The MUSE maps were smoothed to match the resolution of the ALMA data for each object and, using the \textsc{reproject} task from the \textsc{Astropy} package, we resampled the ALMA data onto the MUSE data pixel grids. We made a small WCS shifts in the ALMA maps, of below (x,y) = ($\pm$8, $\pm$8) pixels (or $\sim$1$\arcsec$) to match the MUSE intensity image, assuming that the CO and H$\alpha$ are aligned morphologically. Depending on science application, the re-projected ALMA image was then either divided directly by the MUSE map, or divided after normalization (e.g., the normalized flux) or rescaling by some other factor (for example, to convert pixel units). In the case of PKS\,0745-191, RXJ0820.9+0752 and 2A~0335+096, two molecular emission lines are available : CO(1-0) and CO(3-2). In order to compare with MUSE at larger scales, we choose the CO(1-0) emission line as the most reliable tracer of the molecular gas.


\noindent{\bf Molecular and ionized nebulae are co-spatial and co-moving.} The molecular gas is generally spatially distributed along the brightest emission from the warm ionized nebula (Section~\ref{sec:hatoco}). It also appears that the cold molecular and warm ionized gas also share the same overall velocity structure (Fig.~\ref{fig:haco}). The velocity difference maps, in general, are filled with velocity offsets well below 100~km~s$^{-1}$, which indicate that the H$\alpha$ and CO emissions regions are likely co-moving. Indeed, 100~km~s$^{-1}$ difference may be explained by the different Halpha and ALMA spatial resolutions (ALMA beam can go from 0.3 to 3 arcsec, i.e. from 0.3 to 4kpc),
and the V-gradient averaged out in a beam, can already account for 50-100~km~s$^{-1}$. The moment 2 CO  maps show maxima at $\sim$ 100~km~s$^{-1}$.

\noindent{\bf Molecular clouds have lower velocity dispersions than the ionized nebula.} The comparison between H$\alpha$ and CO(1-0) emissions indicates that the H$\alpha$ velocity dispersion is broader than the CO one by a factor of $\sim$2 (Fig.~\ref{fig:haco} and \ref{fig:haco1}). \citet{tremblay18} suggested that lines-of-sight are likely to intersect more warm gas than cold clouds, which will lead to a broader velocity distribution in the warm ionized gas than in the cold molecular gas. \citet{Gaspari_2018} have shown that warm gas is more likely to be turbulent with higher velocities compared to the cold molecular gas.
This is compatible with the millimeter and the optical emissions arising from the same structures: the filaments around BCGs are made of a large number of dense molecular clumps (and clouds). The cloud surfaces are then heated and excited by various mechanisms (intracluster UV radiation field, X-rays and ionizing photons from the AGN, X-ray emission from the ICM, etc) and are bright in optical emission lines. In the following, we argue that the excitation is due to processes that are local to the environment of the filament.

\noindent{\bf The ratio of the H$\alpha$ and CO fluxes is constant along the nebula.} The normalized H$\alpha$-to-CO flux ratios are close to unity all along the nebula (Fig.~\ref{fig:haco} and \ref{fig:haco1}). There is not a large radial variation of this ratio for all the BCGs studied apart from a few systems like Abell\,1795, Abell\,2597 and Abell\,3581 (Fig. \ref{fig:ratioshaco}). The lack of significant radial gradients in the flux ratios indicates a local excitation mechanism \citep{lim12}.

In a few rare cases, some variations are seen in the H$\alpha$-to-CO flux ratios, in a few cases the peak of the CO is not located at the same position as the peak in H$\alpha$, which is offset from the BCG center, having a high H$\alpha$-to-CO flux ratio$\approx$2--3 at this position. Otherwise, we observe a lack of H$\alpha$ emission in regions of bright CO emission. Recent star formation can consume the molecular gas and also increase the surface brightness of H$\alpha$ thereby increasing the H$\alpha$-to-CO flux ratio. Other mechanisms, such as photo-dissociation or evaporation by collisions with hot electrons, could reduce the amount of molecular gas available. Dust attenuation of the optical emission lines can reduce the H$\alpha$ flux, decreasing the H$\alpha$-to-CO flux ratio.

\noindent{\bf Could the molecular nebulae be as extended as the ionized filaments.} It is possible that the molecular filaments are as long as those seen in H$\alpha$, as previously observed in Perseus \citep{salome06,salome11}. ALMA observations detect about 50--80\% of the emission measured by single dish observations (e.g., IRAM 30m) and it is clear that the ALMA observations may be missing molecular gas.

\begin{figure}
    \resizebox{\hsize}{!}{\includegraphics{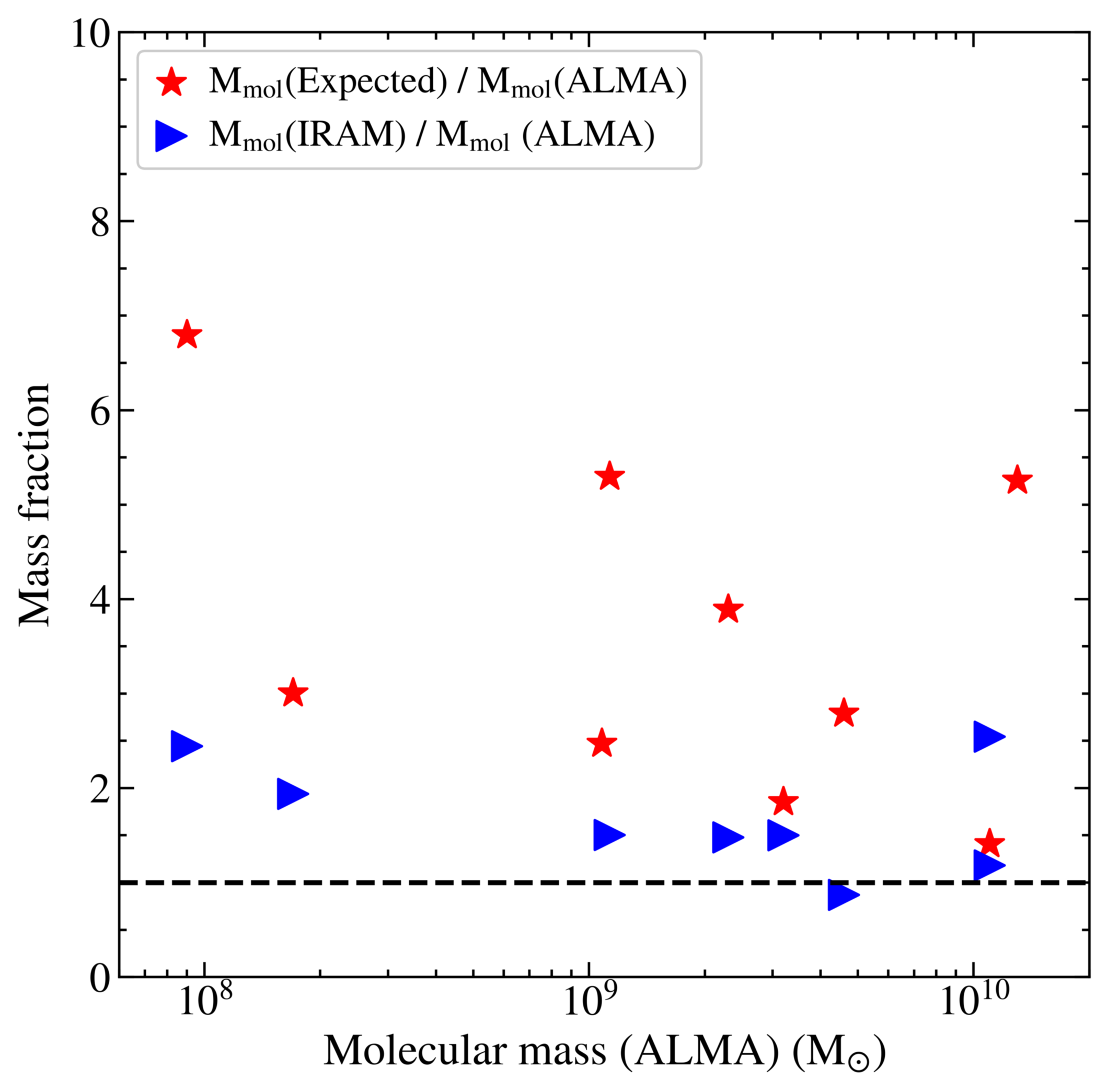}}
    \caption{Molecular mass fraction from IRAM observations (blue leftward pointing triangles) and those based extrapolating the CO measurements assuming a constant H$\alpha$-to-CO flux ratio (red stars; see text for details) over total molecular masses from ALMA. The IRAM measurements are taken from \citet{edge01,salome03,pulido18}. The measured and expected mass of the molecular gas in RXJ0821+0752 is almost the same. Indeed the expected over measured molecular mass from ALMA and IRAM is close to unity, between 1.2 and 1.4, suggesting that ALMA has detected a significant fraction of the molecular mass. 
    These mass comparisons are all based on the same fluxes, so are not affected by the CO-to-H$_{2}$ conversion factor.
}
    \label{fig:expected_molecularmass}
\end{figure}

Based on the un-normalized flux maps of the MUSE and ALMA observations, we derived the total expected molecular mass by assuming that: (i) molecular emission should be detected over all ionized nebulae, (ii) the flux ratio of the molecular and ionized gas is constant for all radii. Note that the second statement is not true for all the sources, where the un-normalized flux ratio decrease at larger radii in some objects, see Fig.~\ref{fig:ratioshaco}. In order to compute the expected total molecular mass, we thus used the most common value of the H$\alpha$ to CO flux ratios.
We then find that the expected total molecular mass is a factor $\sim$1.2 to 7 times higher than the one derived from ALMA observations, with a median of 3.4, see Figure~\ref{fig:expected_molecularmass}. This means that the total amount of gas cycling around the BCGs could be of the order of 10$^{9}$--10$^{11}$~M$_{\odot}$. Future high sensitivity ALMA observations are needed to confirm this, in order to detect even the faintest filamentary emission.

Such correlations can also be interpreted as the manifestation of a common origin, like the condensation of low-entropy gas via the top-down multiphase condensation cascade through thermal instabilities. The conversion of the hot gas into a multiphase medium, regulated by the formation of thermal instabilities, has been discussed frequently in the last few years.

\subsection{Origin of the Cooling Gas.}

Cooling from the hot ICM atmosphere can naturally explain the spatial correlation of the different temperature phases of the gas, owing to an extremely short cooling phase time: only a few \mbox{10~Myr} for low density, 10$^{-2}$~cm$^{-3}$, intermediate temperature, \mbox{10$^6$~K}, with a cooling rate of $\dot{M}_{\rm cool}$ = 1--10 M$\sun$~yr$^{-1}$.

The thermal plasma of the ICM gas appears to be much more complex and out of equilibrium than the first cooling flow models \citep[e.g.]{fabian12}, and the observed turbulence, bubbles, shocks are the likely manifestation of a self-regulation between the radiative cooling and the AGN heating. Their blurring of the picture prevents a clear grasp of what is really going on, and numerical simulations have brought a very helpful insight on the phenomena. Already \citet{pizzolato06} have established through equations that the bubbles are subject to Rayleigh-Taylor and Kelvin-Helmoltz instabilities.
\citet{Revaz_2008} have simulated an AGN-driven hot plasma bubble rising buoyantly in the intra-cluster thermal medium. They have shown that the bubble was efficiently lifting some low-entropy gas present at the center of clusters to a higher radius where the cooling conditions were more favorable. The gas then cooled radiatively in a few tens of Myr and became neutral. Losing its thermal support, it then started to inflow in cold filamentary structures around and below the bubbles. 
The magnetic field in the filaments could contribute to the pressure or the order of a few of the thermal pressure, and it could approach equipartition around the filaments, this could also help to stabilize the filaments against collapse and star formation \citep{fabian16}.
Thermal and magnetothermal instabilities have been studied in detailed simulations of the intra-cluster plasma by \citet{mccourt12,sharma12} showing that local instabilities could trigger the formation of a multi-phase medium, in which density contrast then make cold filaments condense out of the hot medium. For this cooling to occur, it is sufficient that the thermal instability time-scale t$_{\rm TI}$ falls below 10 times the free-fall time, or t$_{\rm TI}$/t$_{\rm ff}$$\sim$10.


The role of AGN feedback on the turbulent and chaotic behaviour of the intra-cluster plasma was emphasized by \cite{gaspari11} and \cite{gaspari12}. They showed how the AGN jets and bubbles create the turbulence and the instabilities (though large-scale cosmological infall can also produce such turbulence in the ICM \citep{dubois11}), accompanied by density contrast and radiative cooling, making the gas condense. The chaotic cold accretion (CCA) is then the consequence of the self-regulation of the cooling and heating by the AGN.

\citet{voit15} noticed also that cold gas condenses in the center of clusters when a threshold for instability is reached, and propose a precipitation model. They remark that the ratio t$_{\rm cool}$ /t$_{\rm ff}$ remains constant in a central floor, when the entropy profile is $K \propto$ r$^{2/3}$ as observed \citep{panagoulia14}, and the potential due essentially to the BCG stellar component is isothermal \citep{hogan17b}. This floor might be due to self-regulation by turbulent driving of gravity wave oscillations \citep{voit18}.

\citet{mcnamara16} promote a stimulated feedback mechanism. They use an infall time-scale t$_{inf}$, which is longer than the free-fall time by a factor of a few. Together with the classical thermal instability theory, they propose that it is the uplifted low-entropy gas, trapped with the rising bubble that can eventually cool and condense in cold filaments. The significance of uplift is now an integral aspect of the precipitation model \citep{voit17} as well as of the Chaotic cold Accretion model \citep{Gaspari_2018}. 

\subsubsection{Are molecular gas clumps formed in low-entropy gas dragged-up by radio lobes?}

Molecular filaments are often trailing the X-ray cavities or aligned in the direction of the radio jets. The drag of molecular gas by the radio jet is not expected to be very significant on large spatial scales, principally because it requires moving very dense small clumps. Models indicate that the acceleration timescale of the dense clumps are generally longer than the dynamical timescale of the system \citep{wagner12,li12,li14,gaibler14}. A preferred scenario is cold gas (re-)formation from a hotter but low entropy gas within the inner radii of the ICM that is lifted by the AGN jets and bubbles \citep{Revaz_2008}. X-ray cavities are often found to be surrounded (in projection) by cold and dense molecular filaments. The cold gas and radio jets and lobes would have to be strongly coupled to displace the large molecular masses found in those systems \citep{russell17b}. Hot diffuse gas that fills the ICM is much easier to lift and may cool via thermal instabilities to form cold clouds \citep{li14}. Eventually the cold gas will decouple from the hot gas and the expanding cavities, and will fall back towards the center of the potential and the AGN.

In Centaurus, the molecular filaments are draped around the exterior of the radio bubble. The southern filament shows a shallow velocity gradient over 100~km~s$^{-1}$ and is not purely radial. The last middle section of the filament is situated in projection along the edge of the eastern radio lobe. The non-radial directionality of the filament may arise because the inner portion of the filament is located behind the X-ray cavity dragged by the radio lobes. Therefore, both features, i.e., bulk motions along the filament plus those closely associated with the radio lobes, indicate that the filament are flowing, being either gas entrained by the expanding radio bubbles or gas falling back onto the central BCG. In RXJ1539.5 there is not obvious connection between any cavity and cold filaments, even though there is a good overall morphological correlation between the X-ray and cold/warm filaments, which might suggest a different mechanism for the formation of the cold gas in this system. In addition, some filaments detected in CO and H$\alpha$ do not show any close connection with X-ray cavities: the northern network of filaments in Abell\,S1101 is a good example of this phenomenon.

Several ALMA data sets have shown that X-ray cavities are capable of lifting enough low entropy gas to form their trailing molecular filaments \citep[e.g.,]{mcnamara14, russell16, russell17, russell17b, vantyghem16, tremblay18, vantyghem18}. However, in RXJ0820.9+0752, the power of the X-ray cavity is too feeble to account for the current gas distribution suggesting perhaps that the association between X-ray cavities and molecular gas is not always water tight. It is also likely that the condensation of the hot atmosphere has been triggered by sloshing motions induced by the interaction with a nearby galaxy group, rather than by uplifting low-entropy gas \citep{bayer-kim02,vantyghem19}. Alternatively, the age of the bubble is an important factor, as well as projection effects, to evaluate the association of the molecular gas with X-ray cavities.


We estimated that 10--50\% of the low-entropy gas displaced by the radio bubbles in Centaurus and Abell\,S1101 could end up in the form of molecular clouds. Simple energetic arguments suggest that radio bubbles are powerful enough, in the two new ALMA sources (Centaurus and Abell\,S1101) to lift the low-entropy gas in order to form the cold molecular along the cavity. In RXJ1539.5, no clear X-ray cavities have been found to date.

In Centaurus, the hot gas displaced ($M_{\rm disp}=\mu (m_{\rm p} + m_{\rm e}) n_{\rm e} V$) by the two cavities associated with the molecular nebula during their inflation is at least $\sim$6$\times$10$^{8}$~M$_{\odot}$ \citep[using the X-ray gas density and cavity sizes from ][]{rafferty06}. The molecular gas in the filaments has a lower mass, 0.45$\times$10$^{8}$~M$\sun$, than the low-entropy gas displaced by the cavities. The cavity has an estimated $4 P V$ work, $E_{\rm cav}\sim1.5\times10^{57}$~erg, and the total kinetic energy in the cold molecular gas, $E_{\rm kin}=\frac{1}{2}M_{\rm mol}v^{2}$, is a few of orders of magnitude lower, $\sim5.4\times10^{54}$~erg. 
From this order of magnitude estimate, it is energetically possible for the cavity to have lifted and displaced sufficient low entropy gas to form the cold filaments.

The filament in Abell\,S1101 appears associated with the X-ray cavities located SE of the BCG nucleus. The work, $4 P V$, done by the expansion of the radio bubbles in the hot atmosphere is $\sim$7$\times$10$^{58}$~erg \citep{werner11}. The mass of the hot gas displaced is around $\sim$2$\times$10$^{9}$~M$\sun$, while the molecular mass of the filament is less than this, about 1.0$\times$10$^{9}$~M$_{\odot}$. The kinetic energy of the filament, which is moving with a velocity of 170~km~s$^{-1}$, is $\approx 2\times10^{56}$~erg. Therefore it is energetically feasible to lift and displace enough low-entropy gas to explain the entire mass of the molecular emission line gas.

\subsubsection{Where does the gas cool?}

\begin{figure*}
    \subfigure{\includegraphics[width=0.49\textwidth]{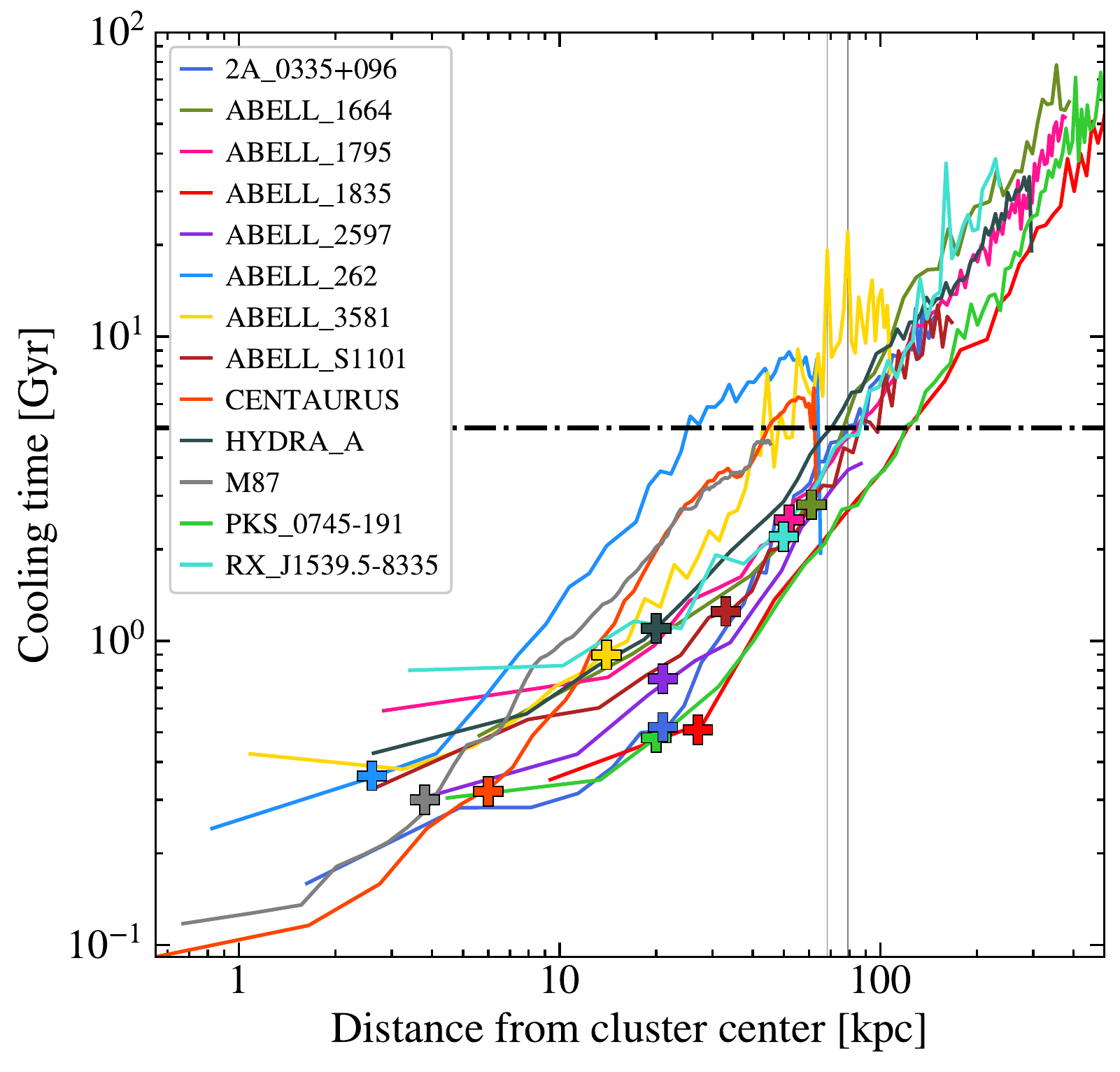}}
    \subfigure{\includegraphics[width=0.49\textwidth]{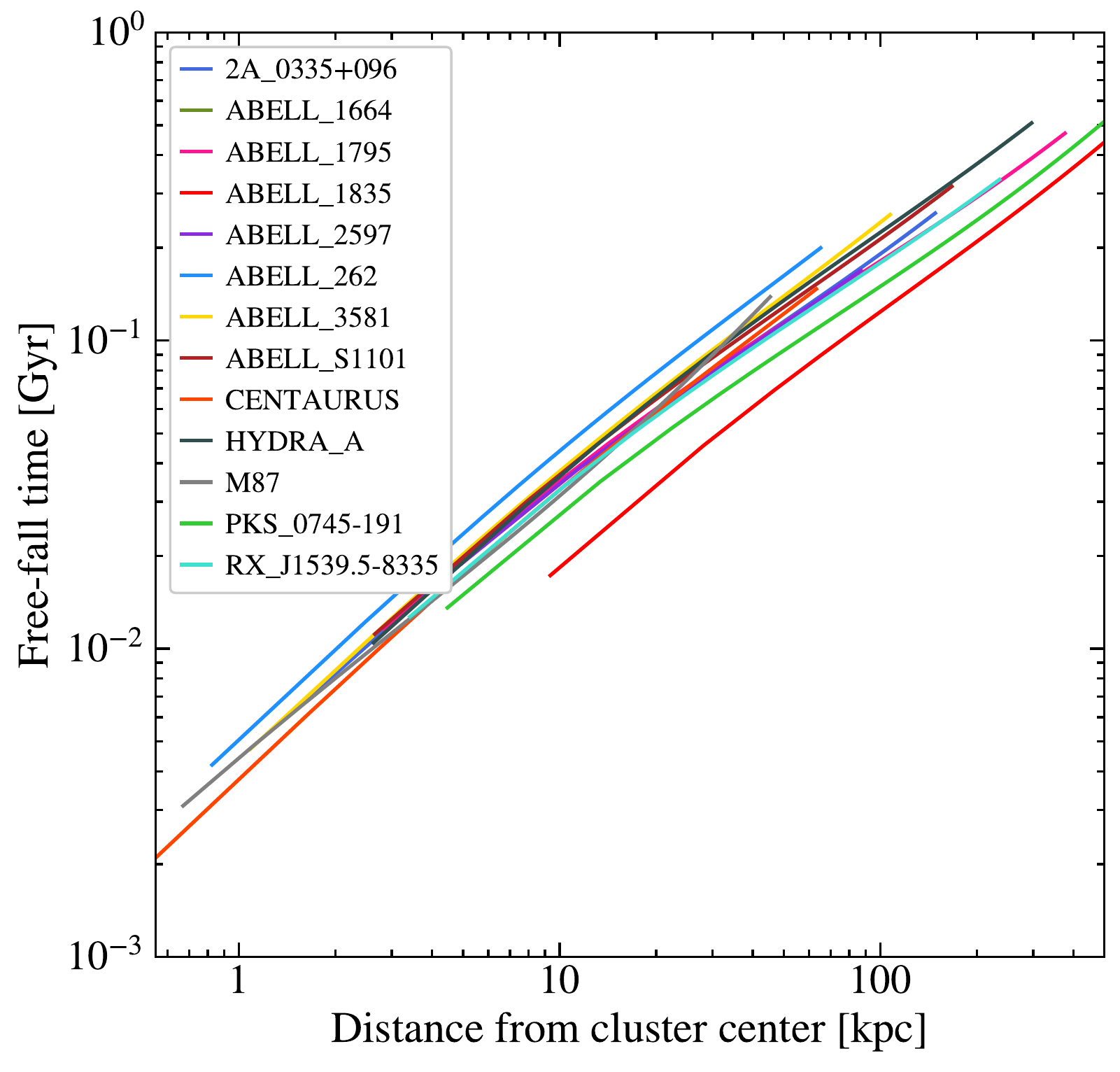}}
    \caption{Left Panel: Deprojected cooling time profiles, t$_{\rm cool}$. The dotted black horizontal lines in panel indicate a constant cooling time of 5$\times$10$^{9}$ yr. The color of each line corresponds to a specific source as given in the legend at the upper left. The radial extent of the filaments are indicated with a crosses and color-coded for each source. Right Panel: Free-fall time versus radius for the different sources of our sample. As shown by \citet{mcnamara16,hogan17b,pulido18}, the free fall profiles in the central region of interest lie on top of each other while the cooling times do not, leading to the conclusion that the driving parameter in the t$_{\rm cool}$/t$_{\rm ff}$ ratio is the cooling time.}
    \label{fig:coolingtime}
\end{figure*}

We investigate where and how the cold gas condensed in the form of filaments by comparing the radial projected length of the filaments, measured from the nucleus of the BCGs to the farthest extension the filaments seen in the warm ionized gas, with a variety of potentially important timescales and ICM properties. These are, the entropy profiles $K$, the ratio of the cooling time and free-fall time, t$_{\rm cool}$/t$_{\rm ff}$, and the eddy turn-over time, t$_{\rm cool}$/t$_{\rm eddy}$. The ICM deprojected radial profiles, entropy and derived cooling times, for 13 sources of the current sample were taken from ACCEPT catalogue (Archive of Chandra Cluster Entropy Profile Tables)\footnote{https://web.pa.msu.edu/astro/MC2/accept/}. Note that the accuracy in central thermodynamical properties of the X-ray emitting gas are limited by the resolution of \textit{Chandra} observations used in making the catalogue. We thus excluded the central bin in the present study (see Fig.~\ref{fig:entropy_profiles}).

The entropy profiles were calculated using the relation, $K = k T_{\rm X} n_{\rm e}^{-2/3}$, where $n_{\rm e}$ and $T_{\rm X}$ are the electron density and the temperature of the X-ray emitting gas, respectively. The inner entropy profiles were fitted with two forms.  The inner part of the profiles were fitted with a power-law, $K \propto r^{\alpha}$, with $\alpha$ varying between 0.65 -- 0.9 \citep[$K \propto r^{3/2}$][]{panagoulia14,babyk18}, in the inner region \citep{Voit_2015,voit17,babyk18,werner18}. For the outer zone, the profile were fitted with a standard cluster entropy profile scaling as $K \propto r^{\beta}$, with $\beta$ ranging 1.1 -- 1.2 \citep{Voit_2015,babyk18}. 

We also compare the deprojected cooling time, $t_{\rm cool}$, with the dynamical time which we assume is the free-fall time, t$_{\rm ff}$. The cooling and free-fall times were then computed as follows:
\begin{equation}
    t_{\rm cool}= \frac{3knT_{\mathrm{X}}} {2n_{\mathrm{e}} n_{\mathrm{H}} \Lambda(Z,T)},
\end{equation}

\noindent
were $n$ is the total number density ($2.3n_{\rm H}$ for a fully ionized plasma), and $n_{\rm e}$ and $n_{\rm H}$ are the electron and proton densities, respectively. $\Lambda(T, Z)$ is the cooling function for a given temperature, $T$, and metal abundance, $Z$. The values of the cooling function were calculated in using the flux of the best-fit spectral model (see \citet{cavagnolo09} for more details).

\begin{equation}
    t_{\rm ff} = \sqrt{\frac{2r}{g}},
\end{equation}
\noindent
where g is the acceleration due to gravity, derived from the total mass of the cluster. The mass profiles contain two components: a NFW (Navarro-Frenk-White) component, \textit{$M_{\rm NFW} = 4 \pi \rho_{\rm 0} r_{\rm s}^{3} \left(ln \frac{r_{\rm s}+r}{r_{\rm s}} - r/(r+r_{\rm s})\right)$}, to account for the majority of the cluster mass on large scales, and an isothermal sphere to account for the stellar mass of the BCG or $M_{\rm ISO} = 2 \sigma_{*}^2 r / G$, where the $\sigma_{*}$ corresponds to stellar velocity dispersion. The $\rho_{0}$ and $r_{s}$ parameters were derived by fitting a NFW profile to the mass, derived through the X-ray emission profiles and by assuming hydrostatic equilibrium. The NFW parameters are similar at those found by \citet{hogan17b,pulido18} (and are listed in the Table~\ref{tab:timescales}).

In Figure~\ref{fig:coolingtime}, we show the central cooling time versus the mean radius of each annulus, R$_{\rm mid}$ $=$ (R$_{\rm in}$ + R$_{\rm out})$ / 2, where R$_{\rm in}$ and R$_{\rm out}$ is the inner and outer radius of the each annulus used to generate each spectrum in the \textit{Chandra} data from the ACCEPT catalogue. The plot shows a clear trend: objects with lower mean radii (i.e., within $\sim$20~kpc) have much lower cooling times \citep[$<$1 Gyr, see also][]{mcnamara16, hogan17b, pulido18}. 

\begin{figure*}[t!]
\centering
    \subfigure{\includegraphics[width=0.49\textwidth]{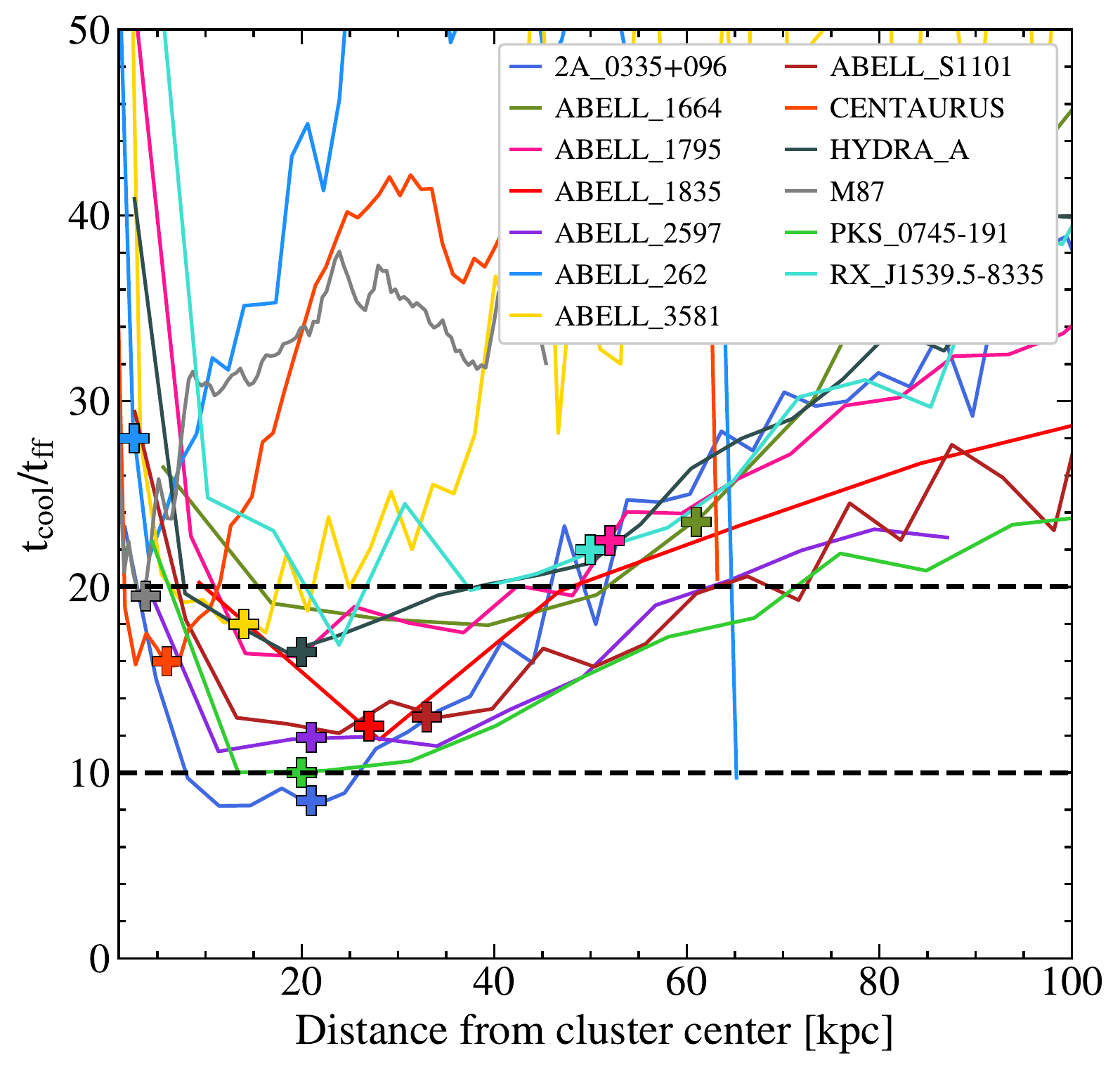}}
    \subfigure{\includegraphics[width=0.49\textwidth]{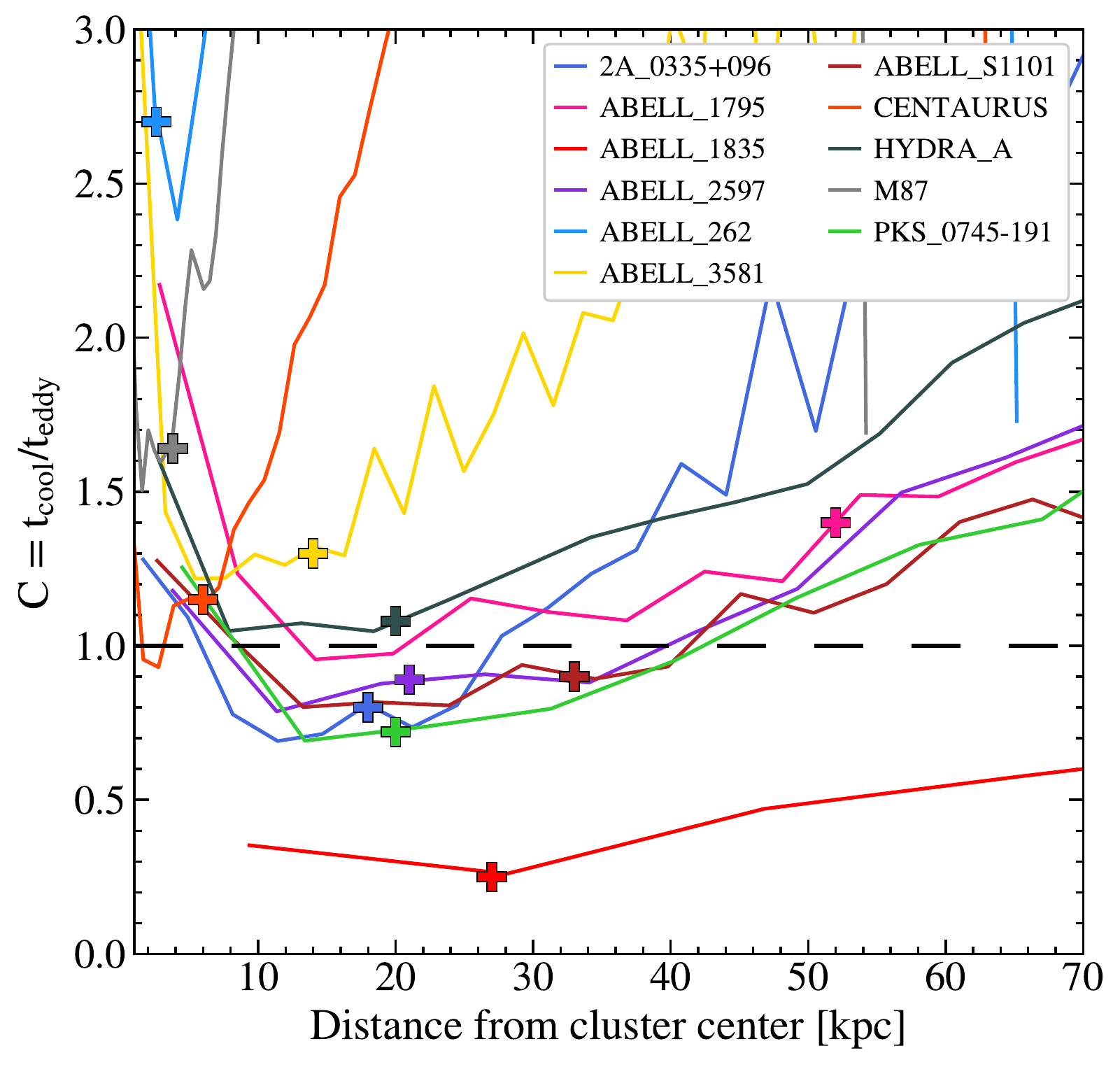}}
    \caption{\textit{Left panel}: The ratio of the cooling time to free-fall time, t$_{\rm cool}$/t$_{\rm ff}$, as a function of radius. The corresponding source for each line is colored according to the legend at the upper left of the panel. We highlighted t$_{\rm cool}$/t$_{\rm ff}$ = 10 and 20 with a horizontal dashed line, which appears to be the approximate threshold for the onset of thermal instabilities \citet[e.g.,][]{li15}. 
    \textit{Right panel}: The ratio of the cooling time over the eddy time as a function of radius. We highlight the C = t$_{\rm cool}$/t$_{\rm eddy}$ = $\sigma_{\rm v}$/v$_{\rm cool}$ = 1 with a horizontal dashed line, which shows the extent of the condensation region and can then be used to assess the multiphase nature of the gas in the BCG systems. We indicated the  maximum radial extent of the filaments with colored crosses for each source (see legend at the upper left).}
    \label{fig:tff-teddy}
\end{figure*}

The radial deprojected entropy profiles show that the filaments are located (in projection) inside the low entropy and short cooling time region (Figure.~\ref{fig:entropy_profiles}). 
On the other hand, we find objects with cold filaments when the ratio of cooling time over free-fall time is t$_{\rm cool}$/t$_{\rm ff}$ $\lesssim$ 20 (Table~\ref{tab:timescales}). We also find that in 9/13 clusters, the observed projected radial sizes of the H$\alpha$ filaments (marked with crosses) often lie at the minimum value of the ratio, t$_{\rm cool}$/t$_{\rm ff}$ (Figure~\ref{fig:tff-teddy}).

The sources, Abell\,1664, Abell\,1795, and RXJ1539.5, exhibit filamentary structures of warm gas, but all have t$_{\rm cool}$/t$_{\rm ff}$ $\gtrsim$ 20 and projected radial size of that are located well beyond their min(t$_{\rm cool}$/t$_{\rm ff}$). These exceptions can be a manifestation of sloshing motions which can affect the cooling time at larger radii (cold front, shocked regions), and also our estimation of t$_{\rm ff}$. X-ray Chandra observations have shown that in Abell\,1664 \citep{calzadilla18} and Abell\,1795 \citep{ehlert15} the hot gas is sloshing in the gravitational potential, creating long, $\sim$50--60~kpc, wakes of dense cooling hot gas. We also emphasize that any sloshing motions affect the distribution of mass, which then could have an impact on the free-fall time estimates.


The filament extents seem to lie at the minimum of the t$_{\rm cool}$/t$_{\rm ff}$, in the central region where it is expected to remain flat \cite{hogan17b}. Those ratio are in general above the canonical value of 10 quoted in CCA and precipitation models. Note however that recent work discusses that the cooling can ensue when t$_{\rm cool}$/t$_{\rm ff}$ is up to 20 : \cite{voit18} describes that as soon as the ratio enters in the 10-20 zone, there is some thermal instability, also fostered by turbulence (i.e. triggered by the AGN bubbles) which introduces cooling, but also heating that prevents the ratio to fall below a lower value.\\
 

{\bf The role of bubbles}. \citet{Gaspari_2018} showed that the ratio between the gas cooling time and the turbulent eddy turnover time,  
C$\equiv$ t$_{\rm cool}$/t$_{\rm eddy}$, is useful for separating when the hot phase may transition from monophase to  multiphase, thus providing some insight as to the physical nature of the hot ICM. The eddy turnover timescale, t$_{\rm eddy}$, is the time at which a turbulent vortex requires to gyrate, producing density fluctuations in the hot atmosphere. The eddy turnover time is calculated as
\begin{equation}
    t_{\rm eddy} = 2\pi \frac{r^{2/3} L^{1/3}}{\sigma_{v}},
\label{eq:eddy}
\end{equation}

\noindent
where $\sigma_{v}$ is the velocity dispersion at the injection scale $L$. This criteria is based on the fact that kinematics of the warm ionized gas correlates linearly with those of the hot turbulent medium. This was recently probed in Perseus by comparing the H$\alpha$+[NII] line width with the FeXXV-XXVI line widths determined in \citet{hitomi16}. The injection scale, $L$, can be estimated using the diameter of the bubble inflated by the jet, which deposits the kinetic energy that is producing the turbulence \citep[see ][]{Gaspari_2018}. Therefore, we only include sources that have measurements of their cavity properties (Table~\ref{tab:timescales}). The values used for the calculation of eddy turnover timescales, i.e., the injection scale and velocity dispersion are given in Table~\ref{tab:timescales}. Given the weak dependence on the injection scale ($L^{1/3}$), the value of the eddy timescale is mostly dependent on the values of $\sigma_{v}$. Note that t$_{\rm eddy}$ estimates have significant uncertainties due to the lack of resolution of \textit{Chandra} observations and the large range of cavity sizes in any given source.

The ratios of the cooling time over the eddy turn-over timescale, t$_{\rm cool}$/t$_{\rm eddy}$, are close to unity, lying in the range 0.3--1.7 (Figure~\ref{fig:tff-teddy}), which is consistent with the range found by \citet[][C $=$ 0.6--1.8]{Gaspari_2018} and suggest a possible role of the bubbles in powering the cooling instability.\\


Note that there is one clear outlier : Abell\,262 for which t$_{\rm cool}$/t$_{\rm eddy}$ is always above 2.5 and t$_{\rm cool}$/t$_{\rm eddy}$ is $\sim$33 at the distance of the CO emission. This source is one the most nearby ones of the current sample and it has a peculiar CO morphology : it is a rotating disk in CO(1-0) also detected in H$\alpha$+N[II] by \citep{hatch07}. This steady disk may thus trace a different and rare stage of the AGN--ICM interaction when the cavity properties have no or less influence on the cold gas and when the cooling of the ICM is no longer triggered sufficently.


\section{Summary and Conclusions}
\label{sec:conclusions}
We used high-resolution ALMA and MUSE observations to study the cold molecular and warm ionized gas in 3 cool core clusters -- Centaurus, Abell\,S1101 and RXJ1539.5 -- for which such data were not previously available. We then extended this investigation to include a total sample of 15 BCGs with similar available data with the goal of studying the origin of filamentary structures of warm ionized and cold molecular gas in cool core clusters. Characterizing the properties of these filaments is an important step to develop a complete census and understanding of galaxy evolution and in particular gauge the importance of AGN feedback in clusters. Our conclusions can be summarized as follows:

\begin{enumerate}[(i.)]
    \item As already shown in other studies of individual objects, we find that most of the BCG have unrelaxed filamentary structure with complex kinematics (13 sources out of 15), while only 2 objects have morphologies and kinematics consistent with a relaxed, central, rotating disk-like gas structure. These disk-like systems thus seem to be a temporary or rare stage in the life cycle of the cool gas. Generally speaking, 13/15 clusters exhibit long, massive 10$^{8}$--10$^{10}$~M$_{\odot}$ cold clumpy molecular filaments, with low velocity dispersions. These extended filaments comprise 20--50\% of the total molecular mass, or even 75--100\% for those objects that are offset relative to the central BCG and with lack of central component. When cavities are present, they usually lie at radii greater than that of the filaments. 
    
    \item For all sources, the optical emission-line and molecular millimeter emission nebula correlate both spatially and kinematically. We conclude that the warm ionized gas and molecular gas are coming from the same structures: most of the mass lies in molecular kpc-size filaments whose surface is warmer and emits optical emission lines. 
    
    In most cases, the ratio of H$\alpha$ to CO is constant along the filament. This suggests that a local mechanism must be responsible for the excitation of the gas (energetic particles, local radiation field, and/or shocks) as the distance to the central AGN does not play any significant role.
    Moreover, we suspect that the molecular emission in the outermost parts of the filaments is likely too faint and/or too diffuse to be detected in the current ALMA data sets. Assuming that the cold molecular filaments follow the whole ionized warm nebula with a constant CO/H$\alpha$ ratio, we have used the H$\alpha$ maps to derive an expected total molecular mass. We conclude that within this hypothesis, the total cold gas in cool core filaments could be as much as $\sim$1.2--7 times larger than the molecular mass derived directly from ALMA data. More sensitive ALMA observations should be able to confirm this prediction.

    \item We compared the projected total extent of the filaments in H$\alpha$ to other characteristic radii for clusters. Using the ACCEPT sample X-ray ICM properties, we found that filaments always lie within the low-entropy and short cooling-time gas. The change in the slope of the entropy K$\propto$r$^{1.1}$ towards K$\propto$r$^{2/3}$ marks the boundary where filaments can start to form. The filaments nevertheless have a range of sizes inside this region. As described in the Chaotic Cold Accretion \citep{Gaspari_2013} or precipitation models \citep{voit17}, an important radius is when the cooling time exceeds the free-fall time by no more than a factor of 10. We find that in 9/13 clusters the extent of the filaments roughly corresponds to the radius where t$\rm_{cool}$/t$\rm_{ff}$ is close to its minimum which is rather always between 10 and 20.
    We also compared the ratio, t$\rm_{cool}$/t$\rm_{eddy}$ as a function of the extent of the filaments, where the eddy turnover time is related to the turbulence injection scale and directly proportional to the diameter of the radio bubble. We find that for 8/10 BCGs, the filaments lie in regions where this ratio is less than $\sim$1. So the AGN bubbles may be powering the turbulent energy helping trigger gas cooling by compression and thermal instabilities. Finally, we showed that the energy contained in the AGN-cavities is enough to drag up some low entropy (short t$\rm_{cool}$) gas in a region distant from the center. Assuming that the gas can keep its original t$\rm_{cool}$, the t$\rm_{cool}$/t$\rm_{ff}$ would be low enough and consistent with an efficient cooling and condensation. So the radio-AGN preventing an overcooling on large scales is likely also the engine that triggers a cold accretion along filamentary structures and provides the material to feed a regulated feedback cycle. 
\end{enumerate}

\begin{table*}
\caption{Estimates of various timescales and their ratios.}
\label{tab:timescales}
\centering
\begin{tabular}{lccc|ccccc}
    \noalign{\smallskip}
    \hline \hline
    \noalign{\smallskip}
    \rm Source                    &
    \multicolumn{3}{c}{Cooling time / Free-fall time}  &
    \multicolumn{5}{c}{Cooling time / Eddy Turnover time}\\
    \noalign{\smallskip}
    \hline
    \noalign{\smallskip}
    \rm      & $\sigma_{*}$  &   R$_{\rm min}$	&    t$_{\rm cool}$/t$_{\rm ff}$ &
    \small Length &   a$_{\rm maj}$  &$\sigma_{v}$ & t$_{\rm eddy}$ & t$_{\rm cool}$/t$_{\rm eddy}$\\

    \rm  	            	& (km~s$^{-1}$) &(kpc)	& [at~R$_{\rm min}$]  & (kpc)   &(kpc)  & (km~s$^{-1}$) & (10$^{8}$ yr)  &	\\
    \rm (1)                 &(2)       &(3)    &(4)	& (5)	       & (6)            & (7)           &   (8) &   (9) \\
    \noalign{\smallskip}
    \hline
    \noalign{\smallskip}
	M87         &189$^{5}$ 		&3.8  	 & 19.5     &3.8    &	0.7--1.1    &153$\pm$11		&1.62   	&	1.64 \\
    Centaurus   &254$^{1}$ 		&6.0	 & 15.7	    &6     	&	1.2--2.9	&127$\pm$10		&2.88   	&	1.10 \\
	Abell\,262   &189$^{4}$ 	&5.8     & 21.2     &4     	&	3.7--4.2	&180$\pm$10		&1.34       &	2.70 \\
	Abell\,3581  &195$^{4}$ 	&16.2    & 17.5     &14    	&	3--4.2		&104$\pm$8		&7.01		&	1.30 \\
	2A0335+096  &220$^{4}$ 		&21.2    & 8.0      &18    	&	6.3--9.6    &183$\pm$3		&6.16		&	0.80 \\
    Hydra-A     &237$^{4}$ 		&18.4    & 16.5     &20    	&	15.3--16.4  &142$\pm$4		&9.65		&	1.08 \\
    Abell\,S1101 &219$^{3}$ 	&30.0    & 14.0	    &33    	&	6.2--20.5   &158$\pm$10		&14.21      &  0.89 \\
	Abell\,1795  &221$^{3}$ 	&20.1  	 & 17.8     &54	    &	18.5        &151$\pm$10		&19.33	    &	1.40 \\
  	RXJ1539.5   &242$^{2}$ 		&44.4   & 16.8	    &50    	&   --          & --             & --    	&   -- \\
	Abell\,2597  &218$^{4}$  	&25.0   & 12.0	    &23		&	3.1--14.0   &175$\pm$15		&11.28	&	0.89 \\
	PKS\,0745-191 &290$^{4}$ 	&14.0	 & 10.0     &20		&	9.9--11.6	&196$\pm$10		&7.09		&	0.72 \\
	RXJ0821+0752&247$^{4}$ 		&--		&  --       & --  	 &  --          & --             & --    	&   -- \\
	Abell\,1664  &267$^{4}$		&39.4	& 18.3	    &61    	 &  --          & --             & --    	&   -- \\
	Abell\,1835  &486$^{4}$ 	&28.0 	& 12.6 	    &27		 &	5.8--16.9   &89$\pm$4		&20.63	&	0.25\\
	Phoenix-A   & -- 			&	 --	&   --	    & --   	 &  --          & --             & --    	&   -- \\
   \noalign{\smallskip}
   \hline
   \noalign{\smallskip}\\
\end{tabular}

\raggedright (1) Source name. \\
\raggedright (2) Stellar velocity dispersions of the galaxies. The references are indicated by the superscript by each value and the corresponding references are : (1) \citet{bernardi02}; (2) \citet{hamer16}; (3) \citet{hogan17b}; (4) \citet{pulido18}; (5) \citet{smith90}.\\
\raggedright (3) Radius of the minimum of t$\rm_{cool}$/t$\rm_{ff}$.\\
\raggedright (4) Ratio between t$\rm_{cool}$ and t$\rm_{ff}$ at R$\rm_{min}$.\\
\raggedright (5) Maximum radial extent of the filaments, measured from MUSE H$\alpha$ maps or [CII] for M87 \citep{Werner_2014}. \\
\raggedright (6) Half of the value of the injection scale, i.e. X-ray major axis of the cavities taken from the literature \citep{shin16, birzan04}.\\
\raggedright (7) Velocity dispersion of the warm or cold phase. For Centaurus, Abell\,S1101, PKS\,0745-191, Hydra-A, 2A0335+096, Abell\,1795, Abell\,3581 and Abell\,1664 we used H$\alpha$ emission lines from MUSE observations. For Abell\,2597 we used HI broad absorption lines from ALMA and VLBA observations taken from \citep{Gaspari_2018}.For Abell\,1835 and Abell\,262 we used CO(3-2) ALMA observations. For M87 we used [CII] from Herschel \citep{werner13}.\\
\raggedright (8) Eddy turnover time estimated using the Eq.~\ref{eq:eddy}.\\
\raggedright (9) Condensation time, i.e., the ratio of the cooling time over the eddy turnover time at maximum projected radial extension of the filaments \citep{Gaspari_2018}.
\end{table*}

\begin{sidewaystable*}
\label{tab:sourceprop}
\caption{Source Properties.}
\centering
\begin{tabular}{lcccrrrrrrrrrr}
    \noalign{\smallskip} \hline \hline \noalign{\smallskip}
    
    \small Source & 
    \small Cat. & 
    \small Scale        &
    D$_{\rm L}$ 	&
    SFR$_{\rm IR}$  &
    log$_{10}$~(SFR) & 
    M$_{*}$ 	& 
    M$_{\rm dust}$	&
    log$_{10}$($\dot{M}_{\rm cool}$)	&
    P$_{\rm mech}$ 	&
    P$_{\rm cav}$ &
    L$_{\rm X-ray}$ &
    M$\rm_{mol}$ & 
    t$\rm_{dep}$ \\ 
    \rm         & 
    \rm  	    &
    \small (kpc/$\arcsec$)	&  
    \small (Mpc) 		& 
    \small (M$_{\odot}$~yr$^{-1}$)  	& 
    \small (M$_{\odot}$~yr$^{-1}$)  &
    \small (10$^{11}$~M$_{\odot}$) &	
    \small (M$_{\odot}$) & 
    \small (M$_{\odot}$~yr$^{-1}$) &
    \small (10$^{42}$~erg~s$^{-1}$)	 &
    \small (10$^{42}$~erg~s$^{-1}$)	&
    \small (10$^{44}$~erg~s$^{-1}$) &
    \small (M$_{\odot}$)	&
    \small 10$^{8}$~yr	\\
    \small  (1)& 
    \small  (2)&
    \small  (3)&  
    \small  (4)& 
    \small  (5)&    
    \small  (6)&
    \small  (7)&
    \small  (8)&	
    \small  (9)& 
    \small  (10)&
    \small  (11)&
    \small  (12)&
    \small  (13)&
    \small  (14)\\
    \noalign{\smallskip} \hline \noalign{\smallskip}
    
M87			 	&1  &0.082	&16.1    					& -- 		&0.14$\pm$0.07 &	0.15	        & <1.3$\times$10$^{8}$& 1.29$\pm$0.00	&124.6$\pm$17.6	    & 6                             & --   &  4.7$\times$10$^{5}$  &   0.03   \\

Centaurus		&1  &0.208	& 43.9  		   	    	&0.13		&0.16$\pm$0.76  &	6.15$\pm$0.14 	& 1.6$\times$10$^{6}$	&0.97$\pm$0.01 	&12.8$\pm$0.8	    & 7.4$^{+5.8}_{-1.8}$           & 1.74 &  0.9$\times$10$^{8}$  &  5.55    \\

Abell\,262		&2  &0.330	&70.2  	    		    	& 0.55 		&0.22$\pm$0.87  &	3.31$\pm$0.08    & 8.7$\times$10$^{6}$	&0.48$\pm$0.04 	&12.4$\pm$2.3		&9.7$^{+7.5}_{-2.6}$            & --   &  3.4$\times$10$^{8}$  &  17.64     \\

Abell\,3581		&1  &0.435	&94.9   				   	&-- 	  	&0.78$\pm$0.33  &	0.86            & $\sim$10$^{6}$		&1.35$\pm$0.22 	&48.2$\pm$0.3		&3.1                            & --  & 5.4$\times$10$^{8}$  &  2.57    \\

2A0335+096		&1  &0.700	& 150.0 		    	    &2.09		&0.46$\pm$0.66 &	5.33$\pm$0.26    & 7.3$\times$10$^{5}$	&2.26$\pm$0.01 	&19.4$\pm$2.5		&24.0$^{+23.0}_{-6.0}$          & 9.77 &  1.1$\times$10$^{9}$  &  23.52    \\

Hydra-A			&2  &1.053	 &242.0  					&4.0		&4.07$\pm$2.81  &	4.18$\pm$0.24	& 5.5$\times$10$^{6}$   &2.04$\pm$0.02	&816.5$\pm$408.9    &430.0$^{+200.0}_{-50.0}$       & -- &5.4$\times$10$^{9}$  &  6.14     \\

Abell\,S1101		&1  &1.094	&251.8	    				 &2.30		&1.02$\pm$2.57  &	6.13$\pm$0.55    & $\sim$10$^{7}$		&2.37$\pm$0.02	& 37.2$\pm$0.9	    &780$^{+820.0}_{-260}$          & 1.10 & 
10.8$\times$10$^{8}$ &  5.96     \\

Abell\,1795		&1  &1.220	& 283.9 			    	& 8.0		& 3.45$\pm$5.01 &   6.99$\pm$0.52    & 6.7$\times$10$^{7}$  &2.27$\pm$0.02 	&      -- 			& 160				            & 7.59 & 3.2$\times$10$^{9}$ &  9.23     \\

RXJ1539.5		&1  &1.437	& 343.0 					&-- 		&1.86$\pm$1.14 &	2.3            	& -- 					&2.19$\pm$0.05	& 5.8$\pm$0.7 		& --                            & -- & 1.3$\times$10$^{10}$&  69.81    \\

Abell\,2597		&1  &1.546	&373.3      		    	&2.93		&3.98$\pm$2.29  &	3.24$\pm$0.31    & 8.4$\times$10$^{7}$ 	&2.49$\pm$0.05	&295.1$\pm$86.7	    &67.0$^{+86.0}_{-29.0}$         & -- & 2.3$\times$10$^{9}$ &  5.78     \\

PKS\,0745-191 	&1  &1.890	&474.1                 	    &17.07		&13.48$\pm$1.73  & 	5	             & 3.5$\times$10${^7}$	&2.89$\pm$0.01	&400.2$\pm$141.1	&1700.0$^{+1400.0}_{-300.0}$    & 18.41 & 4.9$\times$10$^{9}$&3.63     \\

RXJ0821+0752	&1  &2.000	& 510.0 		    	    &36.91 		&36.30	        &	1.4              &2.2$\times$10$^{7}$	& 	--			&19.7$\pm$2.5		&13                             & -- & 1.1$\times$10$^{10}$ & 3.03       \\

Abell\,1664		&1  &2.302	&604.2  	 		    	&14.54		&13.18$\pm$1.09  &	2.5             & 1.3$\times$10${^7}$  	&2.21$\pm$0.04	&68.1 $\pm$3.0	    &95.2$\pm$74.0                  & 2.59 & 1.1$\times$10$^{10}$ & 8.34     \\

Abell\,1835		&1  &3.966	&1260.1	        			&138    	&117.48$\pm$1.58 & 	5.7	            & 1.0$\times$10${^8}$ 	&3.07$\pm$0.06	&129.8$\pm$19.2	    &1800.0$^{+1900.0}_{-600.0}$    & 39.38 & 3.0$\times$10$^{10}$ & 2.55  \\

Phoenix-A		&1  &6.750	& 3501.5 	              	&530$\pm$53 & 616.59$\pm$2.29  & $\sim$8        &	-- 			    	&3.23$\pm$0.08 	&10000.0            & 800--4700                     & -- & 2.1$\times$10$^{10}$ & 1.70      \\
    \noalign{\smallskip} \hline \noalign{\smallskip}
\end{tabular}
\tablefoot{(1) Source name. (2) Category. Cat. 1: Extended filaments, Cat 2: Nuclear emission with disk-like morphologies and kinematics (see text for details). (3) Scale conversion of angular size to physical size. (4) Luminosity distance. (5) The star-formation rate estimated from infrared observations as given in the literature. References: Centaurus \citet{mittal11} from Herschel (36$\arcsec$) and Spitzer (65$\arcsec$) observations, Abell\,262 \citep{O_Dea_2008}, 2A0335+096 \citep{O_Dea_2008} from IR photometry within a 6\arcsec aperture, which excludes both the companion galaxy and the filament, PKS\,0745-191, RXJ0821+0752, Abell\,1664, Abell\,1835 using Spitzer photometry using different apertures depending the band from 12$\arcsec$, 26$\arcsec$, 35$\arcsec$, see \citep{O_Dea_2008, quillen08}, Abell\,S1101 from Herschel PACS and SPIRE imaging \cite{mcdonald15}, Hydra-A and Abell\,1795 from \citep{hoffer12}, Abell\,1835 \citep{mcnamara06}, Abell\,2597 \citep{donahue07}, Phoenix-A \citep{Tozzi_2015}. 
(6) SFR from \citep{McDonald_2018}, they estimated the SFR values in a variety of robust methods taking into account the AGN contamination. Phoenix: The spectrally decomposed the AGN component. For the rest of the sample they determine SFRs based on an ensemble of measurements from the literature, principally from IR dust emission, optical lines, stellar continuum and optical SED fitting.
(7) Mass of the stellar population collected from literature, by using apparent K-band luminosities taken from the 2MASS Extended Source Catalog, it was applied galactic extinction, evolution and K-corrections. Then they used the M/L$_{K}$ relation \citep{bell03} for Centaurus, Abell\,262, Hydra-A, 2A0335+096, Abell\,S1101, Abell\,1796, Abell\,2597 from \citep{main17}, for Phoenix from \citep{lidman12}, PKS\,0745-191 from \citep{donahue11}. Finally for Abell\,3581, Abell\,1664, Abell\,1835, PKS\,0745-191, RXJ0821+0752 WISE observations.\\
(8) Dust mass collected from the literature, for Centaurus from \citep{mittal11}, Hydra-A, 2A0335+096, RXJ0821+0752, Abell\,2597, Abell\,1664, Abell\,1835 from \citep{edge01}, Abell\,262, Abell\,1795, PKS\,0745-191 from \citep{salome03} using $T_{dust}$ of 35~K, M87 and Abell\,S1101 \citep{mcdonald15}, Phoenix-A.
(9) Classical cooling rate are from \citep{McDonald_2018}, calculated inside the $r_{cool}$ (radius within the cooling time is less than 3~Gyr). Integrating the gas density within a sphere bounded by this radius provides an estimate of the total mass available for cooling, which is then  divided by the cooling time. (10) Mechanical Power: Inferred from the AGN radio luminosity correlation \citep{birzan04, pulido18}. (11) The cavity power were obtained from \citet{rafferty06,Hlavacek-Larrondo15,pulido18}. (12) Bolometric X-ray luminosity (0.1 -- 100 keV) from CHANDRA observations \citep{cavagnolo09}. (13) Molecular mass estimated from ALMA observations, assuming a X$_{\rm CO}$ factor of 2$\times$10$^{20}$~cm$^{-2}$ (K~km~s$^{-1}$)$^{-1}$ for all the systems. (14) Depletion time are calculated based on SFR and the molecular mass from ALMA observations.}
\end{sidewaystable*}

\begin{table*}[h]
\label{tab:filament_properties}
\caption{Properties of the cold molecular filaments.}
\centering
\begin{tabular}{lcrrrccc}
    \noalign{\smallskip} \hline \hline \noalign{\smallskip}
    \rm  Source						& V$_{\rm grad}$    & V$_{\rm grad}$/size	& Ang. Res.      & Size 	& M$_{\rm mol}$   &$z$  \\
    \rm  							& (km~s$^{-1}$) 	&(km~s$^{-1}$/kpc)&(kpc)      &(kpc)     &M$_{\odot}$       & 10$^{7}$~yr		        &	     \\
    \noalign{\smallskip} \hline \noalign{\smallskip}
    \textit{Category extended} &&&&&&& \\
    \noalign{\smallskip} \hline \noalign{\smallskip}
M87 &&&&&&& \\
\hspace{2.0mm}Clump  CO(2-1)	             &40 	&36	&0.01   &0.07  &(4.7$\pm$0.5)$\times10^{5}$	    & 0.00428		   \\
\noalign{\smallskip} \hline \noalign{\smallskip}	
Centaurus &&&&&&& \\
\hspace{2.0mm}S Filament CO(1-0)	    	&100    &23	&0.4    &4.2   &(0.30$\pm$0.02)$\times10^{8}$	    & 0.01016		 \\
\noalign{\smallskip} \hline \noalign{\smallskip}	
Abell\,3581 &&&&&&& \\
\hspace{2.0mm} SN Filament CO(2-1)			&170 	&65	&0.3    &2.6& (1.6$\pm$0.1)$\times10^{8}$	  	& 0.02180			\\
\hspace{2.0mm} EW Filament CO(2-1)			&80	 	&18	&0.3    &4.3& (2.7$\pm$0.2)$\times10^{8}$	  	& 0.02180		 \\
\noalign{\smallskip} \hline \noalign{\smallskip}
2A0335+096 &&&&&&& \\
\hspace{2.0mm} N Filament CO(1-0)		    &100 	&33	&0.7    &3	   &(3.7$\pm$0.2)$\times10^{8}$	  	& 0.03634		\\
\hspace{2.0mm} Uplifted Filament  CO(1-0)   &120	&17	&0.7    &7	   &(6.9$\pm$0.2)$\times10^{8}$	  	& 0.03634		\\
\noalign{\smallskip} \hline \noalign{\smallskip}	
Abell\,S1101 &&&&&&& \\
\hspace{2.0mm} Filament CO(1-0) 			&160 	&6	&1.9    &11     &(10.8$\pm$0.7)$\times10^{8}$	    & 0.05639		\\
\noalign{\smallskip} \hline \noalign{\smallskip}	
Abell\,1795 &&&&&&& \\
\hspace{2.0mm} S Filament CO(2-1)		     &50 	&8	&0.7   &6	    &(3.0$\pm$0.2)$\times10^{8}$	    & 0.06326		\\
\hspace{2.0mm} Curved Filament CO(2-1)		 &200	&20 &0.7   &10	    &(1.7$\pm$0.1)$\times10^{9}$	 	& 0.06326 		\\
\noalign{\smallskip} \hline \noalign{\smallskip}
RXJ1539.5 &&&&&&& \\
\hspace{2.0mm}E Filament CO(1-0)			&100 	&5	&2.6   &21	    &(2.4$\pm$0.3)$\times10^{9}$	  	& 0.07576		 \\
\hspace{2.0mm}N Filament CO(1-0)			&110	&6   &2.6   &16	    &(6.2$\pm$1.3)$\times10^{8}$	  	&0.07576		 \\
\hspace{2.0mm}W Filament  CO(1-0)	    	&400 	&22	&2.6   &18 	    &(2.5$\pm$1.2)$\times10^{8}$	  	&0.07576		 \\
\noalign{\smallskip} \hline \noalign{\smallskip}
Abell\,2597 &&&&&&& \\
\hspace{2.0mm}Nuclear CO(2-1)		   	     &150 	&19	&1.2    &7.7    &(2.2$\pm$0.1)$\times10^{9}$	  	&0.08210		\\
\hspace{2.0mm}S Filament CO(2-1)		 	 &200 	&18	&1.2    &10.8  	&(7.6$\pm$0.5)$\times10^{8}$ 	   	&0.08210		 \\
\noalign{\smallskip} \hline \noalign{\smallskip}
PKS\,0745-191 &&&&&&& \\
\hspace{2.0mm} Compact   CO(1-0)		    &100 	&10	&2.5   &9.5 	&(4.6$\pm$0.4)$\times10^{9}$	   	&0.10280		\\
\hspace{2.0mm} Filament  N CO(3-2)			&200 	&35	&0.4   &5.7	 	&(2.0$\pm$0.3)$\times10^{9}$	    	&0.10280		\\
\hspace{2.0mm} Filament  SW CO(3-2)			&60	 	&15	&0.4   &3.8	 	&(2.1$\pm$0.8)$\times10^{9}$	    	&0.10280		\\
\hspace{2.0mm} Filament  SE CO(3-2)			&80	 	&23	&0.4   &3.4		&(1.7$\pm$0.5)$\times10^{9}$	    	&0.10280		\\
\noalign{\smallskip} \hline \noalign{\smallskip}
RXJ0821+0752 &&&&&&& \\
\hspace{2.0mm} Two Clumps offset 	        &200 	&24	&8.2    &8.2    &(1.1$\pm$0.4)$\times10^{10}$		&0.10900		\\
\noalign{\smallskip} \hline \noalign{\smallskip}
Abell\,1664 &&&&&&& \\
\hspace{2.0mm} Nuclear CO(1-0)		        &110 	&48	&2.0   &2.3	    &(4.5$\pm$0.2)$\times10^{9}$	  	&0.12797		\\
\hspace{2.0mm} NE filament  CO(1-0)		    &180 	&28	&2.0   &6.6	    &(1.5$\pm$0.7)$\times10^{9}$	  	&0.12797		 \\
\hspace{2.0mm} SE filament (HVS) CO(1-0)	&200 	&30	&2.0   &6.4	    &(5.0$\pm$1.0)$\times10^{9}$	  	&0.12797		\\
\noalign{\smallskip} \hline \noalign{\smallskip}
Abell\,1835 &&&&&&& \\
\hspace{2.0mm}  CO(1-0)	emission		    &100 	&14	&2.4   &25 		&(3.0$\pm$0.7)$\times10^{10}$	  	     &0.25198			 \\
\hspace{2.0mm}  CO(3-2)	emission	        &120 	&8	&0.8   &13.5	&(1.4$\pm$0.6)$\times10^{10}$	           &0.25198			\\
\noalign{\smallskip} \hline \noalign{\smallskip}
Phoenix-A &&&&&&& \\
\hspace{2.0mm}  Nuclear CO(2-1)	            &400	&   &4.8      &10    & (5.0$\pm$0.1)$\times10^{10}$	  		 &0.59600       \\
\hspace{2.0mm}  SE filament  CO(2-1) 	  	&250 	&23 &4.8	&15      & $\sim$2.5$\times10^{10}$	      	     &0.59600		 \\
\hspace{2.0mm}  SW filament  CO(2-1) 	    &250 	&15	&4.8    &16      & $\sim$1.5$\times10^{10}$	      	   	 &0.59600		  \\
\hspace{2.0mm}  NW filament CO(2-1) 	    &200 	&21	&4.8    &22      & $\sim$1.5$\times10^{10}$	      	   	 &0.59600		  \\
\hspace{2.0mm}  NE filament CO(2-1) 	    &280 	&35	&4.8    &10      & $\sim$0.5$\times10^{10}$	      	   	 &0.59600		   \\
    \noalign{\smallskip} \hline \noalign{\smallskip}
    \textit{Category compact} &&&&&&& \\
    \noalign{\smallskip} \hline \noalign{\smallskip}
Abell\,262 &&&&&&& \\
\hspace{2.0mm}  Disk CO(2-1)                &350 	&152	&0.2    &2.3	& (3.4$\pm$0.1)$\times10^{8}$   & 0.01619		 \\
\noalign{\smallskip} \hline \noalign{\smallskip}	
Hydra-A &&&&&&& \\
\hspace{2.0mm}  Disk CO(3-2) 				&600 	&115	&0.5    &5.2	&(5.4$\pm$0.2)$\times10^{9}$  & 0.05487		\\
    \noalign{\smallskip} \hline \noalign{\smallskip}
\end{tabular}
\tablefoot{We assumed a X$\rm_{CO}$ factor of 2$\times$10$^{20}$~cm$^{-2}$ (K~km~s$^{-1}$)$^{-1}$ for all the systems in the sample, including Phoenix-A.}
\end{table*}

\begin{acknowledgements}
This work was supported by the ANR grant LYRICS (ANR-16-CE31-0011). ACF acknowledges support by ERC Advanced Grant 340442. ACE acknowledges support from STFC grant ST/P00541/1. BRM acknowledges support from the Natural Sciences and Engineering Research Council
of Canada, This paper makes use of the following ALMA data: ADS/JAO.ALMA\#2015.1.01198.S (PI: S. Hamer), ADS/JAO.ALMA\#2011.0.00374.S,  ADS/JAO.ALMA\#2012.1.00837.S, ADS/JAO.ALMA\#2012.1.00988.S, ADS/JAO.ALMA\#2013.1.00862.S, ADS/JAO.ALMA\#2013.1.01302.S,  ADS/JAO.ALMA\#2015.1.00623.S, ADS/JAO.ALMA\#2015.1.00644.S, ADS/JAO.ALMA\#2015.1.00598.S, ADS/JAO.ALMA\#2015.1.00598.S. ALMA is a partnership of ESO (representing its member states), NSF (USA) and NINS (Japan), together with NRC (Canada), MOST and ASIAA (Taiwan), and KASI (Republic of Korea), in cooperation with the Republic of Chile. The Joint ALMA Observatory is operated by ESO, AUI/NRAO and NAOJ.

\end{acknowledgements}

\bibliographystyle{aa} 

\bibliography{master_volivares} 
\begin{appendix}
\section{H$\alpha$ (MUSE) to CO (ALMA) Comparison: Line ratios and Velocity Maps}

\begin{figure*}[htbp!]
\label{fig:ratioshaco}
    \centering
    \subfigure{\includegraphics[width=0.45\textwidth]{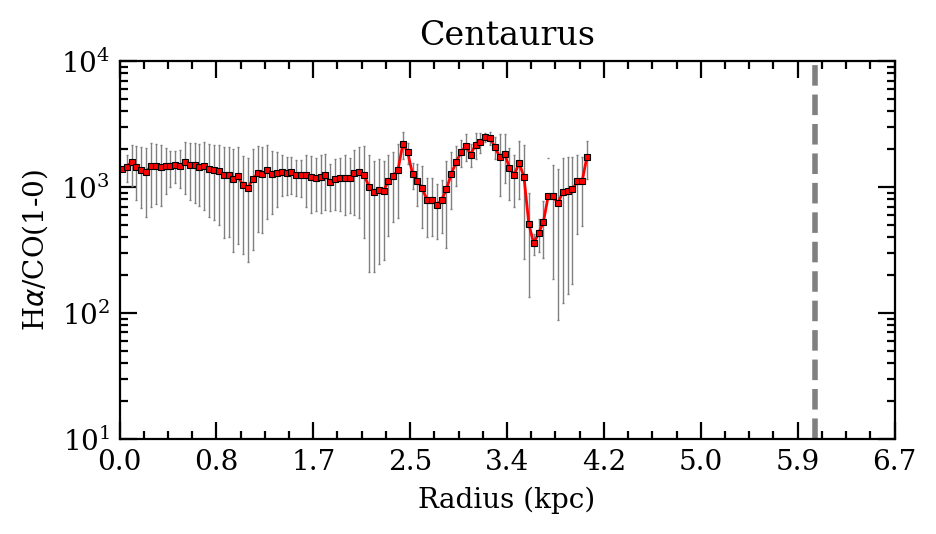}}
    \subfigure{\includegraphics[width=0.45\textwidth]{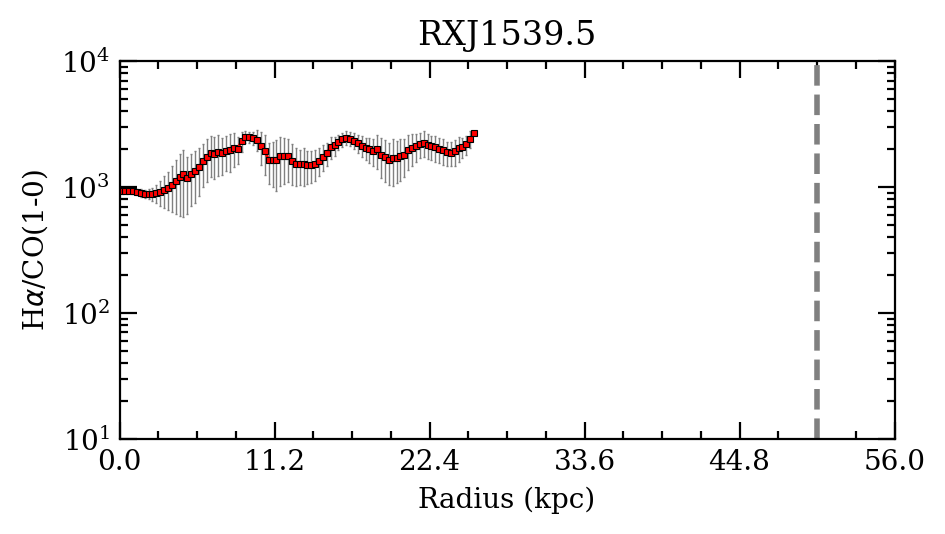}}\\
    \vspace{-4mm}
    \subfigure{\includegraphics[width=0.45\textwidth]{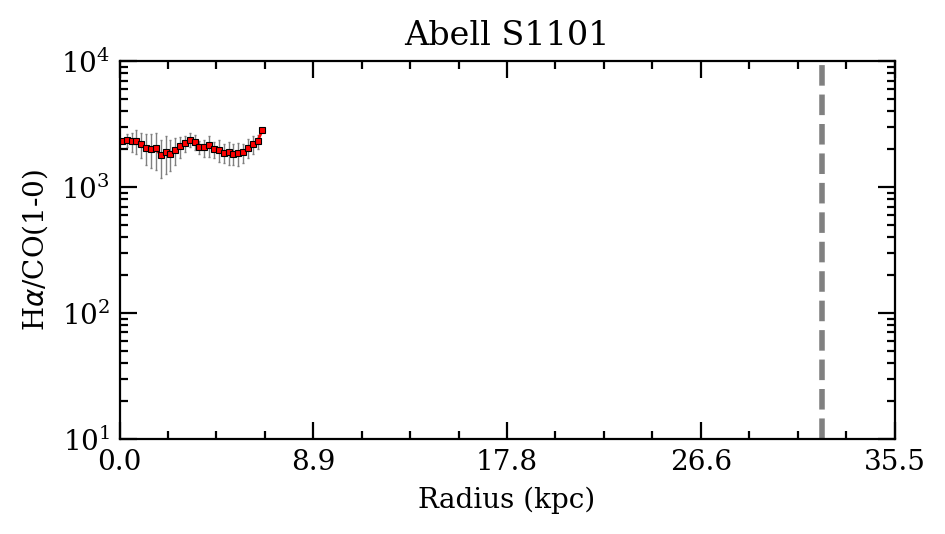}}
    \subfigure{\includegraphics[width=0.45\textwidth]{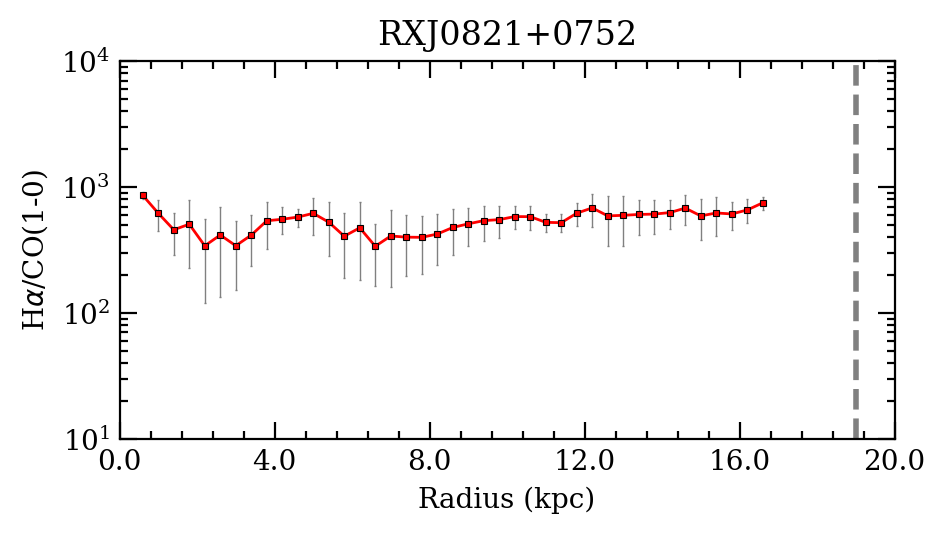}}\\
    \vspace{-4mm}
    \subfigure{\includegraphics[width=0.45\textwidth]{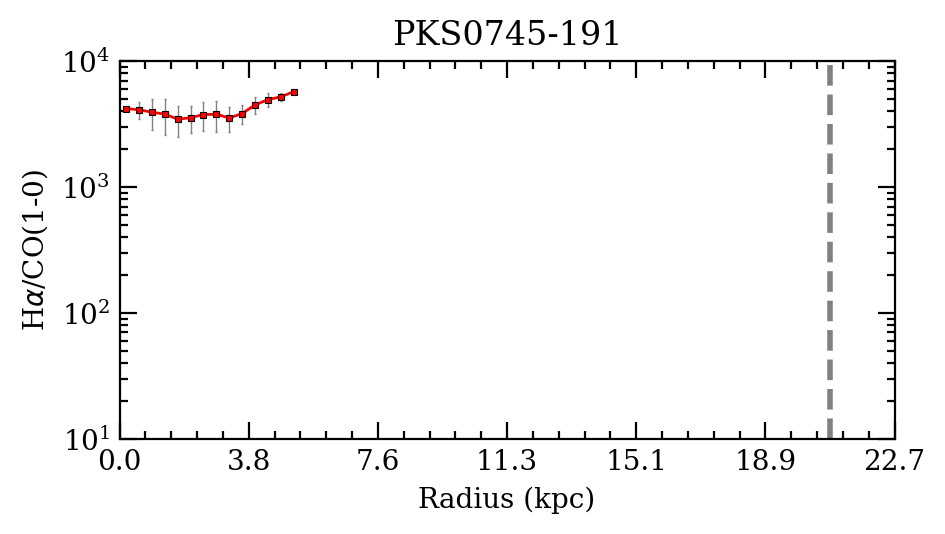}}
    \subfigure{\includegraphics[width=0.45\textwidth]{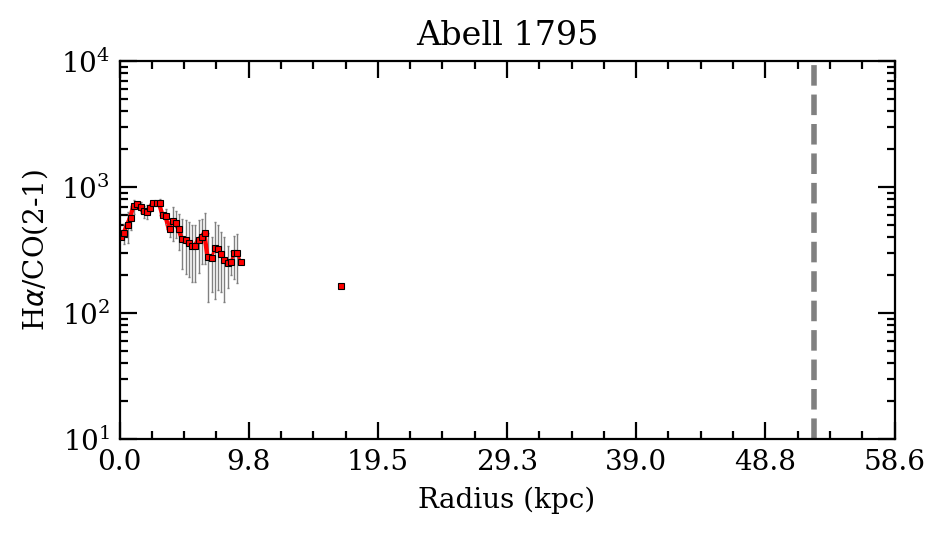}}\\
    \vspace{-4mm}   
    \subfigure{\includegraphics[width=0.45\textwidth]{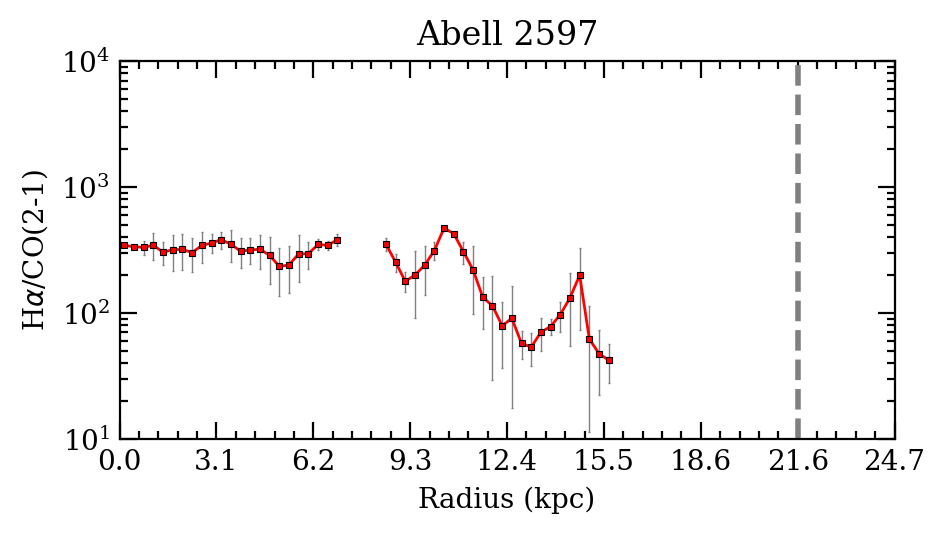}}
    \subfigure{\includegraphics[width=0.45\textwidth]{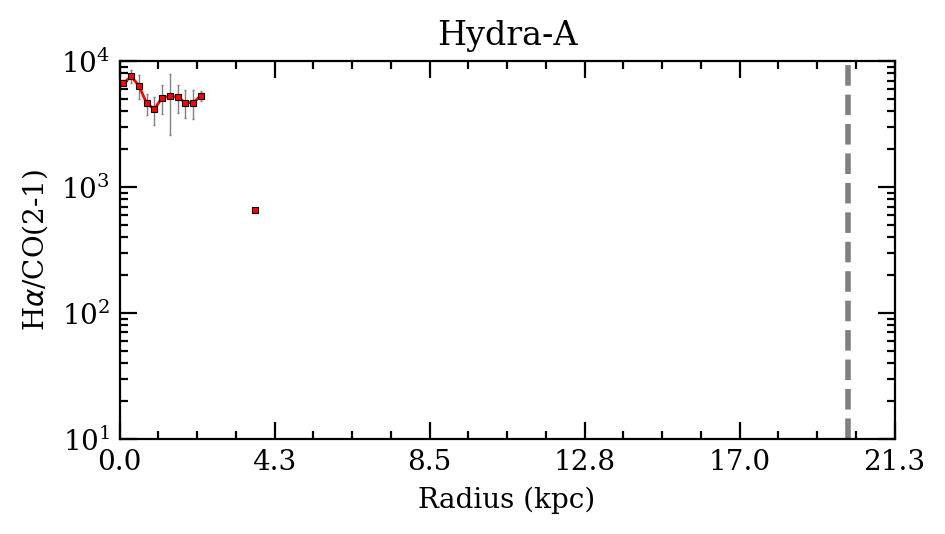}}\\
    \vspace{-4mm}
    \subfigure{\includegraphics[width=0.45\textwidth]{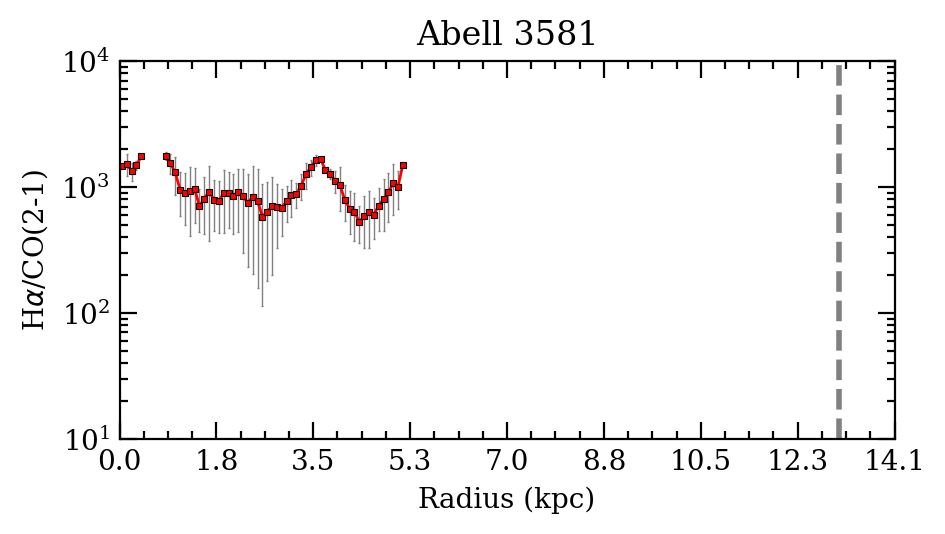}}
    \subfigure{\includegraphics[width=0.45\textwidth]{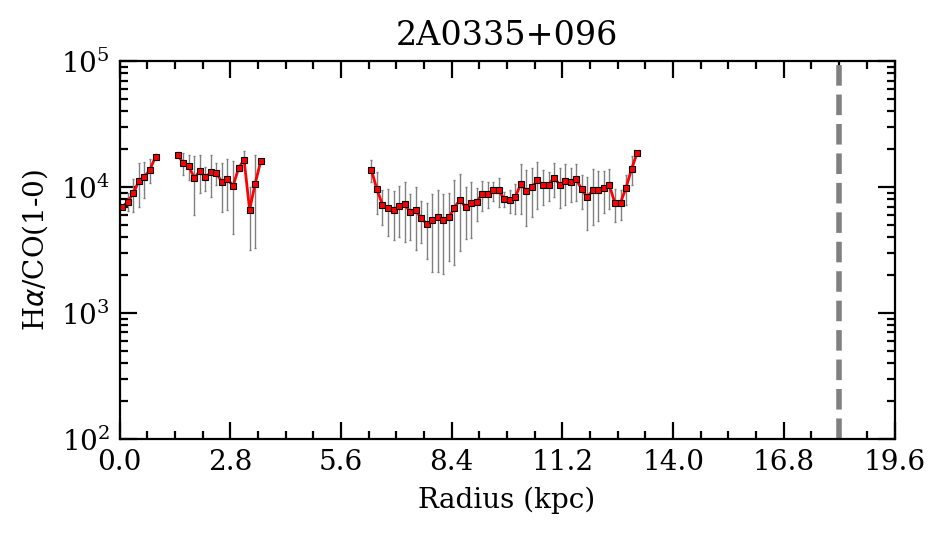}}\\
    \caption{H$\alpha$ to CO ratios as a function of radius in arcsec only where CO 
    is detected. Note that the flux of either CO(1-0) or CO(2-1) were used in estimating the H$\alpha$ to CO line flux ratios. The dashed gray lines denote the maximum radial extent of the warm filaments.}
\end{figure*}

\begin{figure*}[htbp!]
\label{fig:velhaco}
    \centering
    \subfigure{\includegraphics[width=0.49\textwidth]{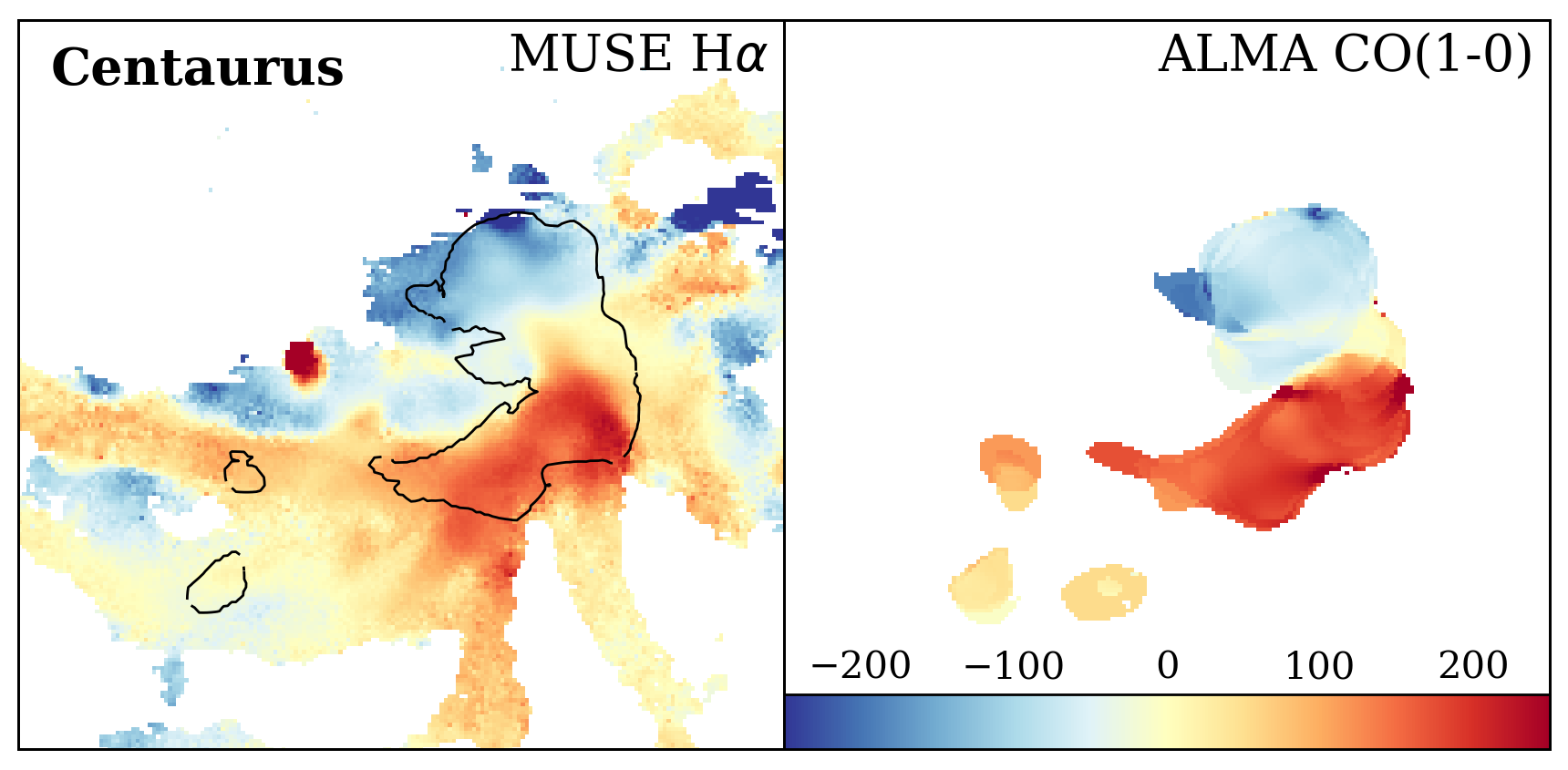}}
    \subfigure{\includegraphics[width=0.49\textwidth]{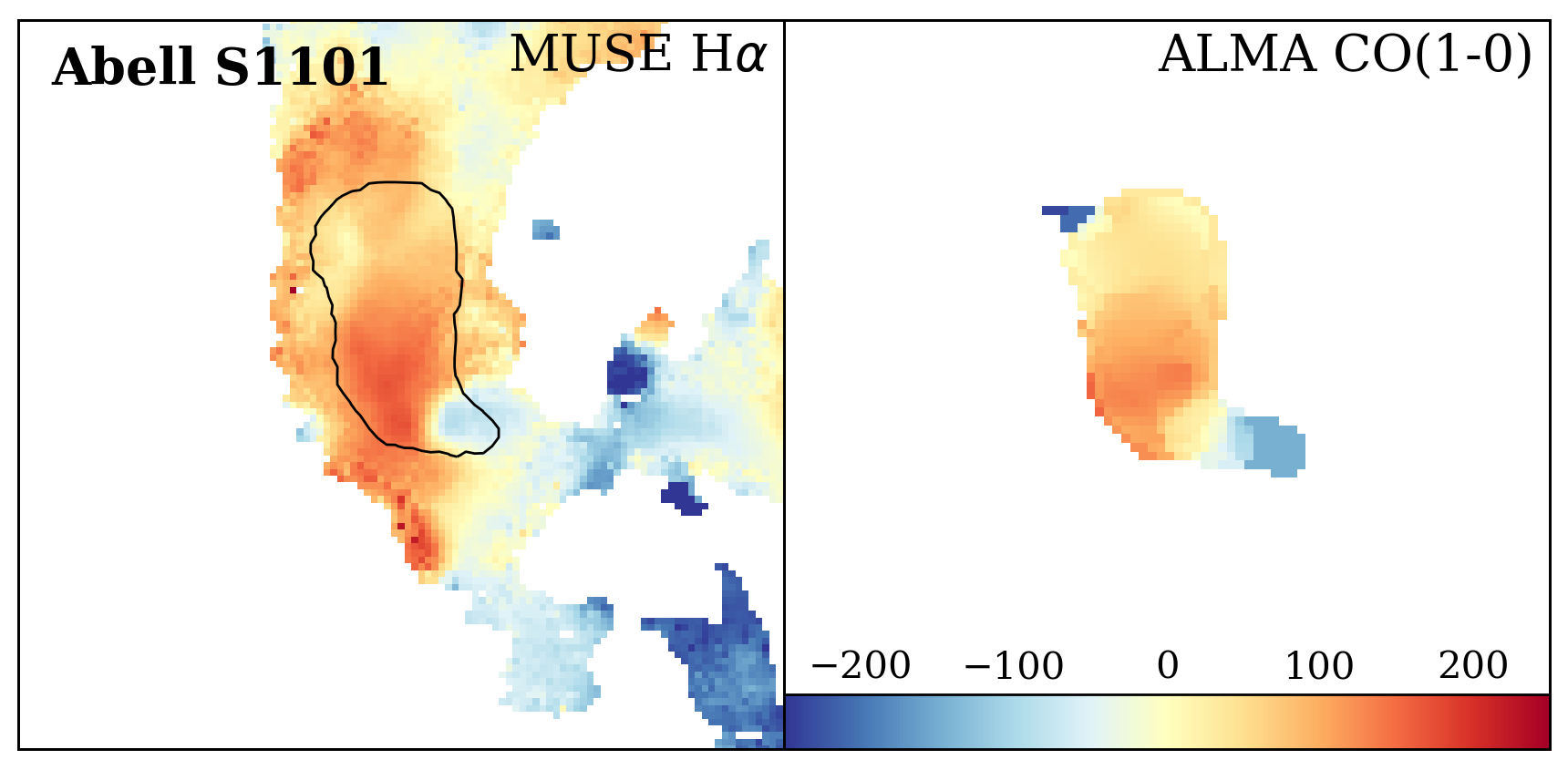}}\\
    \vspace{-2mm}
    \subfigure{\includegraphics[width=0.49\textwidth]{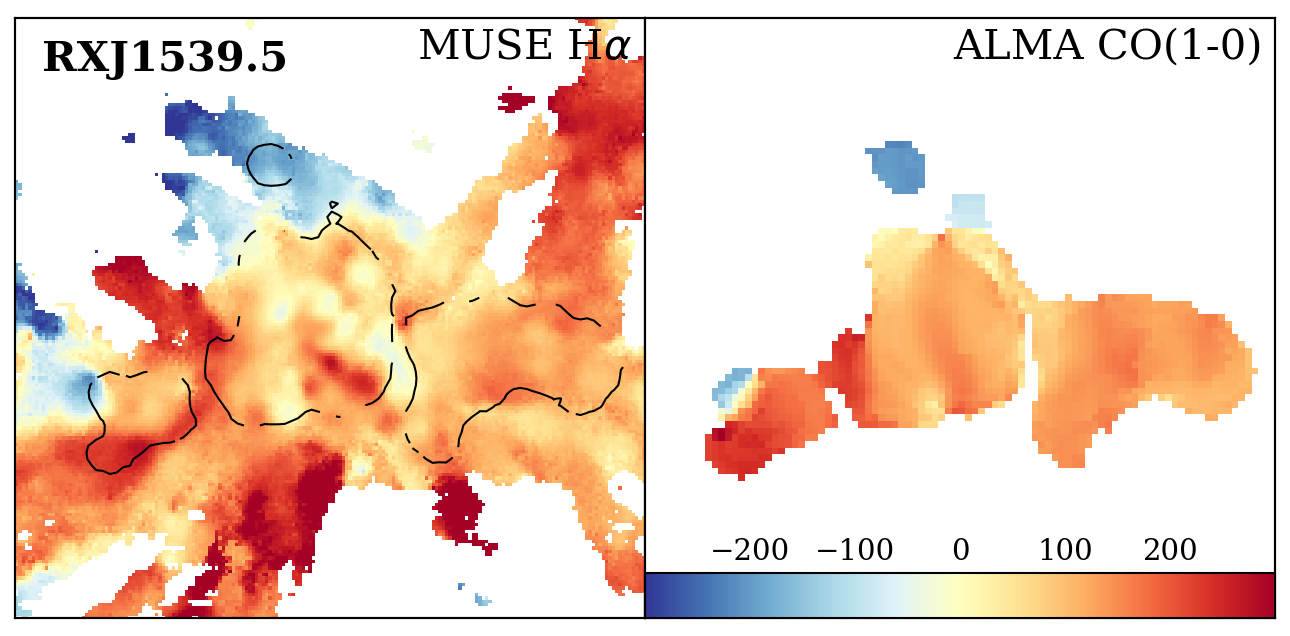}}
    \subfigure{\includegraphics[width=0.49\textwidth]{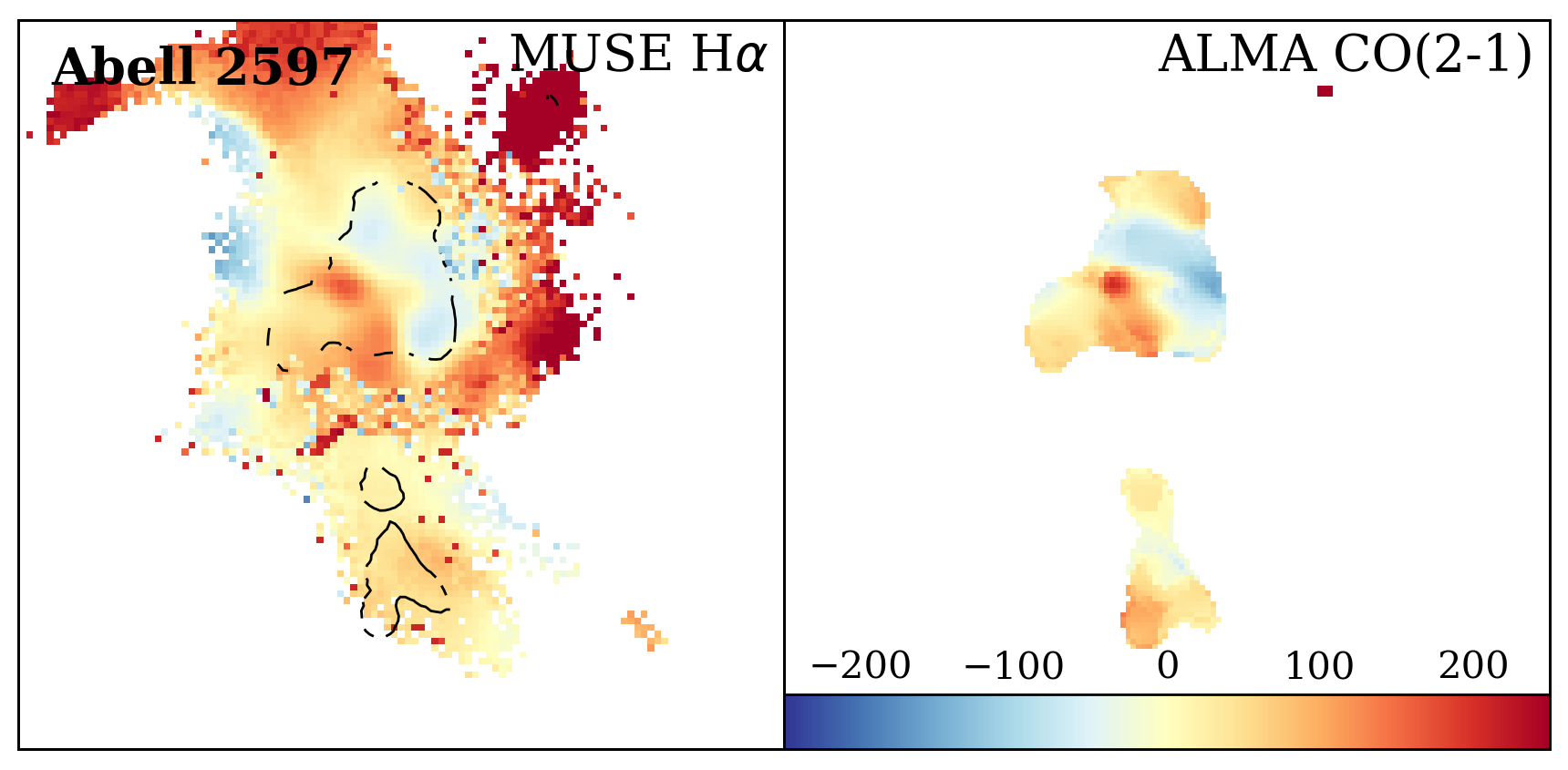}}\\
    \vspace{-2mm}
    \subfigure{\includegraphics[width=0.49\textwidth]{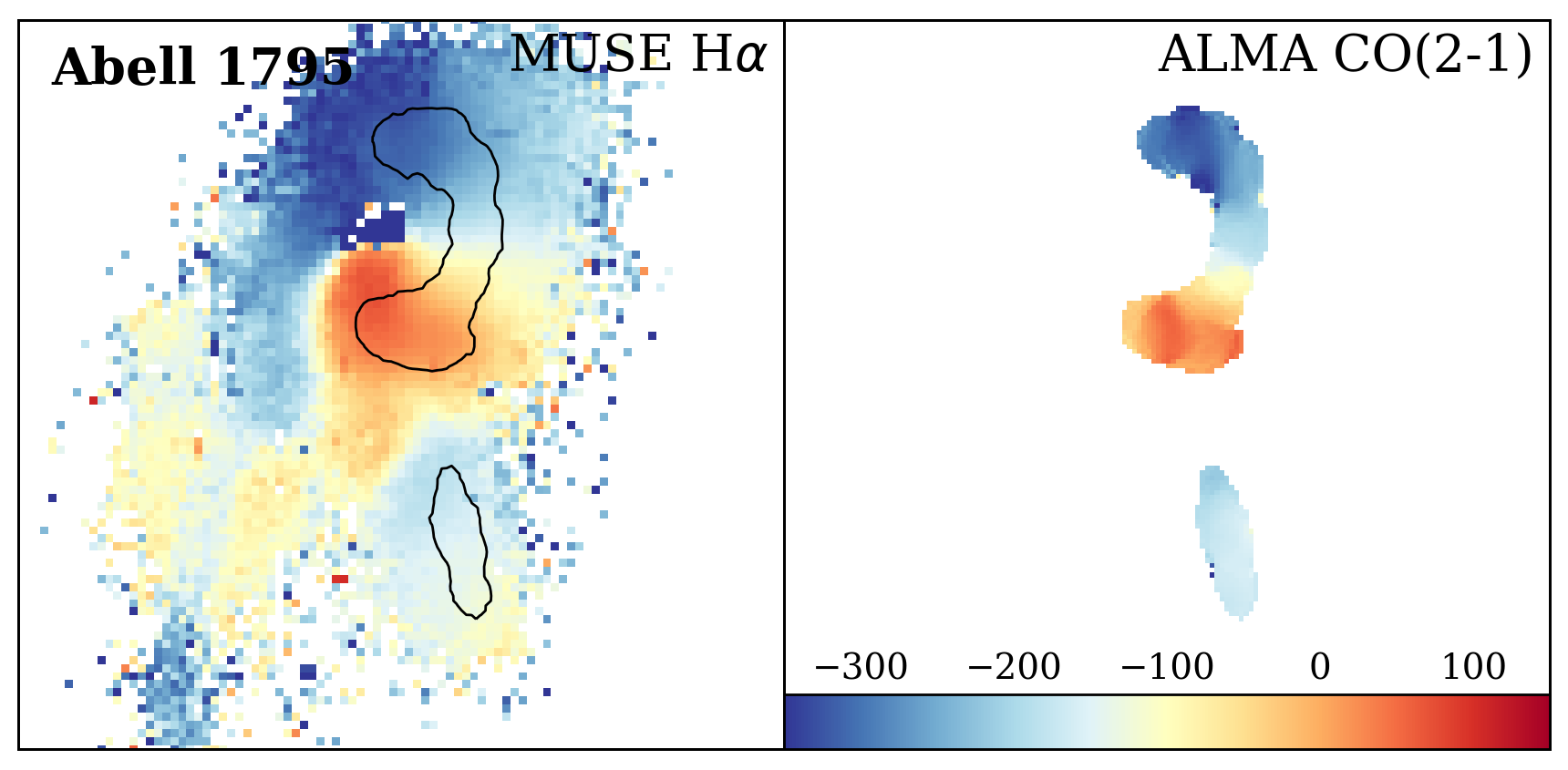}}
    \subfigure{\includegraphics[width=0.49\textwidth]{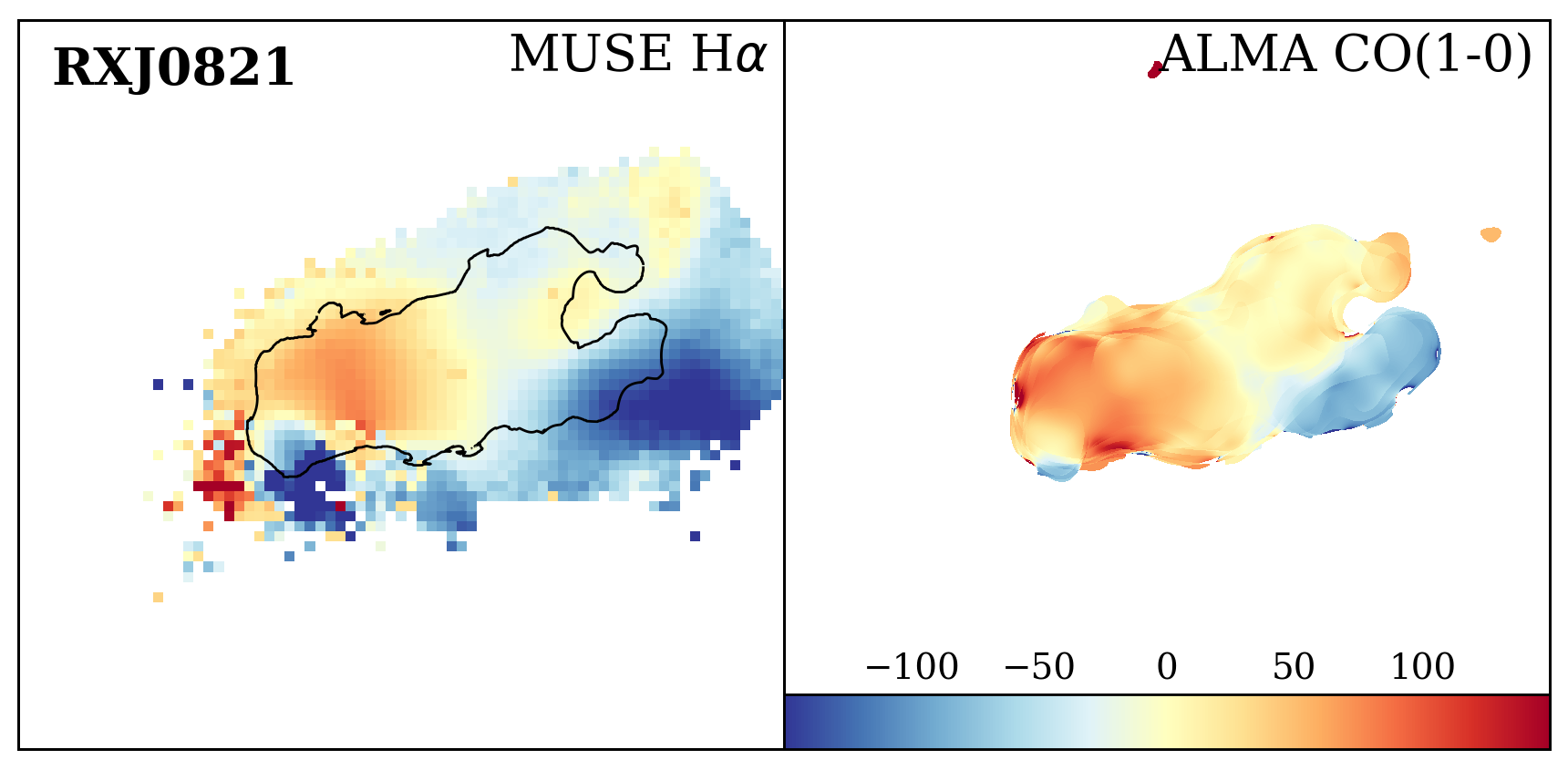}}\\
    \vspace{-2mm}
    \subfigure{\includegraphics[width=0.49\textwidth]{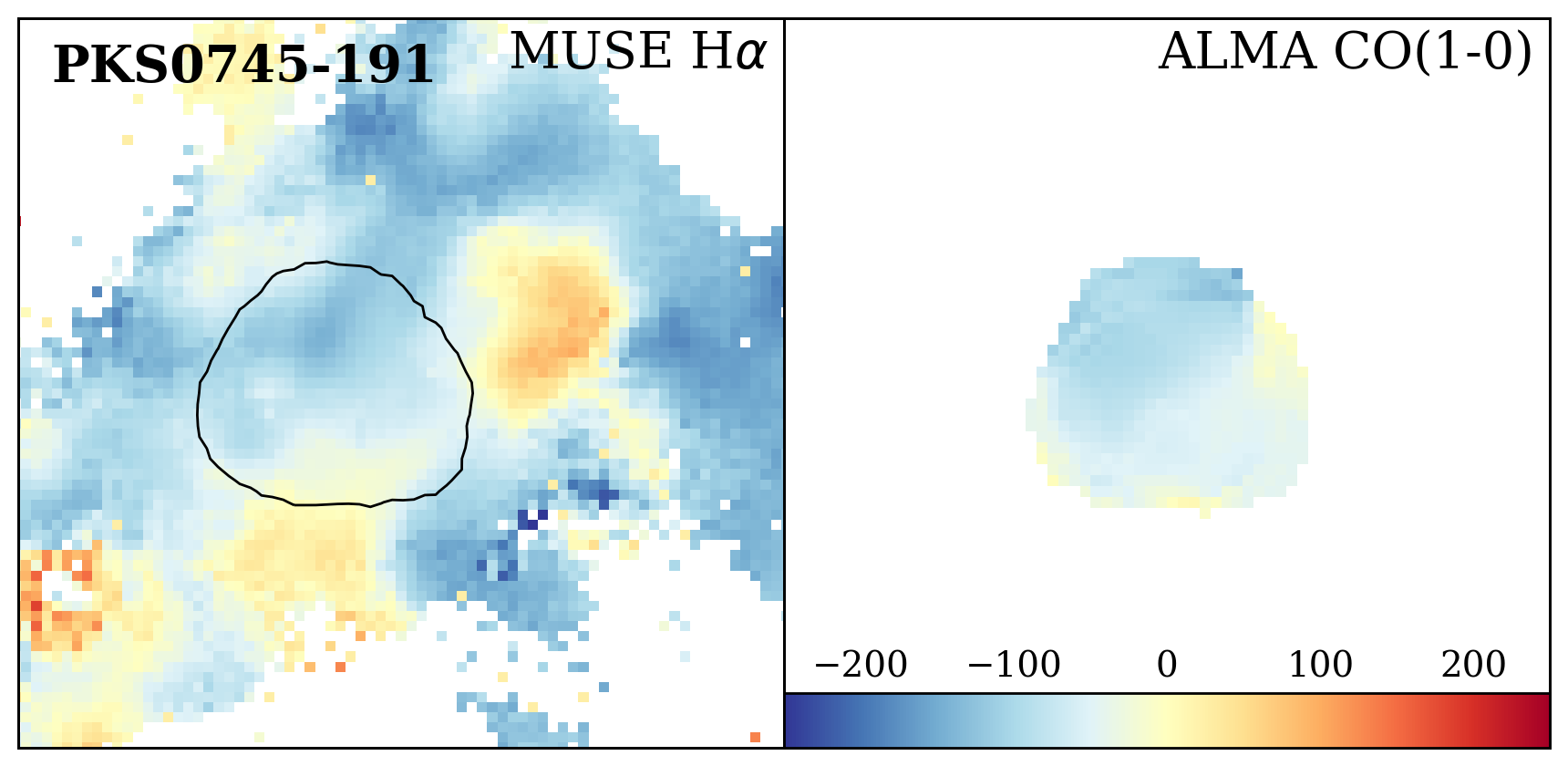}}
    \subfigure{\includegraphics[width=0.49\textwidth]{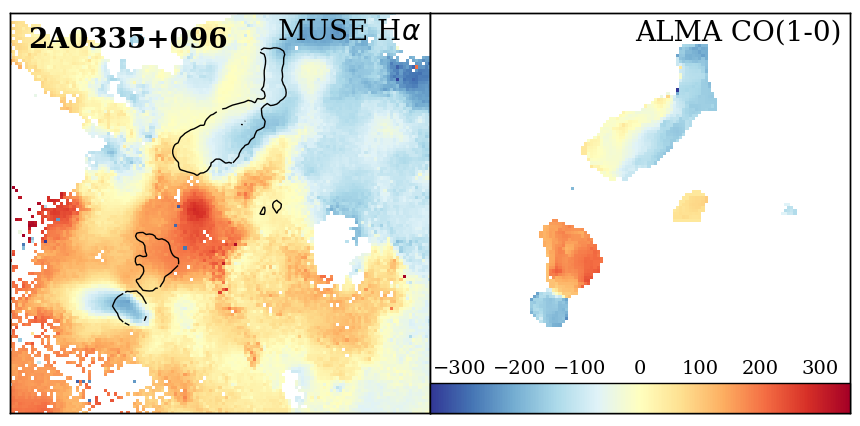}}\\
    \vspace{-2mm}
    \subfigure{\includegraphics[width=0.49\textwidth]{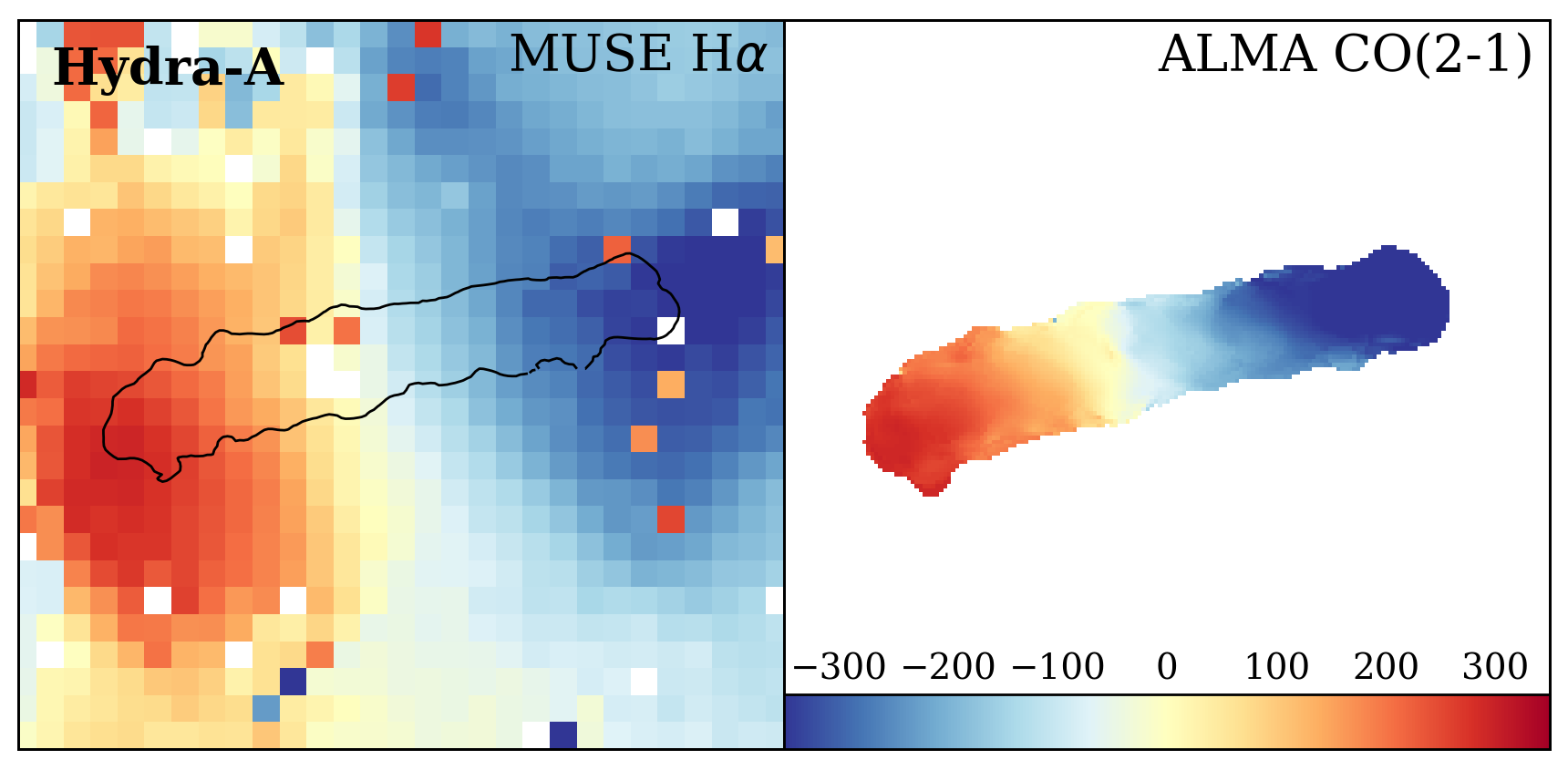}}\\
    \caption{Comparison of the H$\alpha$ MUSE and ALMA CO velocity maps, shown on the same velocity and spatial scales and over the same regions. Both velocity maps look very similar.}
\end{figure*}

\begin{figure*}[htbp!]
\label{fig:entropy_profiles}
    \subfigure{\includegraphics[width=0.33\textwidth]{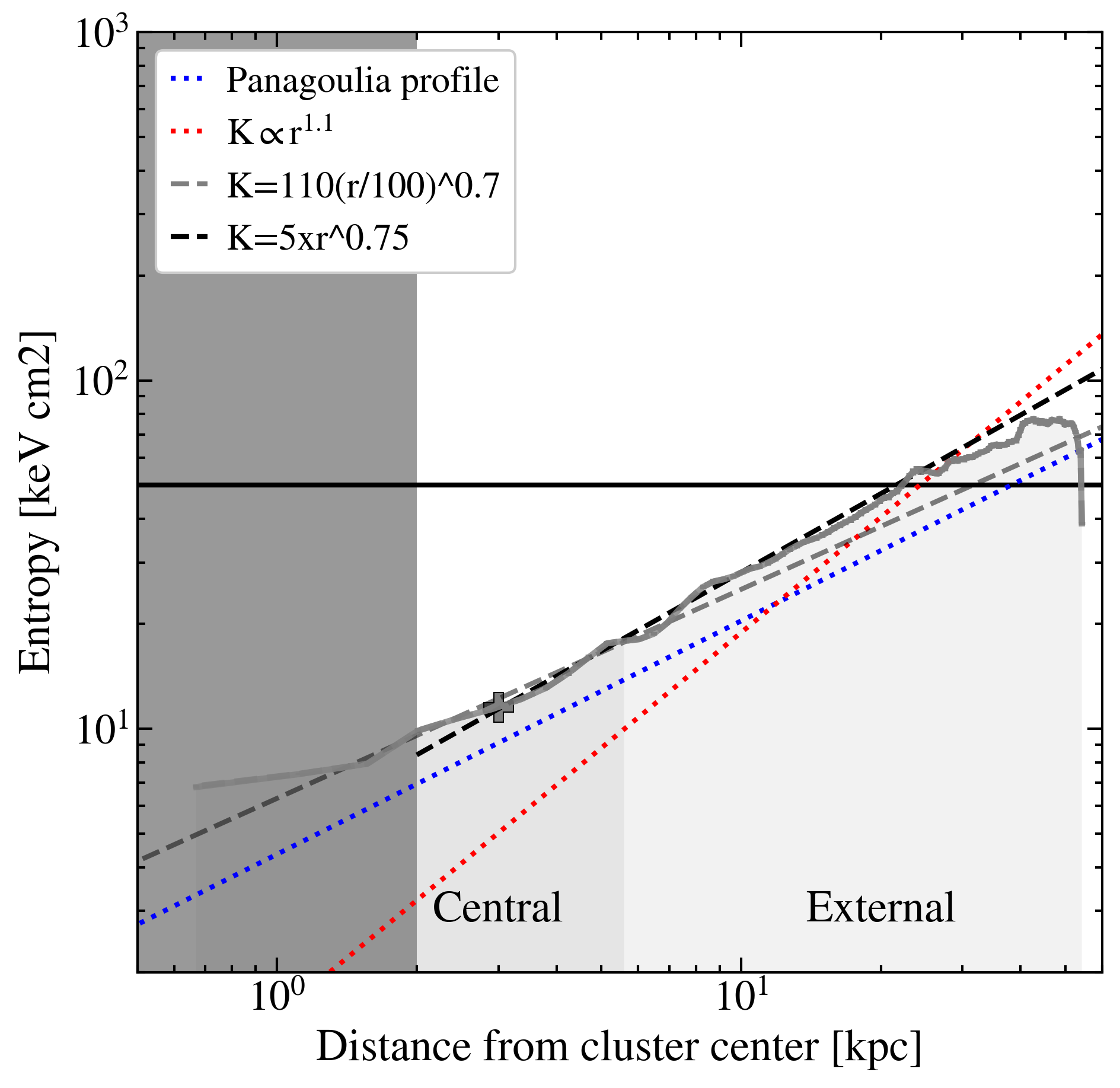}}
    \subfigure{\includegraphics[width=0.33\textwidth]{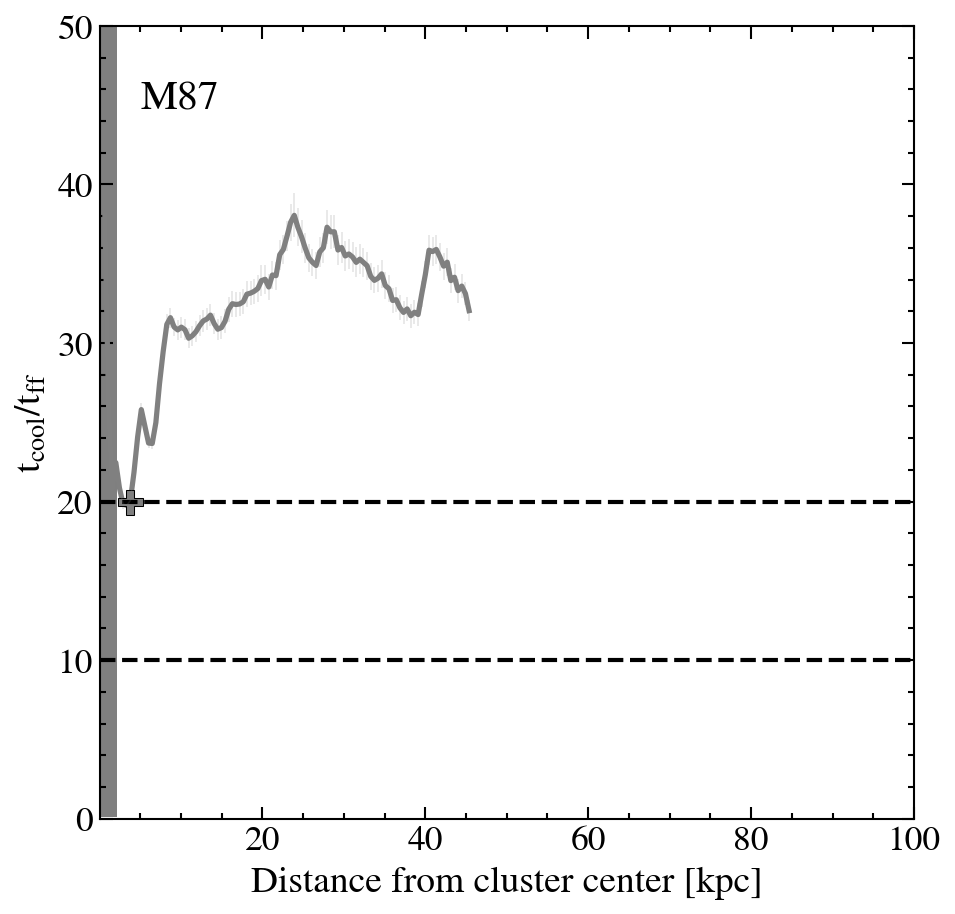}}
    \subfigure{\includegraphics[width=0.33\textwidth]{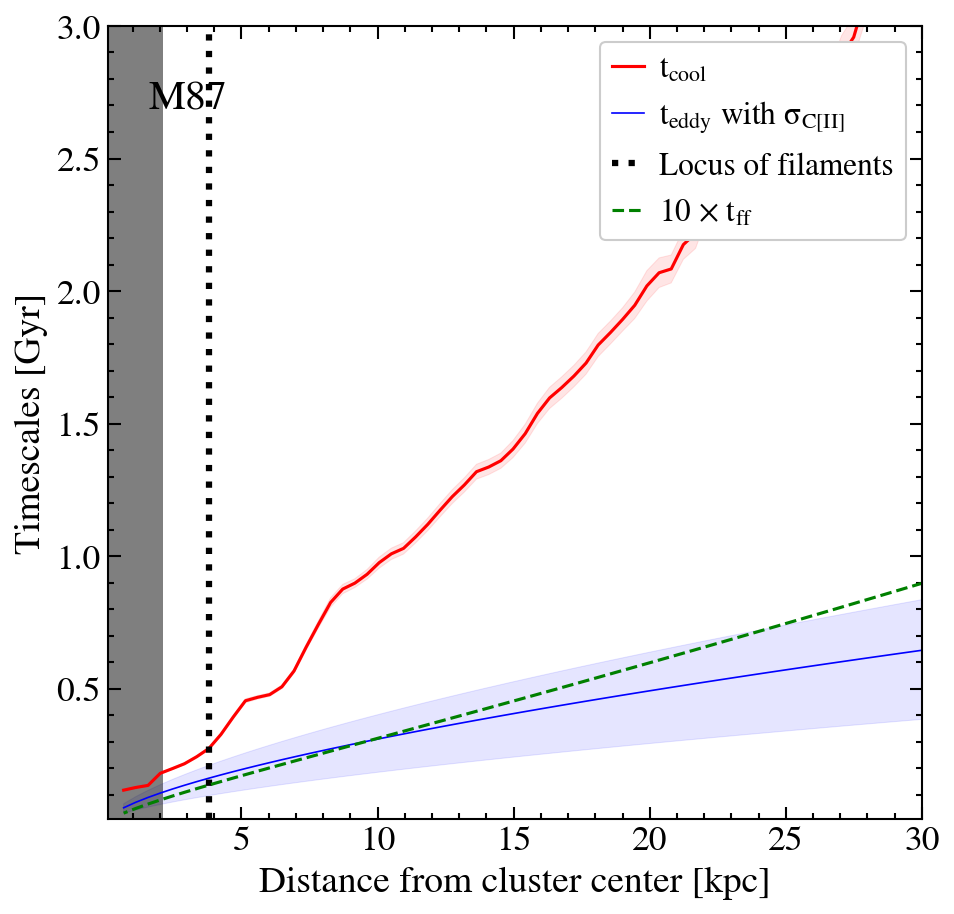}}\\
    \vspace{-4mm}
    \subfigure{\includegraphics[width=0.33\textwidth]{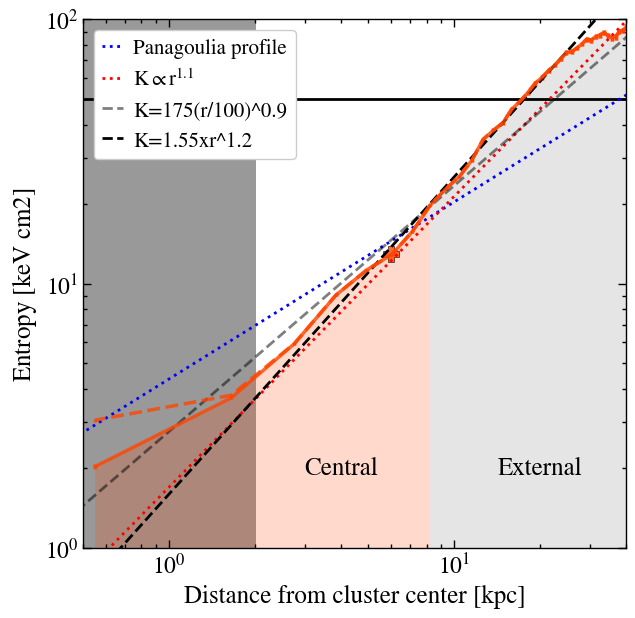}}
    \subfigure{\includegraphics[width=0.33\textwidth]{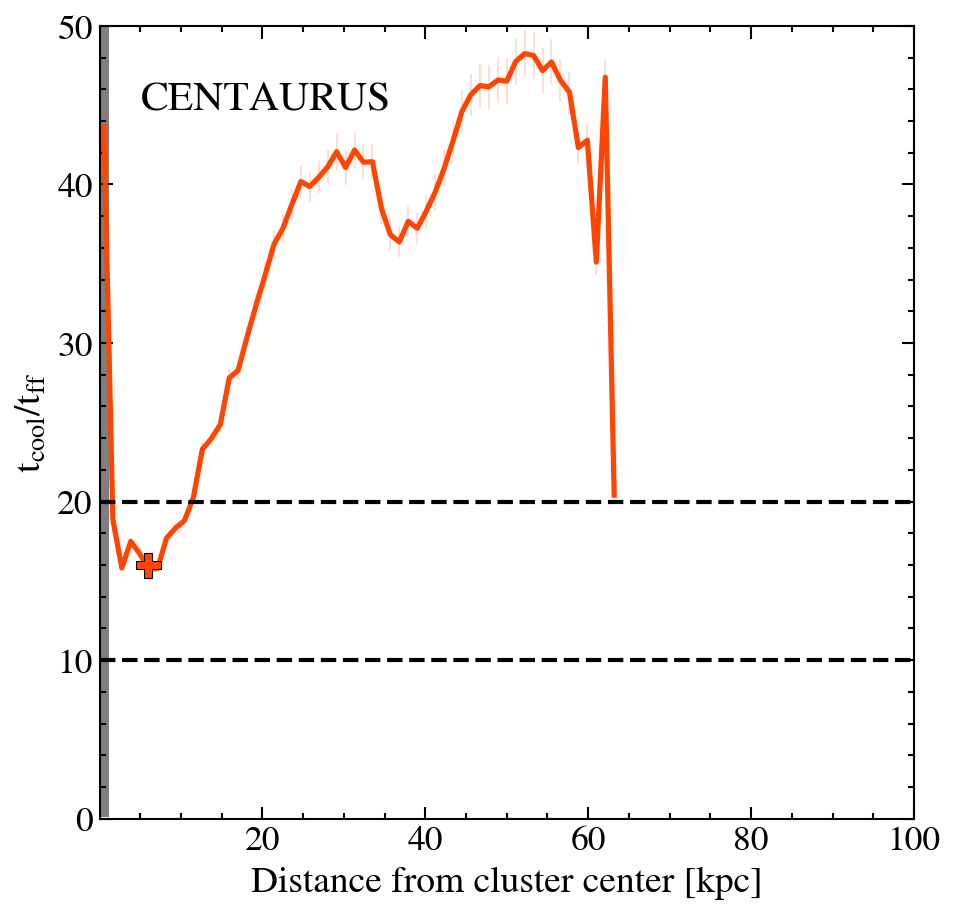}}
    \subfigure{\includegraphics[width=0.33\textwidth]{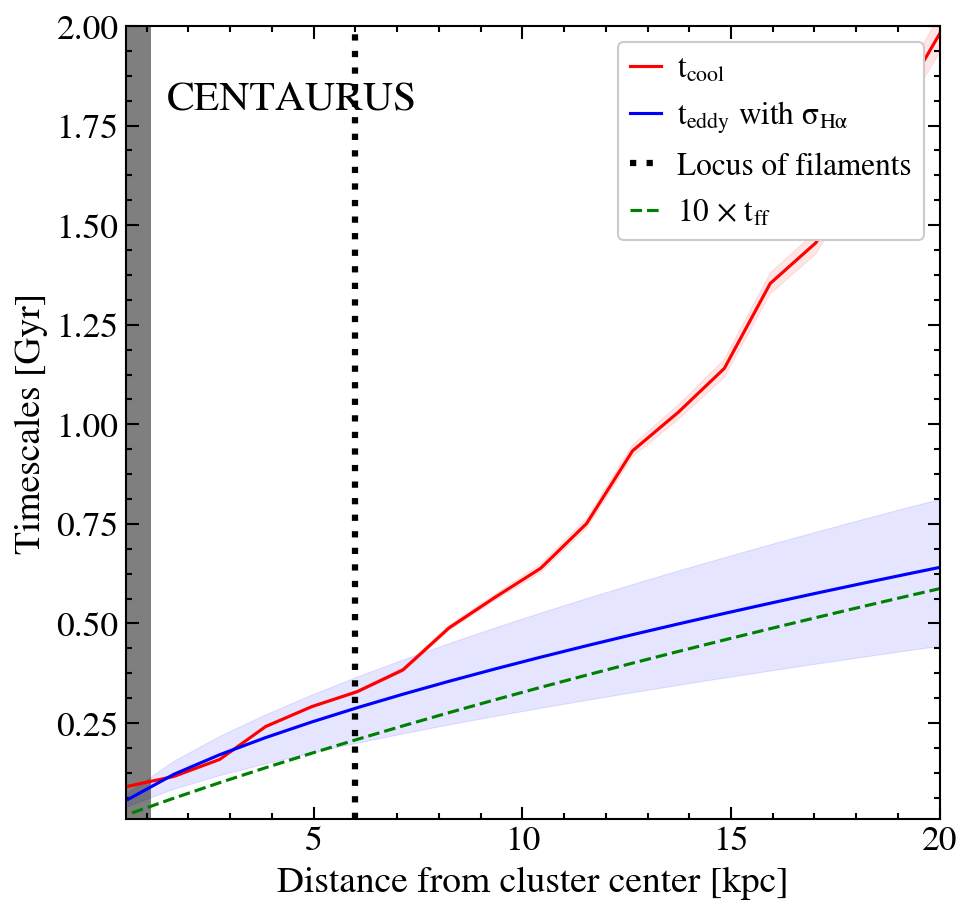}}\\
    \vspace{-4mm}
    \subfigure{\includegraphics[width=0.33\textwidth]{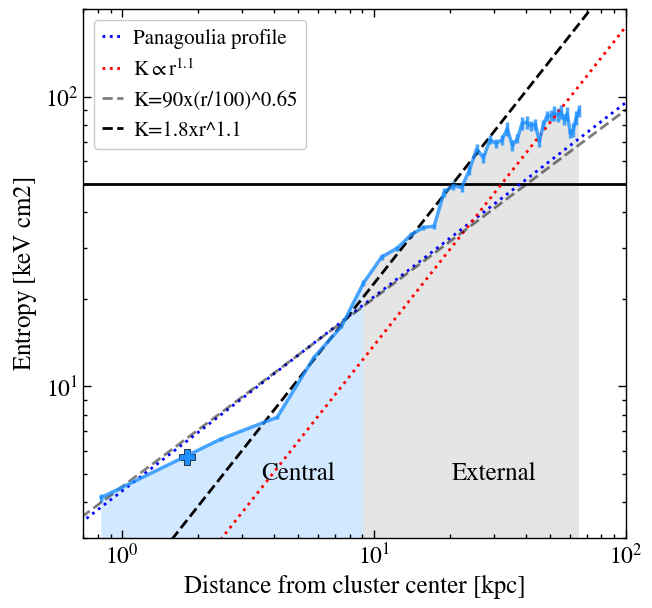}}
    \subfigure{\includegraphics[width=0.33\textwidth]{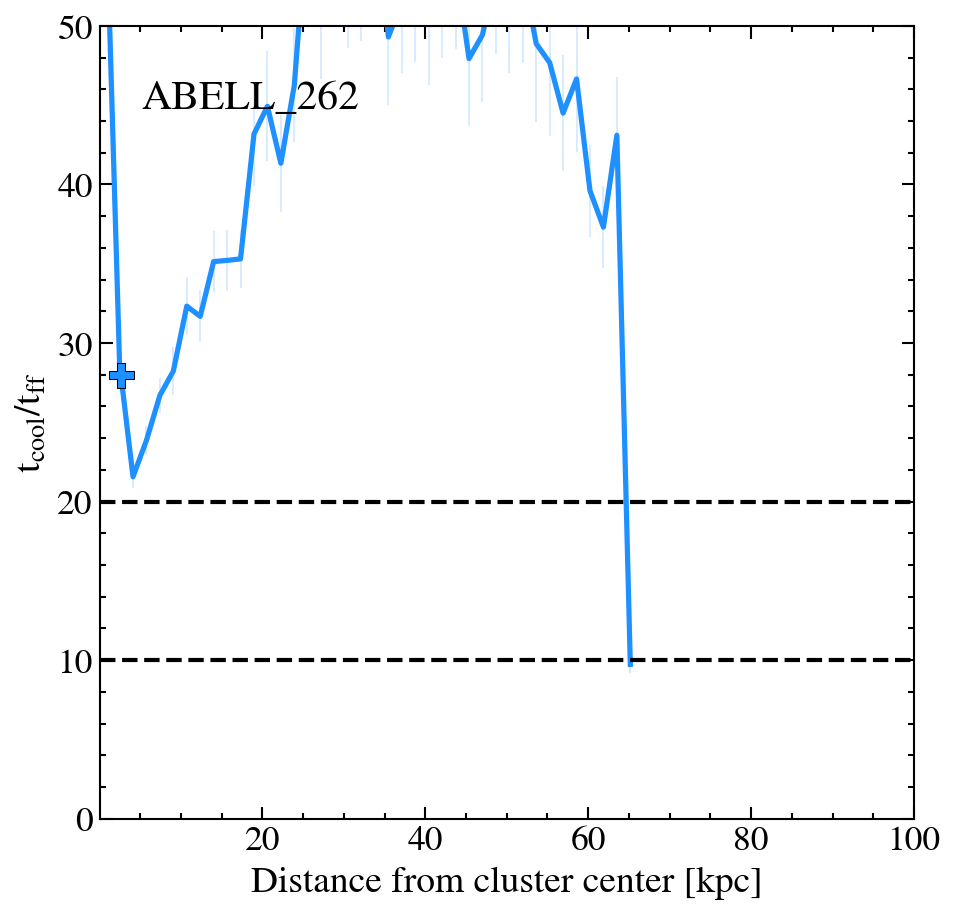}}
    \subfigure{\includegraphics[width=0.33\textwidth]{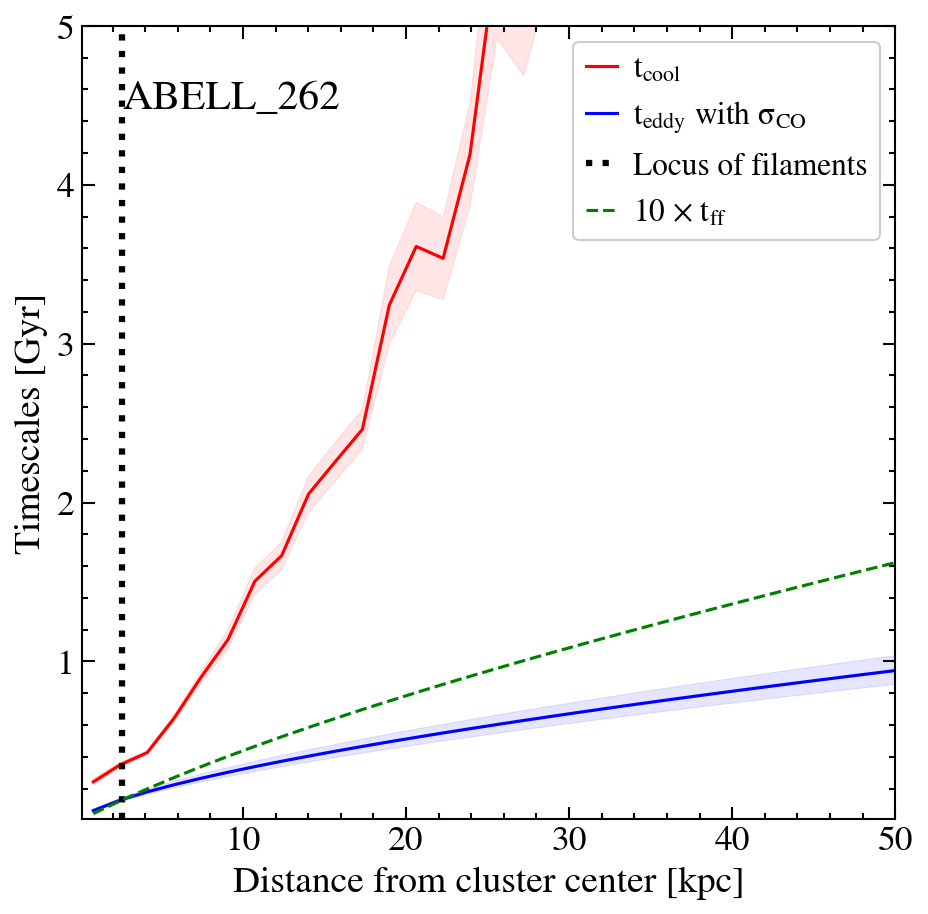}}\\
    \caption{\textit{left panel:} Deprojected entropy profiles, including the values of the innermost entropy done by extrapolating from the fit at large radii from ACCEPT catalogue \citep[][]{cavagnolo09}. We show as a light grey dashed line the best fitting power-law model to the inner entropy profiles as $K \propto r^{\alpha}$, with $\alpha$ varies between 0.65 -- 0.9 \citep[$K \propto r^{3/2}$; ][]{panagoulia14,babyk18}, in the central region \citep[labeled at the bottom left of the panel][]{werner18}. We also overplot as a dark grey dashed line a standard cluster entropy profile (External) with scaling law as $K \propto r^{\beta}$, with $\beta$ ranging from 1.1 -- 1.2. \citep[$K \propto r^{1.1}$; ][]{Voit_2015,babyk18}. The best fits are given in the legend at the upper left of the panel. Plus a Panagoulia profile as been added with a dotted blue line \citep{panagoulia14} and $K \propto r^{1.1}$ profile \citep{hogan17b}. \textit{middle panel:} The ratio of the cooling time to free-fall time, t$_{\rm cool}$/t$_{\rm ff}$, as a function of radius. We highlight t$_{\rm cool}$/t$_{\rm ff}$ = 10 and 20 with horizontal dashed lines because this appears to be the approximate threshold for onset of thermal instabilities in our sample. The source name is given in the legend as is the line style. \textit{right panel:} Cooling (red curve) and turbulence eddy turnover timescales (blue curve).  We also indicate the radius at which the filaments are located (vertical dotted grey line), 10$\times$t$_{\rm ff}$ as a function of radius (dashed green line; see legend at the upper right of the panel). In the left and middle panels, the radial extent of the largest filament is indicated by the colored cross.}
    
\end{figure*}

\begin{figure*}[htbp!]
\label{fig:entropy_profiles2}
    \vspace{-4mm}
    \subfigure{\includegraphics[width=0.33\textwidth]{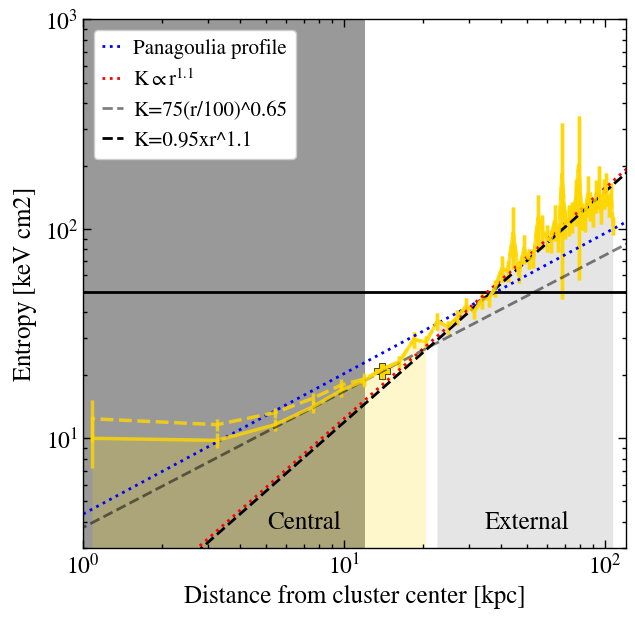}}
    \subfigure{\includegraphics[width=0.33\textwidth]{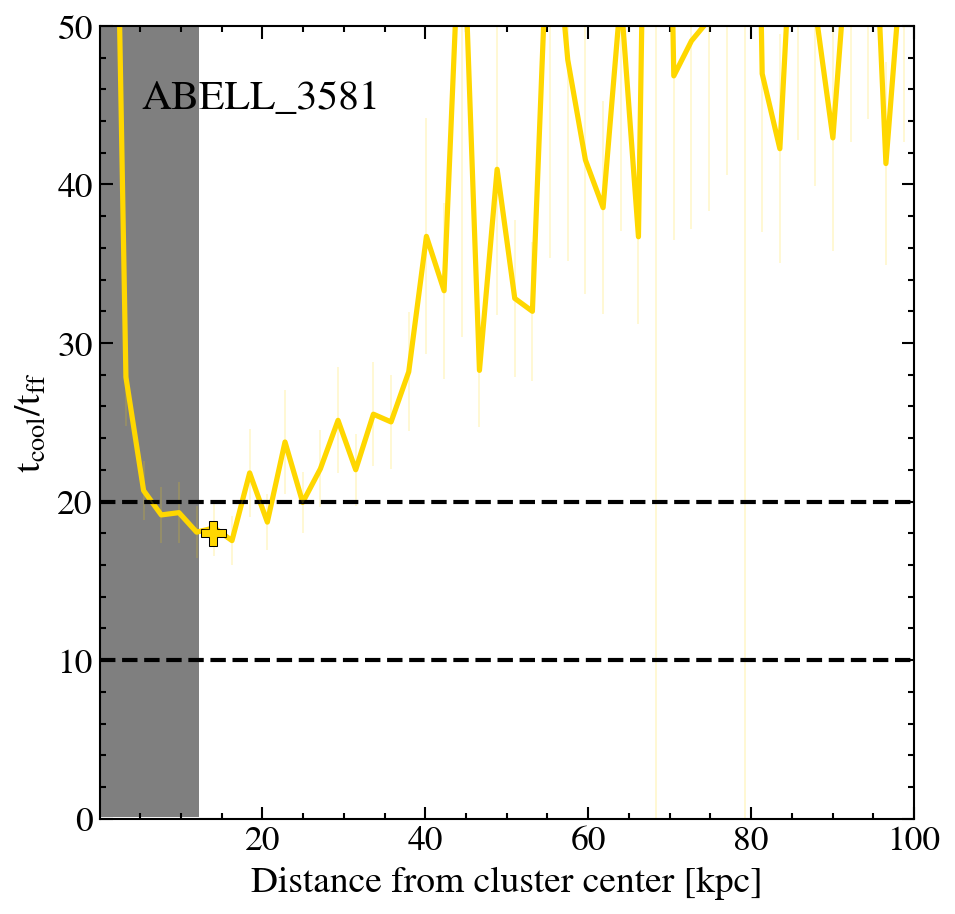}}
    \subfigure{\includegraphics[width=0.33\textwidth]{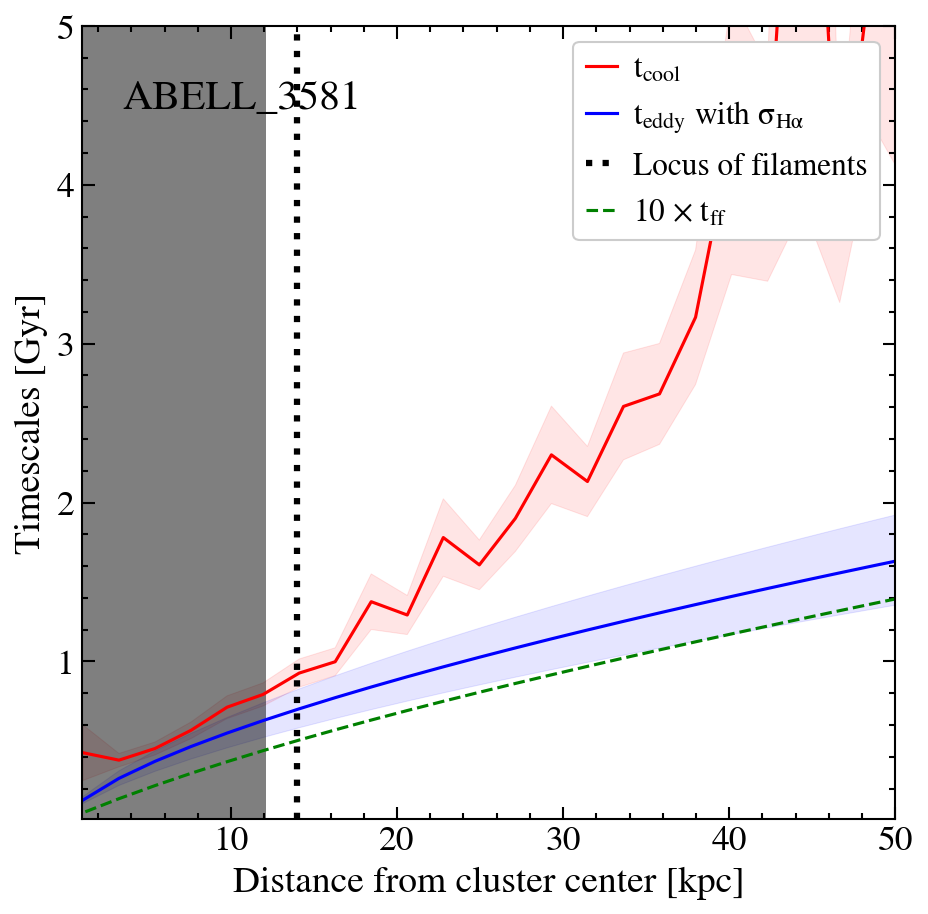}}\\
    \vspace{-4mm}
    \subfigure{\includegraphics[width=0.33\textwidth]{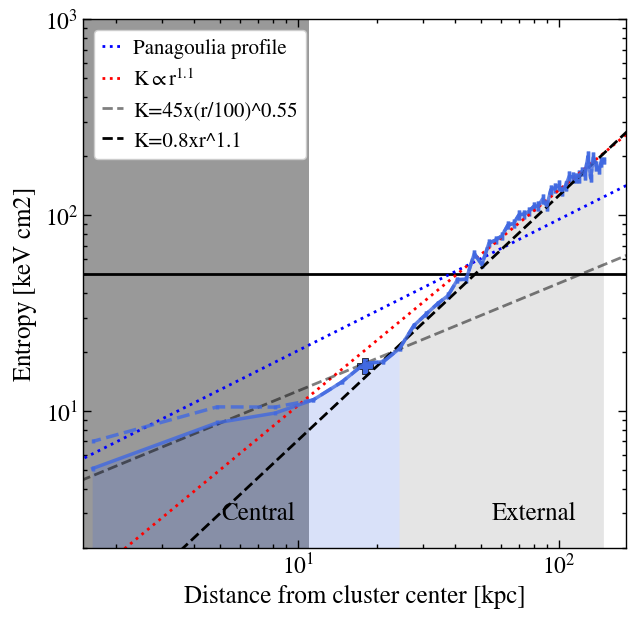}}
    \subfigure{\includegraphics[width=0.33\textwidth]{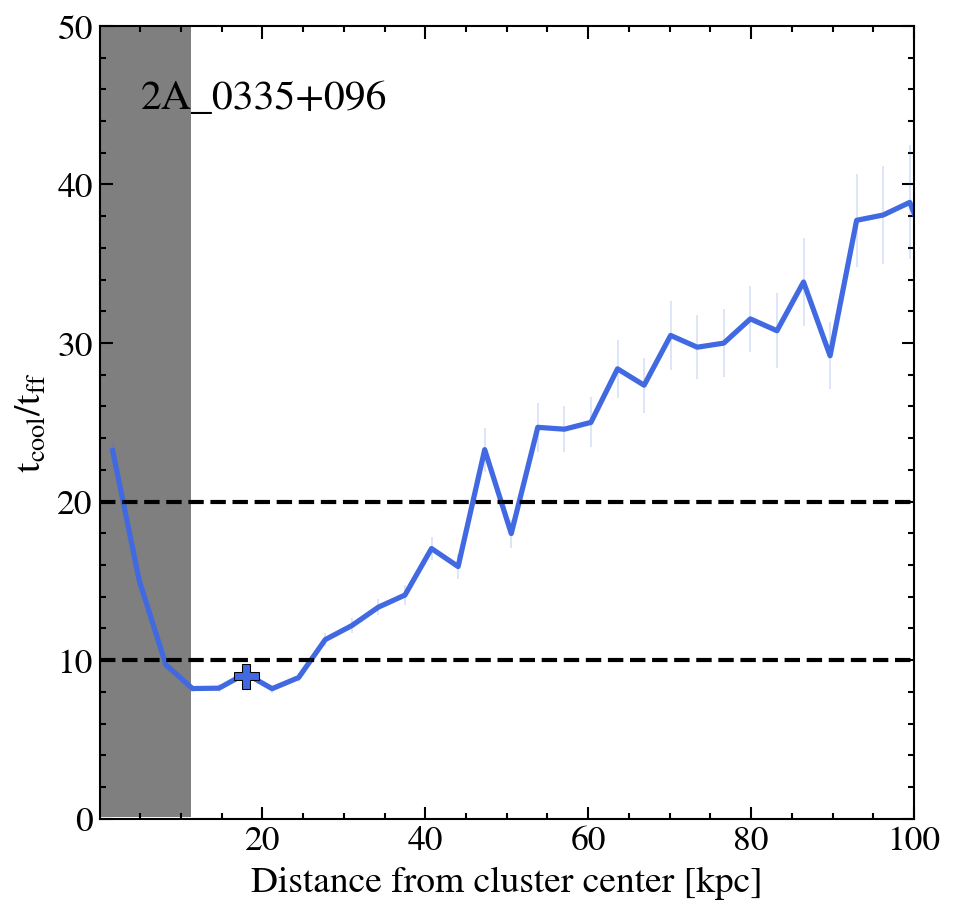}}
    \subfigure{\includegraphics[width=0.33\textwidth]{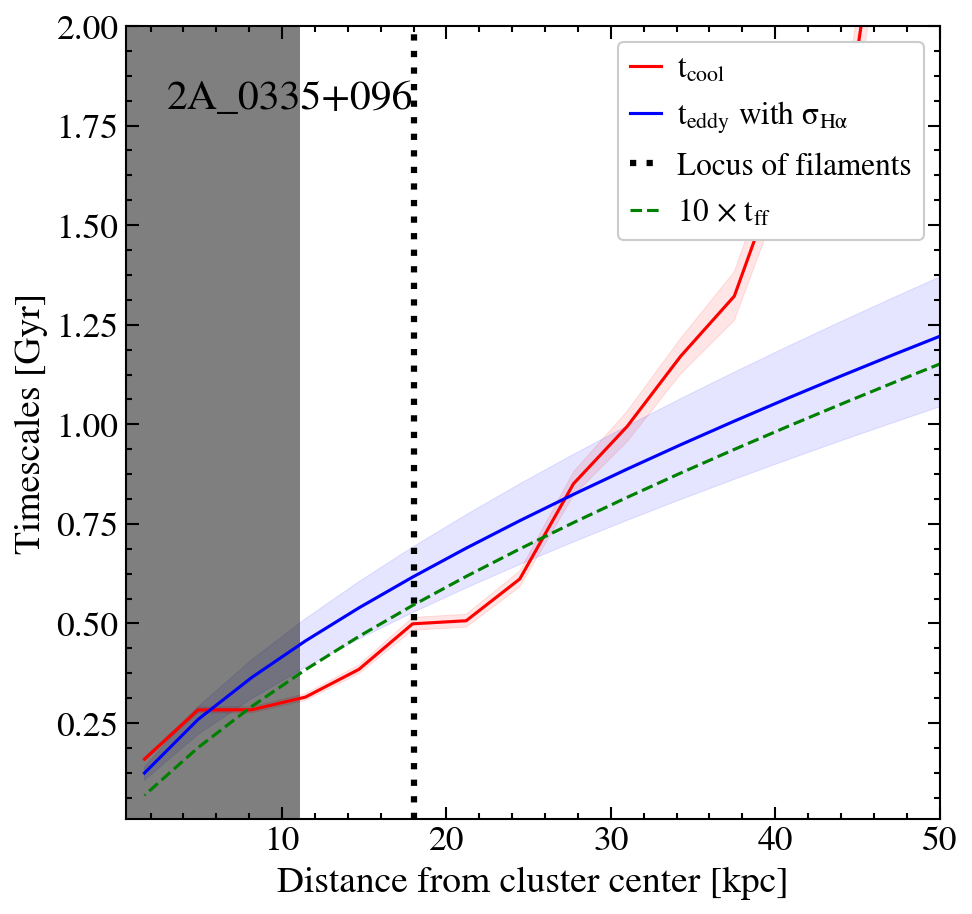}}\\
    \vspace{-4mm}
    \subfigure{\includegraphics[width=0.33\textwidth]{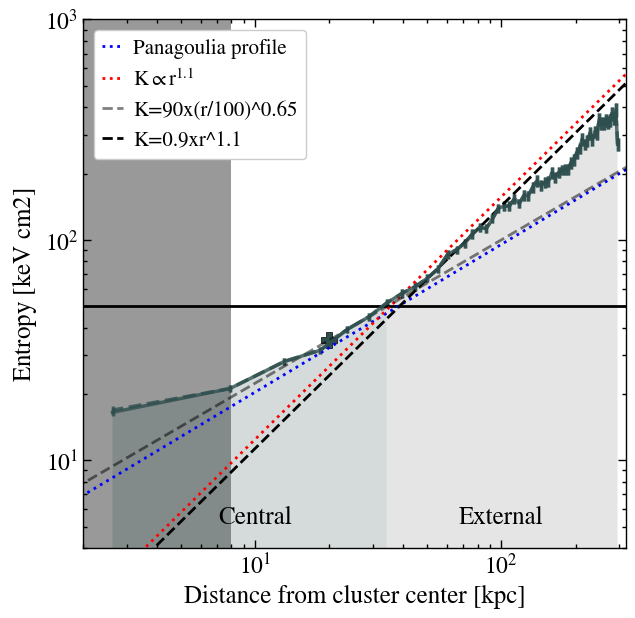}}
    \subfigure{\includegraphics[width=0.33\textwidth]{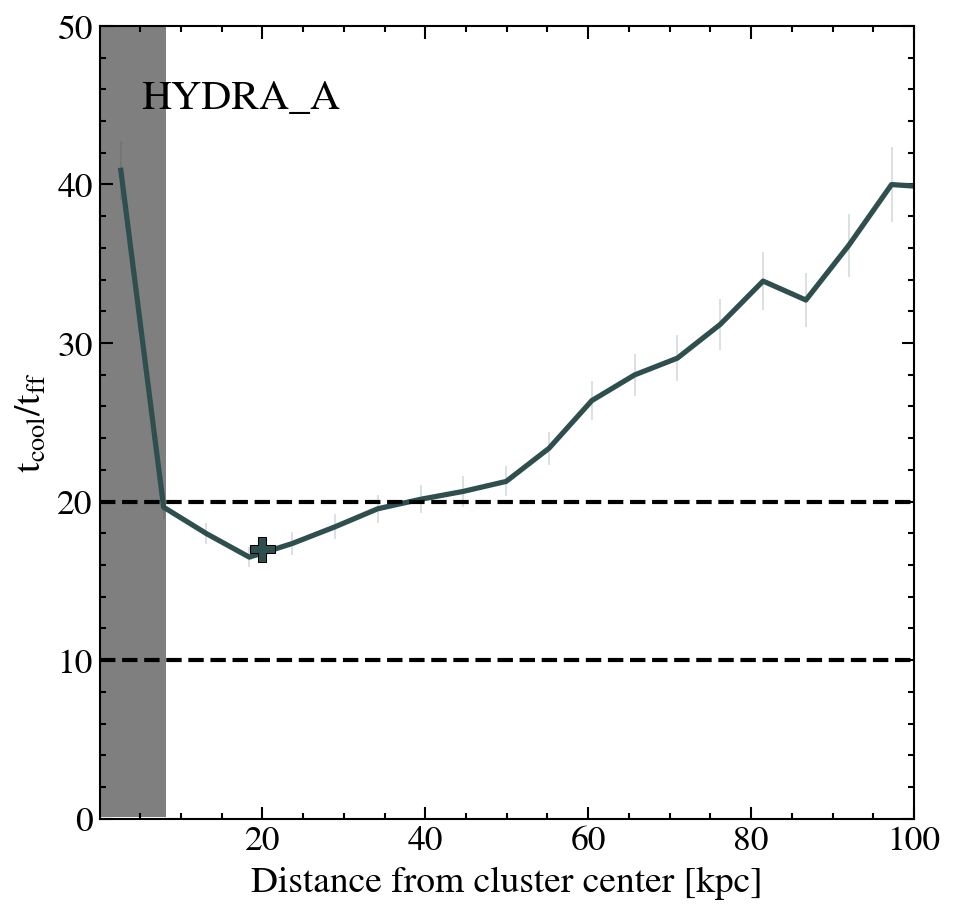}}
    \subfigure{\includegraphics[width=0.33\textwidth]{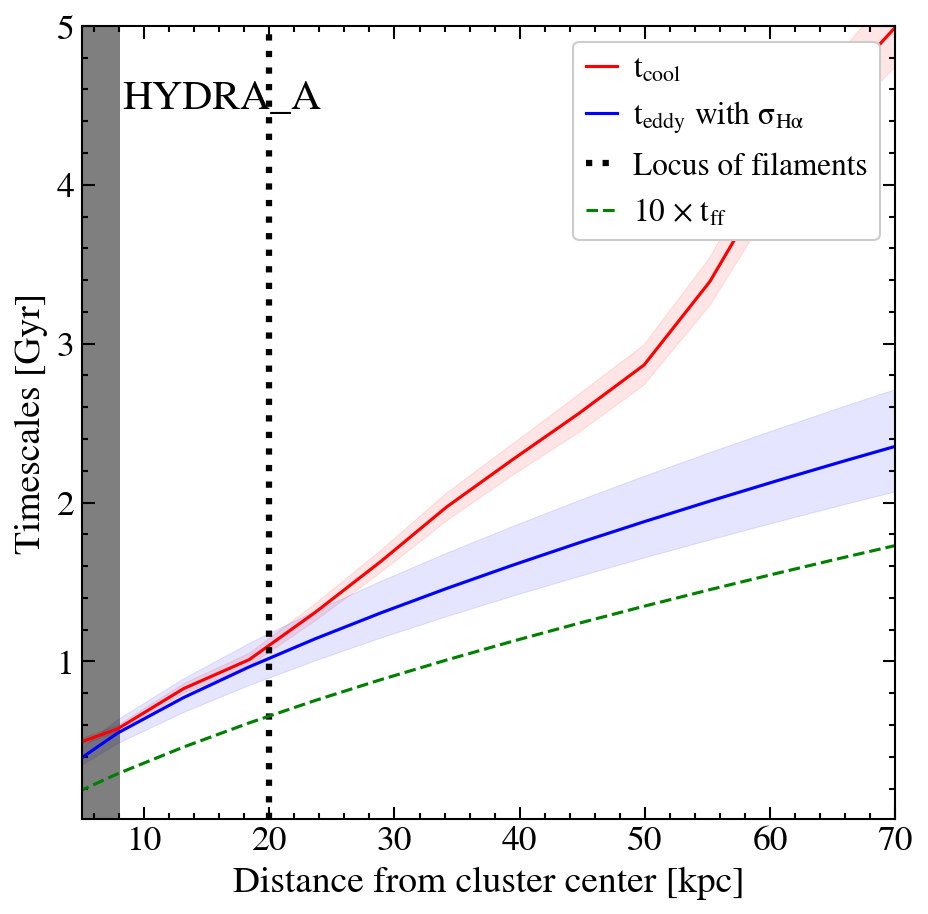}}\\
    \vspace{-4mm}
    \subfigure{\includegraphics[width=0.33\textwidth]{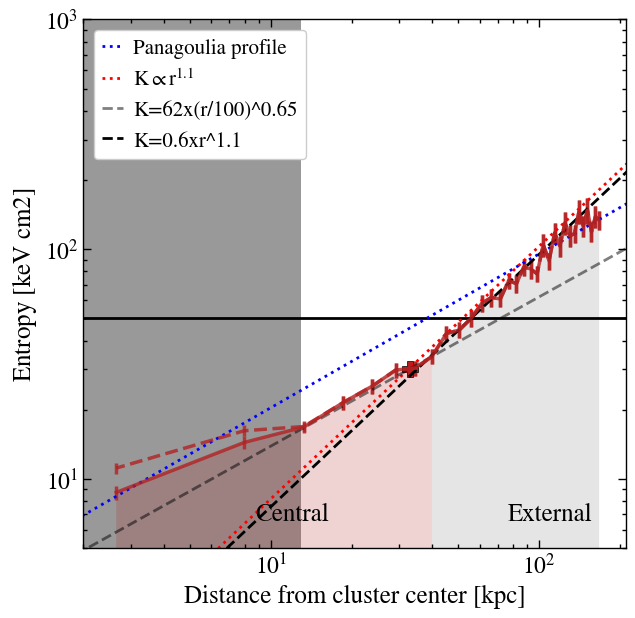}}
    \subfigure{\includegraphics[width=0.33\textwidth]{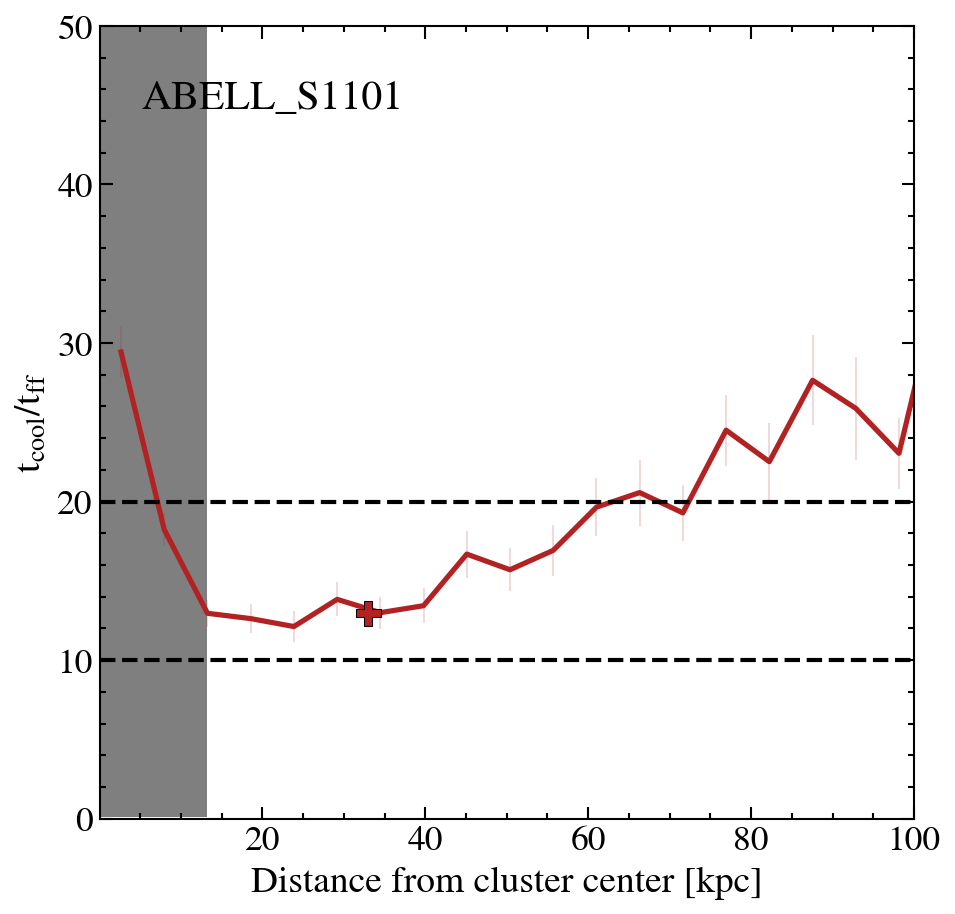}}
    \subfigure{\includegraphics[width=0.33\textwidth]{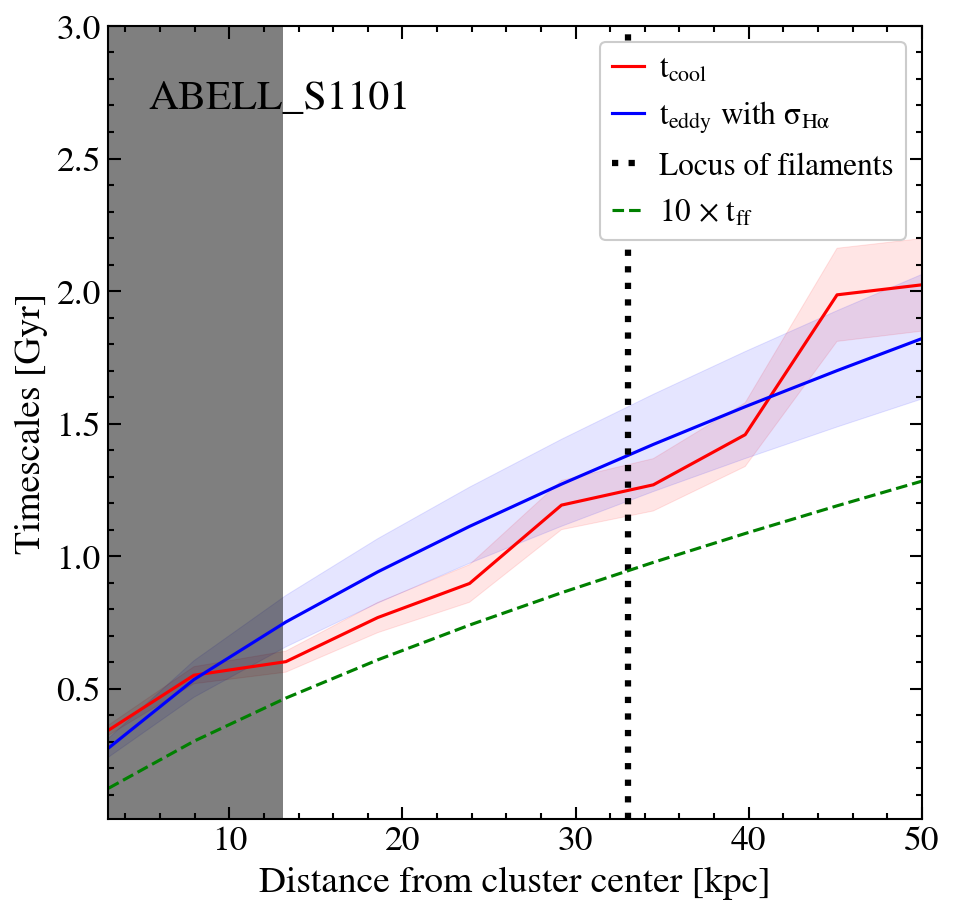}}\\
    \caption{Continuation of Fig.~\ref{fig:entropy_profiles}.}
\end{figure*}

\begin{figure*}[htbp!]
\label{fig:entropy_profiles3}
    \vspace{-4mm}
    \subfigure{\includegraphics[width=0.33\textwidth]{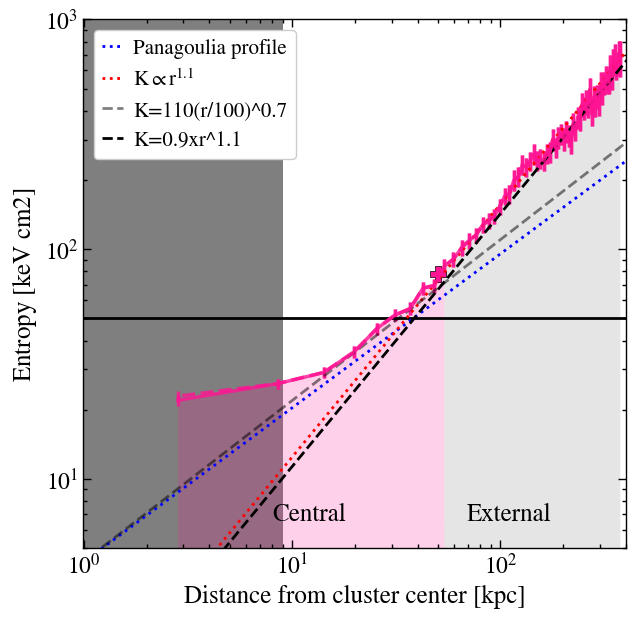}}
    \subfigure{\includegraphics[width=0.33\textwidth]{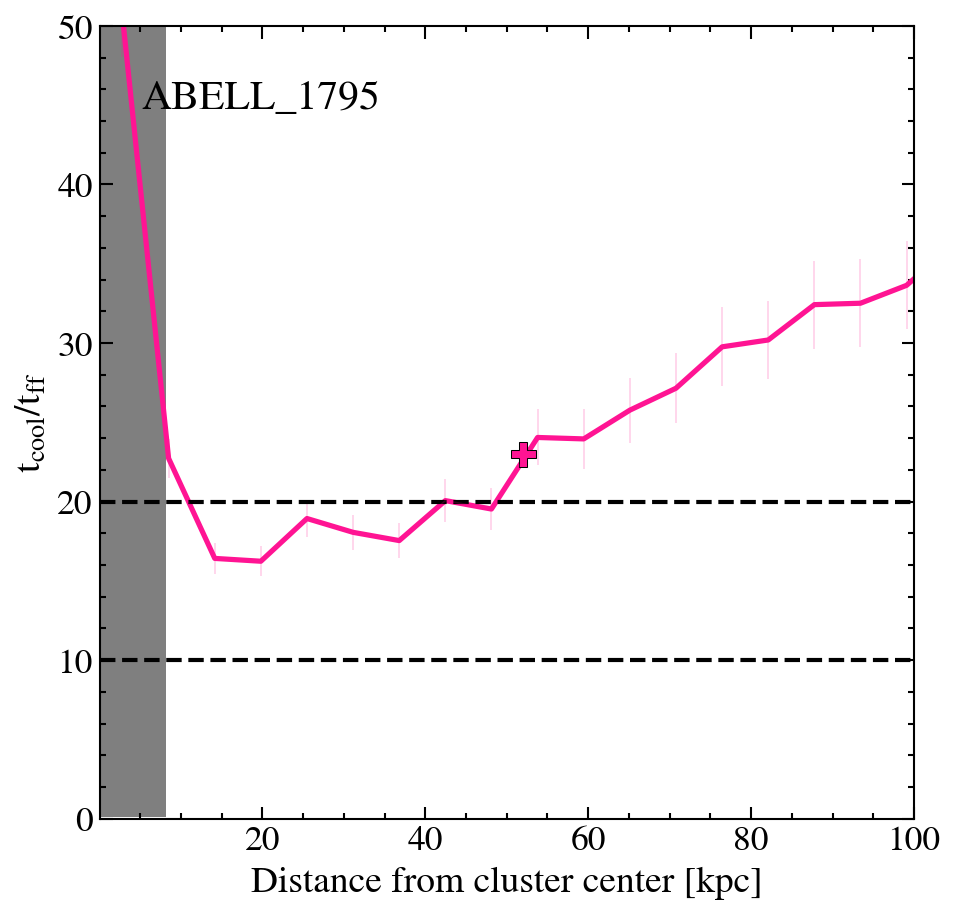}}
    \subfigure{\includegraphics[width=0.33\textwidth]{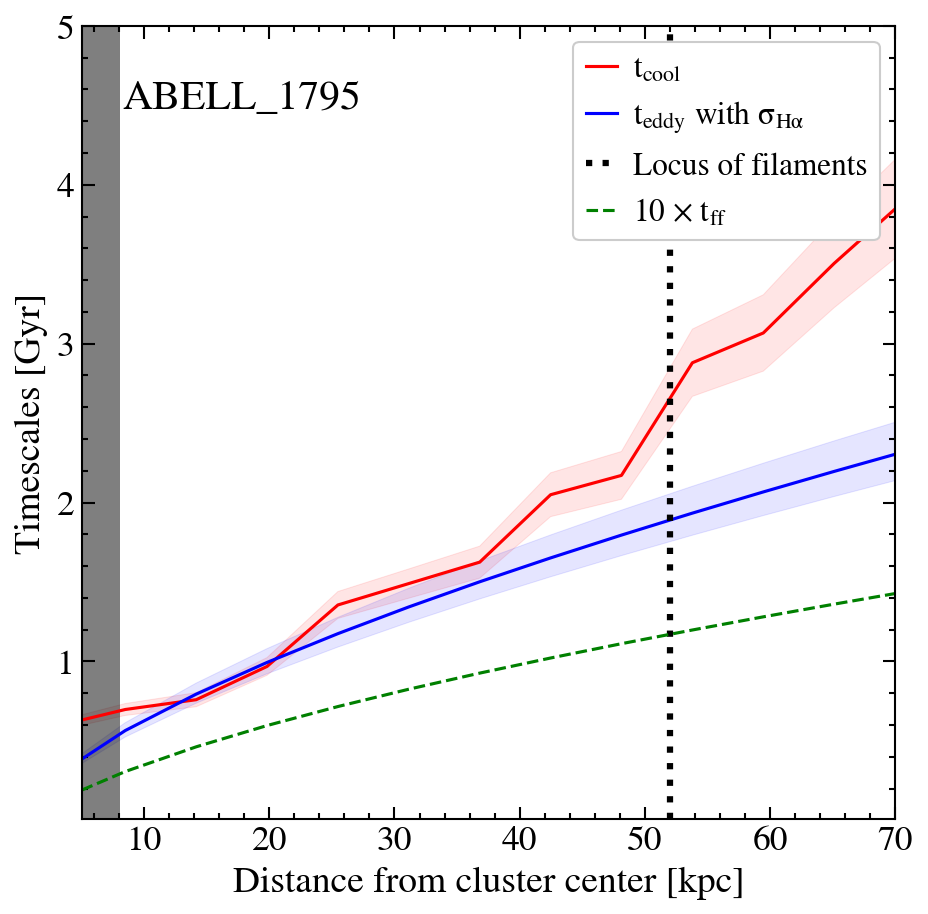}}\\
    \vspace{-4mm}
    \subfigure{\includegraphics[width=0.33\textwidth]{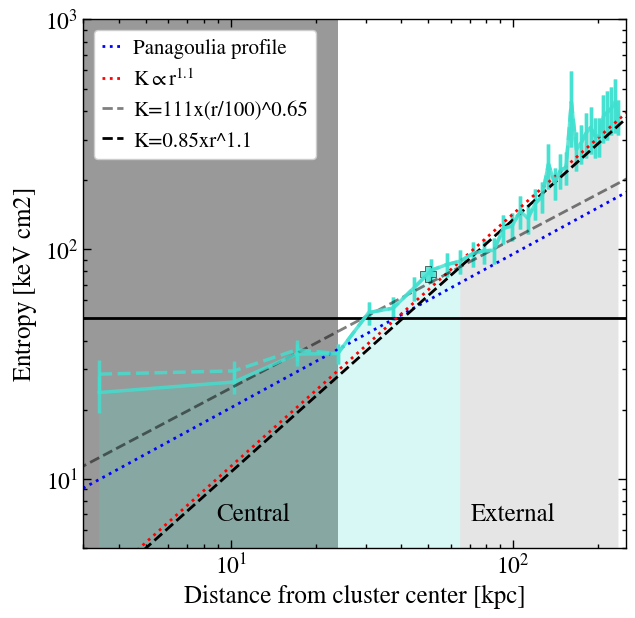}}
    \subfigure{\includegraphics[width=0.33\textwidth]{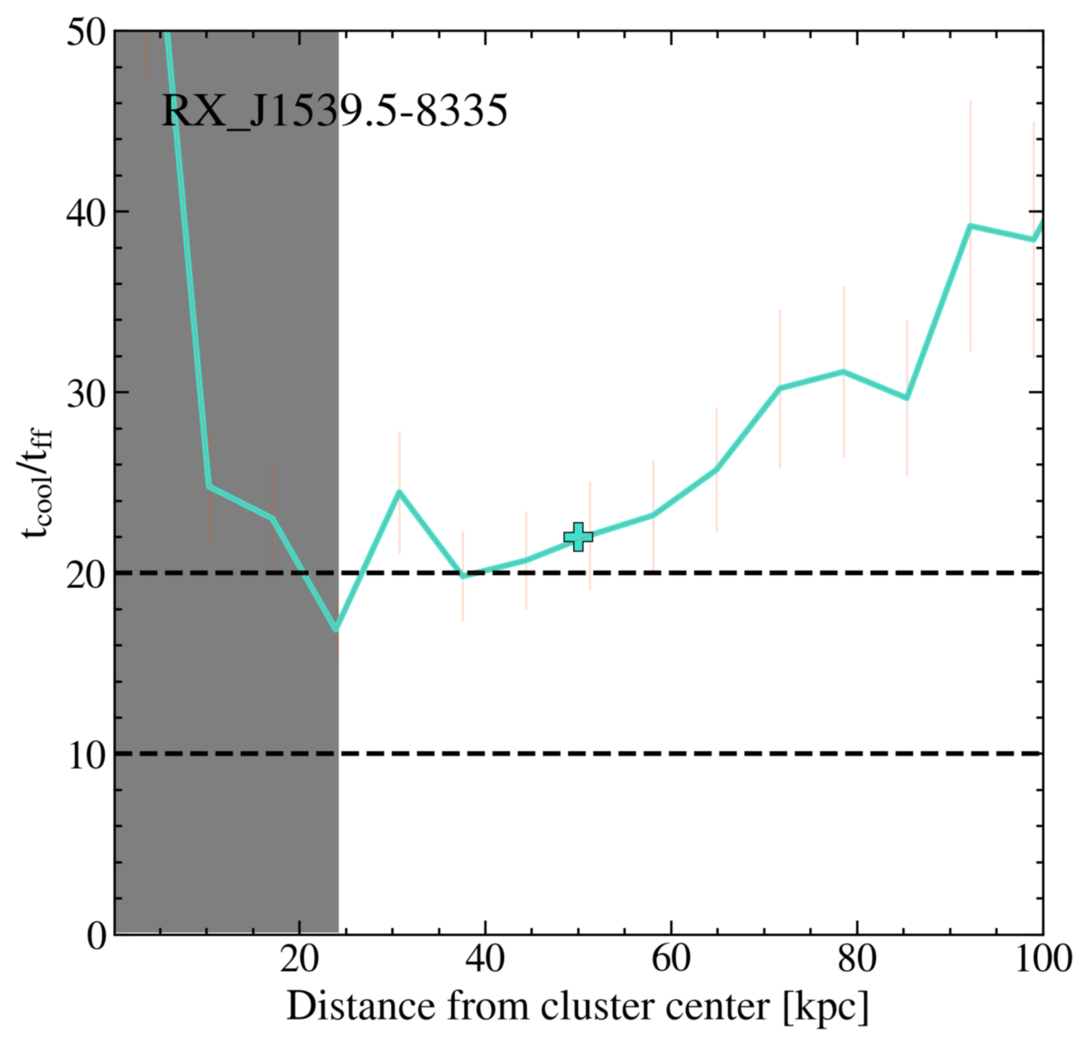}}\\
    \vspace{-4mm}
    \subfigure{\includegraphics[width=0.33\textwidth]{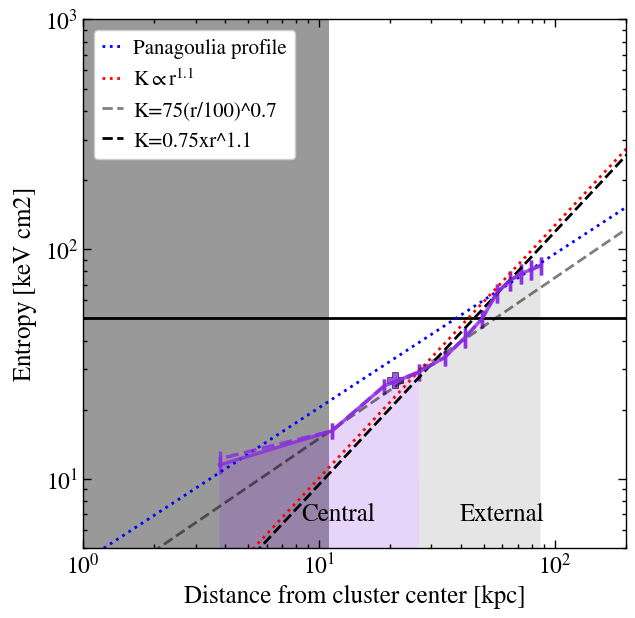}}
    \subfigure{\includegraphics[width=0.33\textwidth]{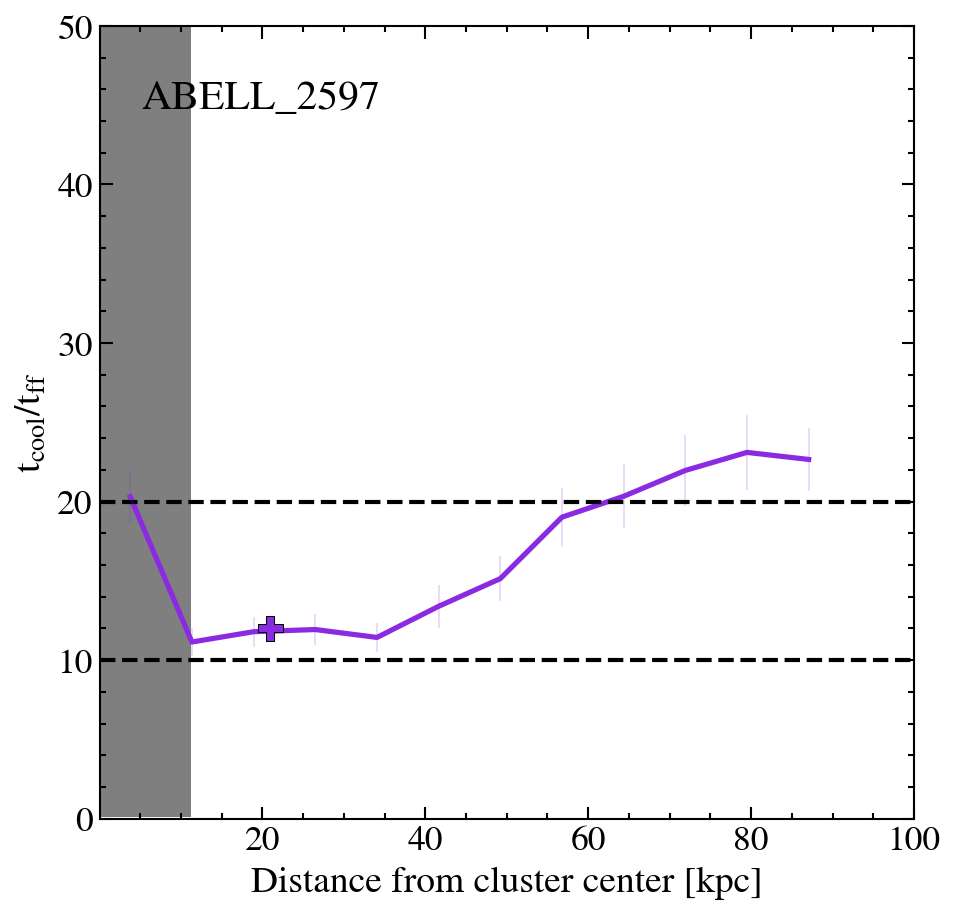}}
    \subfigure{\includegraphics[width=0.33\textwidth]{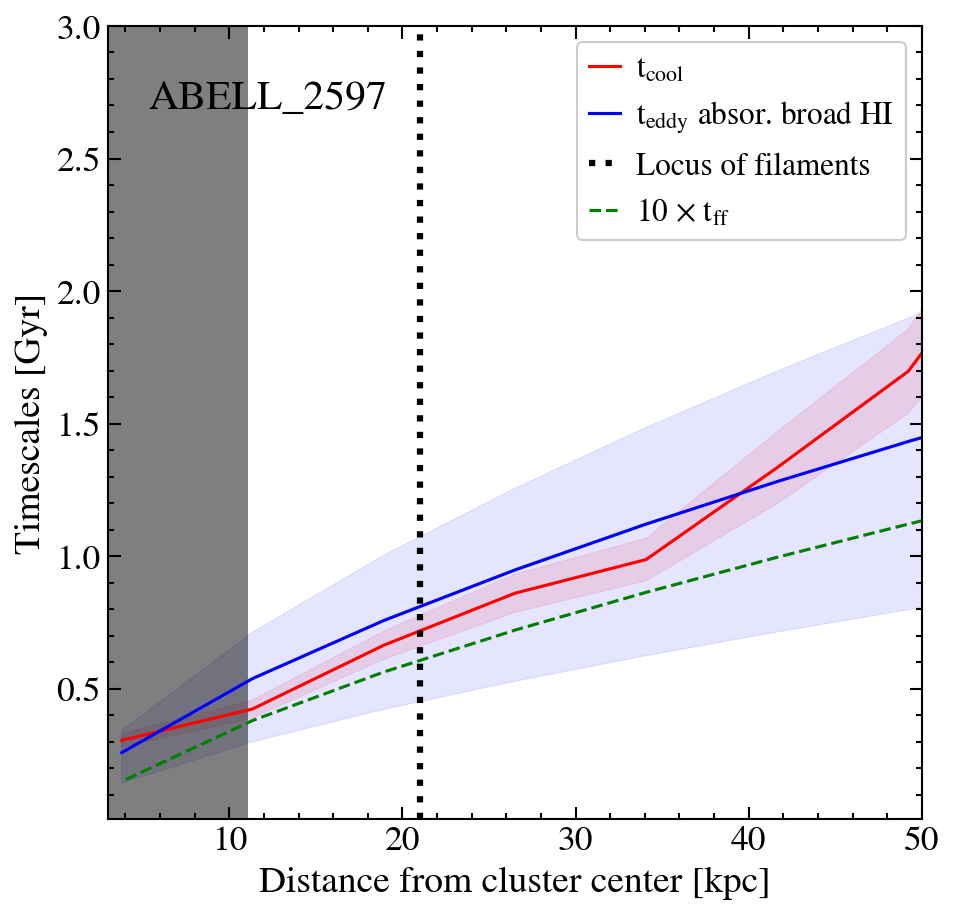}}\\
    \vspace{-4mm}
    \subfigure{\includegraphics[width=0.33\textwidth]{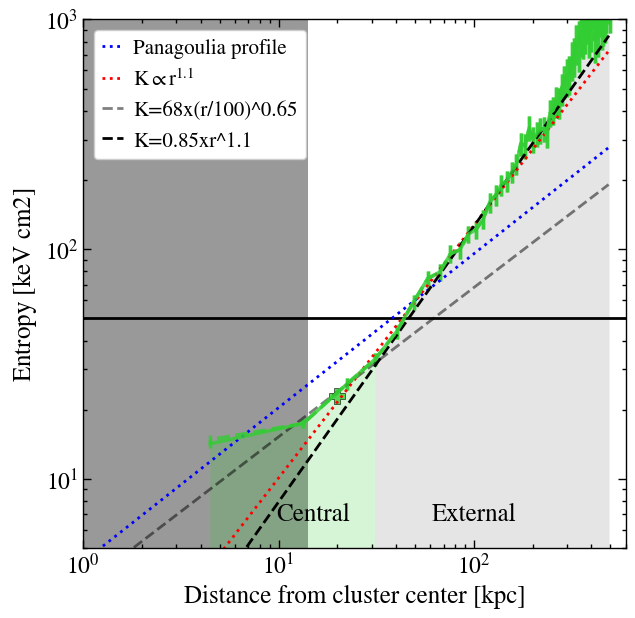}}
    \subfigure{\includegraphics[width=0.33\textwidth]{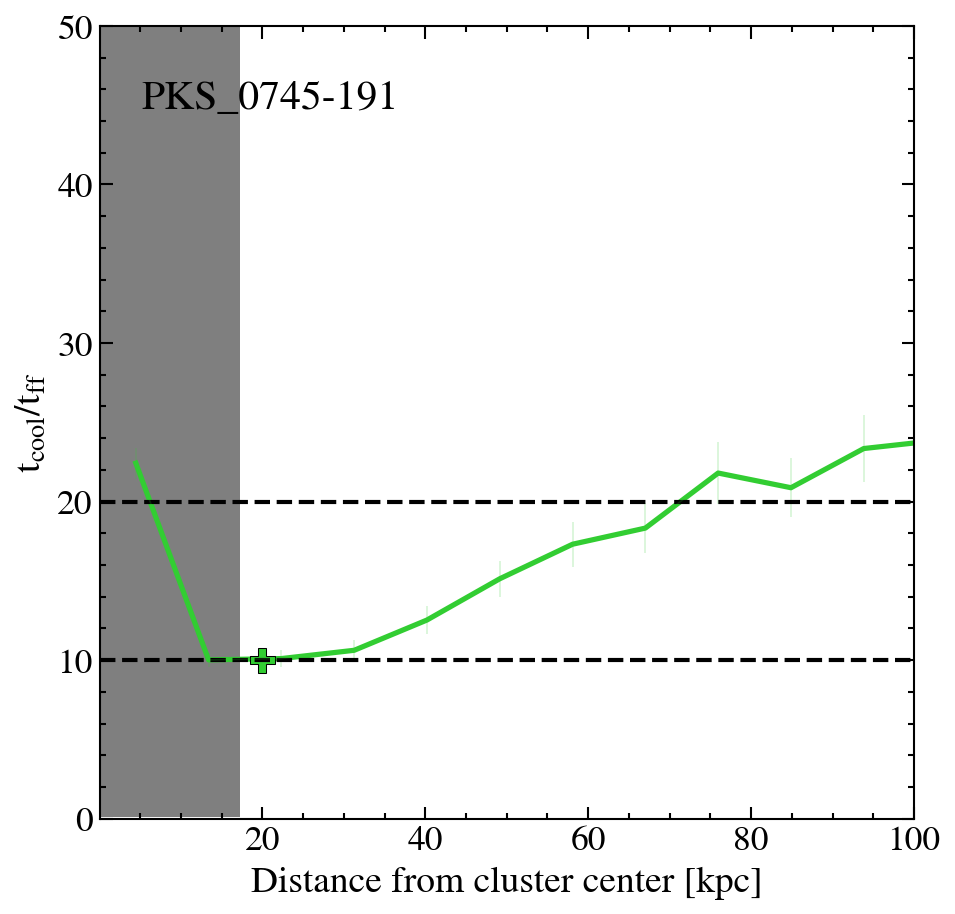}}
    \subfigure{\includegraphics[width=0.32\textwidth]{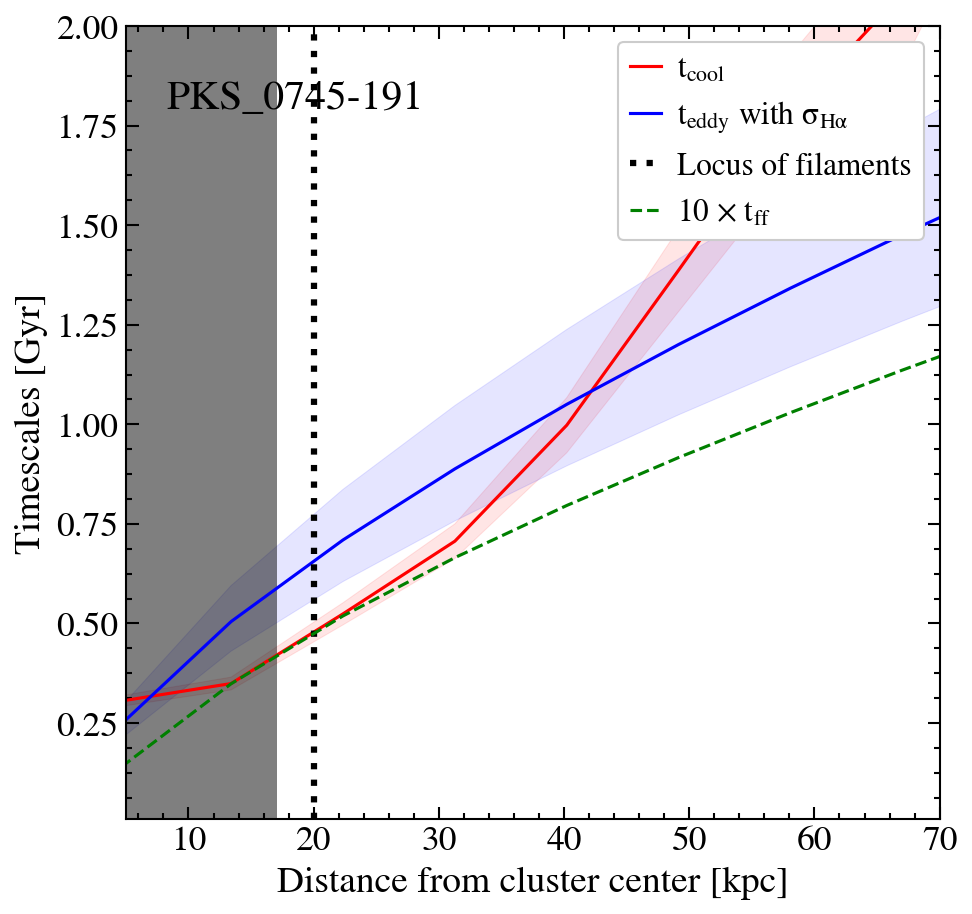}}\\
    \caption{Continuation of Fig.~\ref{fig:entropy_profiles}.}
\end{figure*}

\begin{figure*}[htbp!]
\label{fig:entropy_profiles4}
    \subfigure{\includegraphics[width=0.33\textwidth]{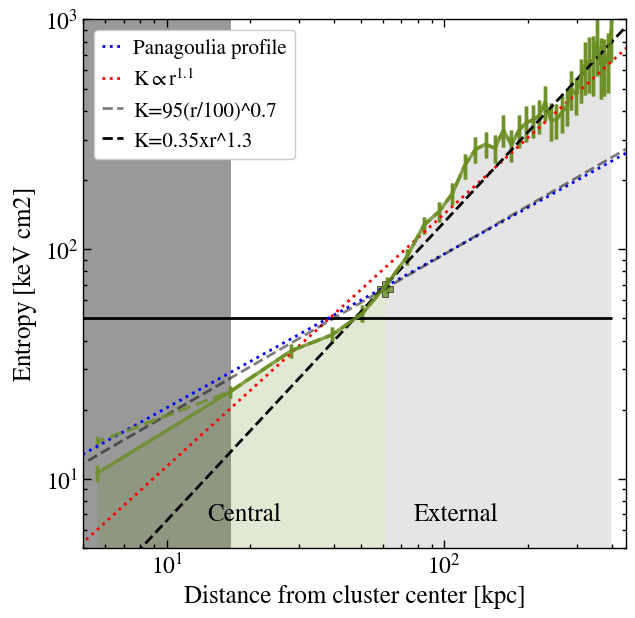}}
    \subfigure{\includegraphics[width=0.33\textwidth]{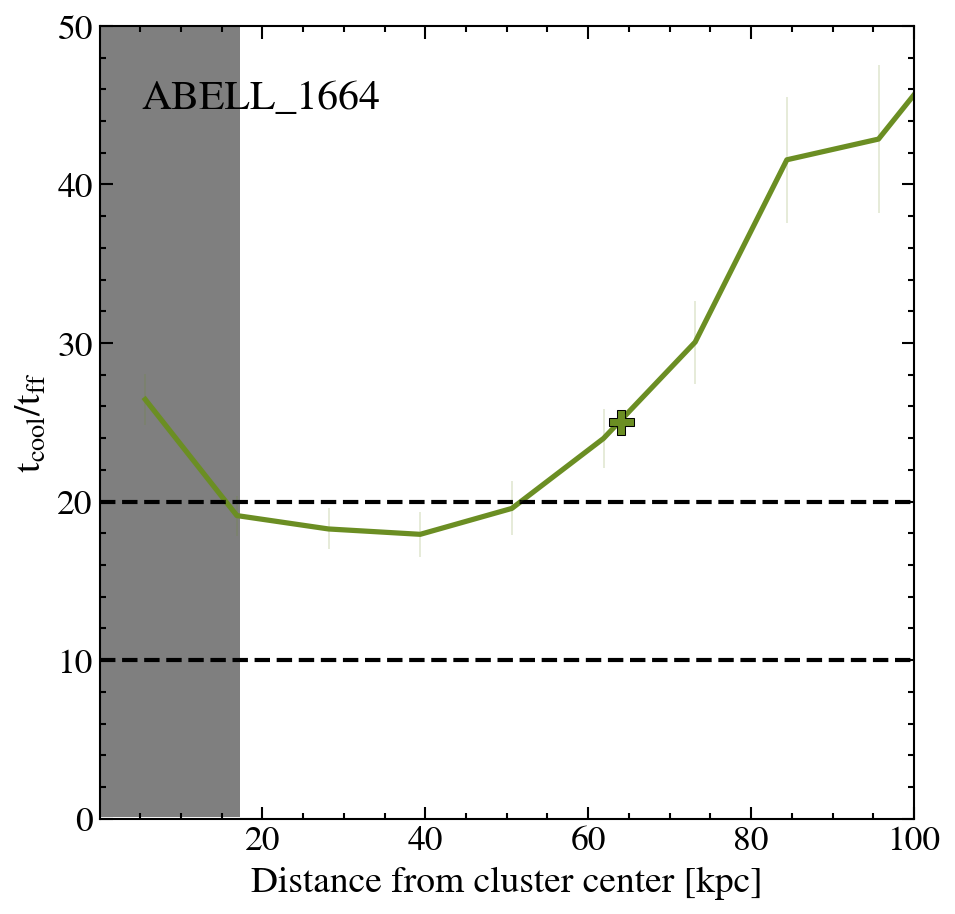}}\\
    \vspace{-4mm}
    \subfigure{\includegraphics[width=0.33\textwidth]{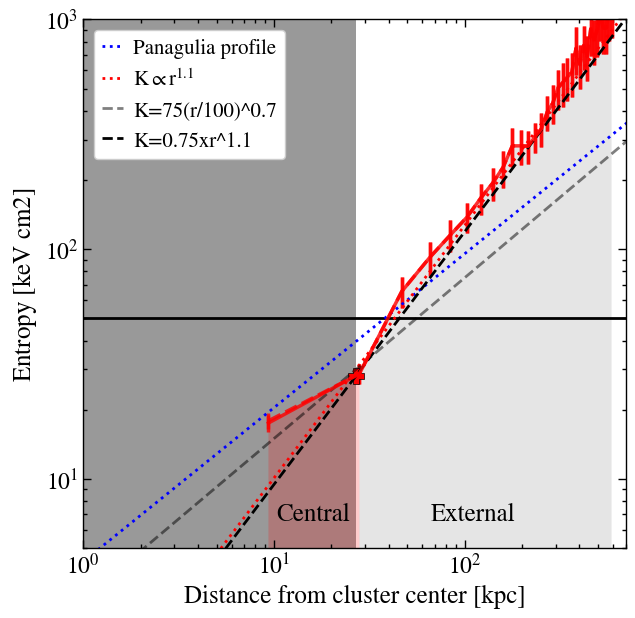}}
    \subfigure{\includegraphics[width=0.33\textwidth]{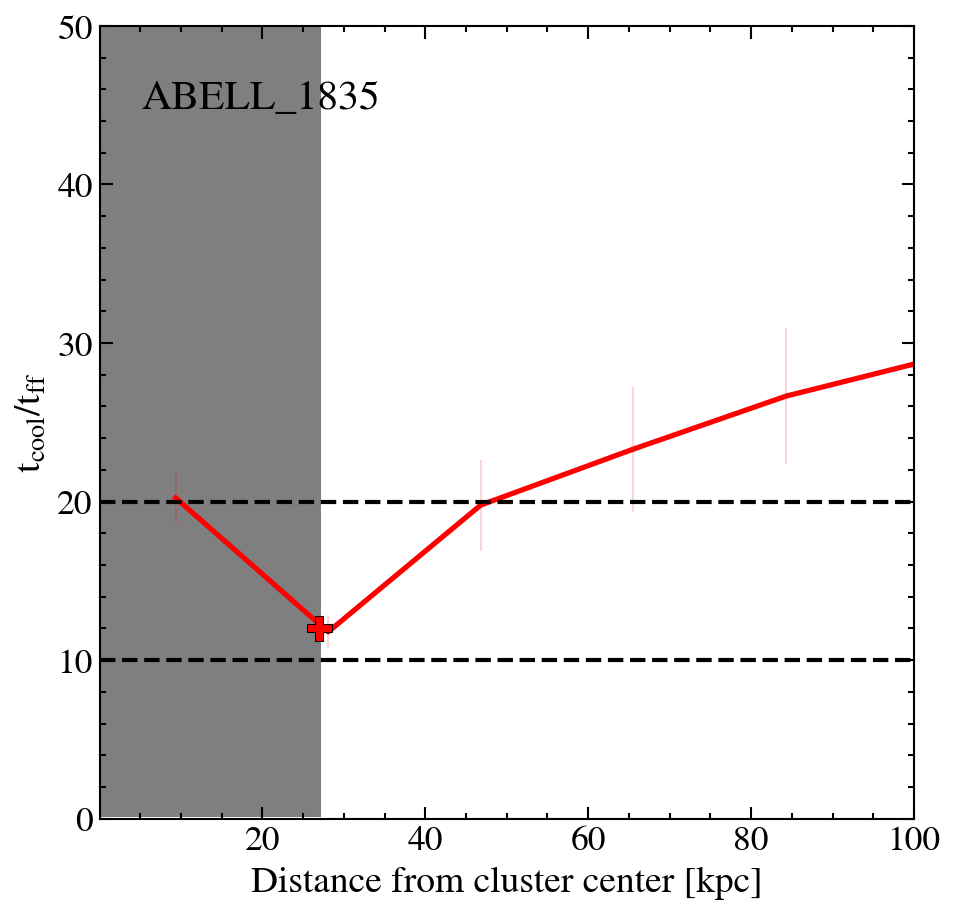}}
    \subfigure{\includegraphics[width=0.33\textwidth]{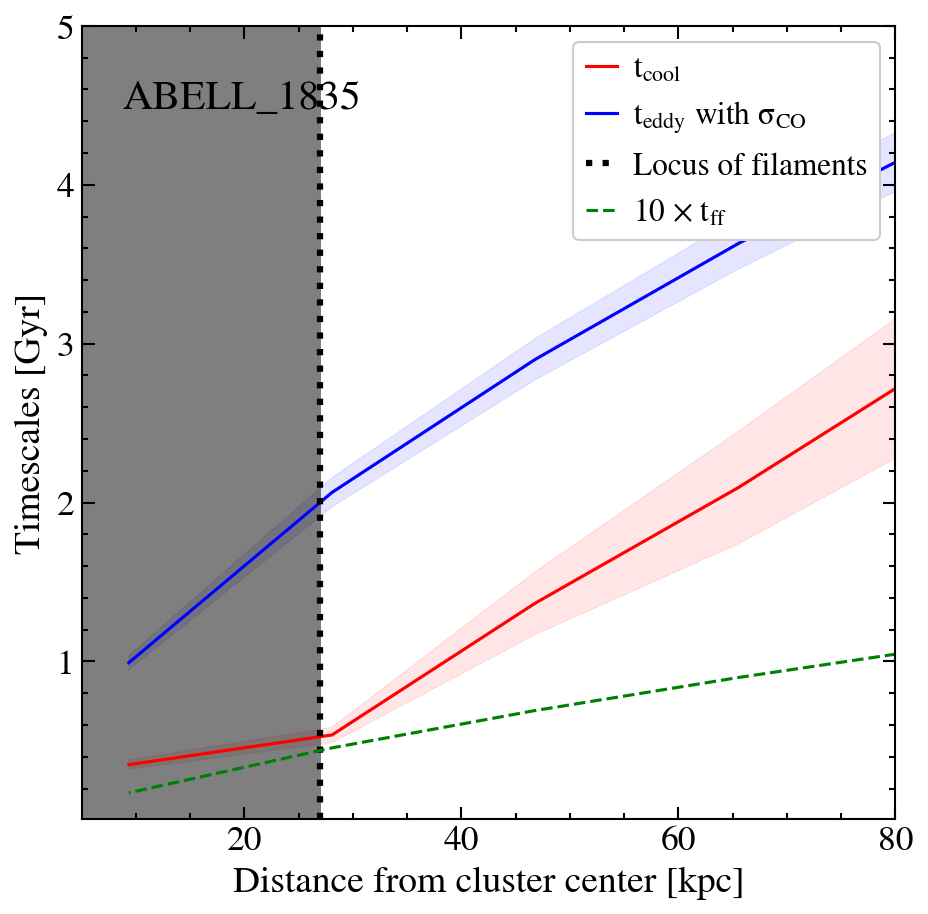}}\\
    \caption{Continuation of Fig.~\ref{fig:entropy_profiles}.}
\end{figure*}

\section{Cold Molecular}
\subsection{Distribution of the Cold Molecular Gas}
\label{appendix:morph}

\noindent{\bf Centaurus} In projection, the 5.6~kpc CO(1-0) emission in Centaurus is distributed in two main structures: a central broad structure with a extension of $\sim$2~kpc followed by a large filament that extends on about $\sim$4.3 kpc to the SE and NW following the ionized gas plume detected in H$\alpha$ emission and also seen in soft X-ray \citep{crawford05}. Several molecular peaks are detected along 4~kpc (20$\arcsec$) inside the southern H$\alpha$ filament. In addition, a small filament is extending to the NE, following the central component with a length of 1~kpc (5$\arcsec$). The width of the molecular filaments in Centaurus are unresolved and smaller than 0.5--5~kpc.\\
From the Fig.~\ref{fig:moments} (Third panel), we note that the molecular gas is distributed preferentially along the peripheries of the radio bubbles from VLA images (red contours) produced by the AGN, and beneath the X-ray cavities detected by \textit{Chandra} observations. We also note that the molecular filaments match the dust lane detected with the HST \citet{fabian16}. This cold molecular nebula is forming stars across its entire detected extent yielding a low star formation rate of $\sim$0.13~M$\odot$yr$^{-1}$ \citep{canning11}.

\noindent{\bf Abell\,S1101} The brightest structures extend over $\sim$8.8~kpc (8$\arcsec$) to the NS (southern clump) while a second structure is extended over $\sim$4.4~kpc (4$\arcsec$) to the NS as well (northern clump). A small unresolved clump is found in the SW (western filament) following the most brightest H$\alpha$ emission from the warm ionized nebulae, see Figure~\ref{fig:haoverco}. Based on \textit{Chandra} observations we can see that the molecular nebulae is located preferentially around the X-ray cavities displaced $\sim$7.5~kpc, which where formed though the inflated radio bubble from the central AGN, see Figure~\ref{fig:moments}. In addition, the molecular clouds appears to be closely associated with the abundant dust lane of $\sim$10$^{7}$~M$\odot$ detected by the HST observations. Furthermore, the CO(1-0) emission is spatially located with FIR and UV emission, indicating that the cold nebular is forming stars but at low rate $\sim$1--3~M$\odot$yr$^{-1}$ \citet{mcdonald15}.

\noindent{\bf RXJ1539.5} The molecular gas nebula is distributed in four main components: a central emission that is extended over $\sim$18~kpc, followed by several filaments. The WE radial filament is extending along $\sim$20 kpc (left filament), while the WN filament has a extension of$\sim$14 kpc (right filament) and finally a third small filament is extending towards the NS of the central emission over $\sim$7 kpc. Unfortunately, VLA images are not available to detect any radio bubbles.

\noindent{\bf Abell\,2597} Recently, \citet{tremblay16, tremblay18} showed ALMA cycle 3 observations of Abell\,2597 and its molecular gas using CO(2-1) emission lines, revealing a filamentary emission consistent in projection with optical emission nebula. Furthermore, they found absorption features that arise from cold molecular clouds moving toward the centre of the galaxy fuelling the black hole accretion, either via radial or in spiralling trajectories. The CO(2-1) continuum-subtracted emission of Abell\,2597, see the second panel of Figure~\ref{fig:moments}, reveals two molecular structures. A central component with showing a ``V'' shape is extending over $\sim$10~kpc (6.5 arcsec), followed in projection by a southern filament of length $\sim$7.7~kpc (5$\arcsec$).

\noindent{\bf PKS\,0745-101} The ALMA Cycle 1 observations of PKS\,0745-191 BCG of CO(1-0) and CO(3-2) emission lines presented by \citet{russell16} resolved the spatial and velocity structure of the cold molecular gas structures at the centre of the BCG revealing extended filaments. The CO(1-0) and CO(3-2) moment zero maps of PKS\,0745-101, are shown in the second panel of figure~\ref{fig:moments}. The CO(1-0) molecular morphology shows a compact and un-resolved nebulae extending over 11~kpc (6 $\arcsec$). While, the CO(3-2) emission resolves three clumpy filaments, with extensions of $\sim$3~kpc (1.6$\arcsec$), $\sim$3.8~kpc (2.0$\arcsec$) and $\sim$5~kpc (2.5$\arcsec$) for the SE, SW and N filaments, respectively. The emission from the SW filaments appears to extend across the nucleus as \citep{russell16} previously found. Here we note that the CO(1-0) intensity map cover a wide FOV with lower resolution than the CO(3-2) intensity map, explaining the lack of molecular filaments on the CO(1-0) map.
We measured a width size for the molecular filaments of 1--2~kpc. Those filaments are projected beneath the X-ray cavities, previously inflated by buoyantly radio bubbles from the central AGN. Furthermore, the cold filaments are coincident with low temperature X-ray gas, bright optical line emission and dust lanes suggesting that the molecular gas could have formed by gas cooling from uplifted warm gas that cooled in situ. One of the filaments appears to be breaking into clumps at large radius, those may eventually fragment with gas clouds falling back onto the BCG centre to fuel the central AGN.

\noindent{\bf Abell\,1795} In the Fig.~\ref{fig:moments} (second panel) we present the CO(2-1) mean flux of Abell\,1795 (moment zero), previously reported by \citet{russell17b}, which reveals a bright peak at the nucleus followed by curved filament extending N to a second emission peak. A southern filament extends to the SW of the nucleus over $\sim$6~kpc (5$\arcsec$). The curved filaments appear separated into two components of $\sim$5 kpc (4$\arcsec$) each. The two filaments are found exclusively around the outer edges of two inner radio bubbles, launched by the central AGN, as shown in red contours in the Figure~\ref{fig:moments}. 
Besides, the N filament is coincident with a strong dust lane revealed in the HST imaging \citet{russell17b}, similar to the molecular gas, the dust appears along the radio lobes and may have been entrained are formed in situ.
The molecular gas emission is also strongly correlated with the H$\alpha$ line emission observed by the MMTF \citet{crawford05, mcdonald12a} and MUSE, both nebulae clearly have similar structures with peaks around the nucleus, at the top of the N filament and along the S filament.\\ 

\noindent{\bf Abell\,1664} ALMA Cycle 0 observations of Abell\,1664 using CO(1-0) and CO(3-2) emission lines were presented in details by \citep{russell14}. In the second panel of Figure~\ref{fig:moments}, the continuum-subtracted CO(1-0) and CO(3-2) integrated emission are shown, in which the CO(1-0) emission is extending over a $\sim$21 kpc (9.5$\arcsec$ in a complex morphology. The CO(1-0) emission line morphology present a central components of 8~kpc (3.5$\arcsec$), followed by two small filaments NE and SW extending over 4.5~kpc and 8~kpc, respectively. Instead, the CO(3-2) observations spatially resolved the inner region of the galaxy cluster, which appears to be very clumpy and disturbed. It is possible to separate the filament into two gas blobs along its length of $\sim$8~kpc (3.5$\arcsec$). These structures are also coincident with low optical--ultraviolet surface brightness, which could indicate dust extinction associated with each clump \citet{russell14}.\\

\noindent{\bf 2A0335+096}
In Fig.~\ref{fig:moments} (second panel) we show the ALMA observations 2A0335+096 (also called RXCJ0338.6+0958) using CO(1-0) and CO(3-2) emission line, reported previously by \citet{vantyghem16}. In both molecular emission we detected similar structures, where the gas appears distributed in several clumpy and dusty filaments. A nucleus emission is present extending over 2$\arcsec$ followed by a dusty NW filament along 3 kpc (4.2$\arcsec$). In CO(1-0) we detect a distant NW clumpy filament of 6.3~kpc (9$\arcsec$) in length, located at 7~kpc (10$\arcsec$) from the nucleus. The cold molecular and ionized warm filaments from H$\alpha$ MUSE observations trails an X-ray cavity, suggesting that the gas has cooled from low-entropy gas that could have been lifted out of the cluster core and become thermally unstable.\\

\noindent{\bf Phoenix-A} \citet{russell17} reported the ALMA observations of CO(3-2) lime emission in Phoenix-A, revealing a massive reservoir of molecular gas, fueling a high star formation rate and a powerful AGN activity. The CO(3-2) molecular emission of Phoenix-A revels a peak at the galaxy centre extending a long NE to SW axis across the nucleus, see Fig.~\ref{fig:moments} (Second panel), followed by two filaments to the NW and SE of $\sim$27~kpc (3.7$\arcsec$) and $\sim$10~kpc (1.5$\arcsec$) and $\sim$20~kpc (3.0$\arcsec$), respectively. The central component is extended to the S of the nucleus, forming a curved filament along $\sim$17~kpc (2.5$\arcsec$). The ALMA observations show that the molecular filaments are located along the peripheries of the huge radio bubbles filled with relativistic plasma into the hot, X-ray atmospheres. Furthermore, the molecular filaments in Phoenix-A appear to be located preferentially along the dust lane detected by the HST broad band imaging. Similar to the molecular gas, the dust appears along the radio lobes and may have been formed in situ within cooling gas.\\

\noindent{\bf Abell\,3581} In Figure~\ref{fig:moments} (Second panel) we can see that the distribution of molecular gas, based on the CO(2-1) emission line for Abell\,3581, is located in several dusty perpendicular filaments. The molecular gas reveals a peak at the centre of the galaxy followed by a northern filament extending over a long of 2.6 kpc (6$\arcsec$). Two EW parallel filaments are found extending along 4.3~kpc (10$\arcsec$), the one at the top is showing a peak at the centre of the filament. In addition, a third high-velocity EW filament is located at the end of the northern filament along 1~kpc (2.5$\arcsec$). We note that the molecular emission from ALMA observations closely coincides with the dust lanes detected by using HST observations \citep{canning13}. In addition, we found that the molecular emission and the dust lanes in the south-east of the nucleus also coincide with the outer edge of the radio bubbles.
The molecular gas emission is also strongly correlated with the brightest H$\alpha$ emission observed by MUSE observations. Both nebulae have very similar structures. However, a significant difference occurs at the peak of the H$\alpha$ emission. Where the central galaxy and thje AGN is located, we found a lack of molecular emission, similar to what was found in Abell\,1795.\\

\noindent{\bf RXJ0820.9+0752}
Recently, \citet{vantyghem17} reported ALMA Cycle 4 observations of CO(1-0), CO(3-2), and $^{13}$CO(3-2) emission lines, one of the first detection of $^{13}$CO line emission in galaxies. The molecular gas originates from two clumps that are spatially offset from the galactic centre, surrounded by a diffuse emission.
In figure~\ref{fig:moments} (Second panel) we can identify two clumps of molecular gas, that are seen in both maps and account for most of the molecular emission extending around $\sim$3~kpc (1.5$\arcsec$) each clump, and surrounded by more diffuse emission. In CO(1-0) emission, we can identify two filaments extending toward the NW and SW along 8~kpc ($\sim$4$\arcsec$) each. The CO(3-2) emission appears to be more clumpy than the CO(1-0) emission (resolution effect). It reveals a small filament that goes from the nucleus to SE along 2~kpc ($\sim$1$\arcsec$).\\

\noindent{\bf Abell\,262 and Hydra-A} We found the presence of a compact molecular disk in two systems of our sample, Abell\,262 (BCG called as NGC0708) and Hydra-A \citep{rose19}, from CO(2-1) observations. Both BCGs show a double-peaked line profile that is typical of disk kinematics. From Fig.~\ref{fig:moments} we can see that in Hydra-A the molecular gas is hosted in a compact disk ($\sim$1.2 kpc) extending over 5.2~kpc (5$\arcsec$) and that coincides with the ionized and warm gas found with SINFONI \citet{hamer14} and MUSE observations.\\
Similarly, the molecular gas in Abell\,262 is hosted in a compact disc extending over 2.4~kpc (7.5$\arcsec$). From the second panel of the Figure~\ref{fig:moments} we note that in both sources the rotation is in a plane perpendicular to the projected orientation of the radio jets from the VLA observations, hinting at a possible connection between the cooling gas and the accretion of material on to the black hole.\\

\noindent{\bf Abell\,1835}
The ALMA cycle 0 observations of CO(1-0) and CO(3-2) emission line of Abell\,1835, reported previously by \citep{mcnamara14}, reveal that molecular gas is found in a compact region with an extent of $\sim$25~kpc (6.5$\arcsec$), based on CO(1-0), see Second panel of Fig.~\ref{fig:moments}.
While, the CO(3-2) emission is extending over $\sim$13.5~kpc (3.5$\arcsec$), showing three radial filaments relative to the nucleus of $\sim$1$\arcsec$ each (3.9~kpc). Given the lack of resolution of the ALMA band 3 they were unresolved in the CO(1-0) emission. Given the limited FOV on ALMA band 7, they look less extended in CO(3-2). A radial variation of the line ratio is also possible but need higher spatial 3mm observations for an accurate comparison.

\noindent{\bf M87} \citet{simionescu18} reported molecular gas using CO(2-1) emission line from ALMA observation in a region located at the very south-east of the nucleus of M87. An extended blue-shifted molecular emission with a lower velocity dispersion is detected there. The CO(2-1) emission is found only outside the radio lobe of the AGN seen in the VLA image, while the filament seen in other wavelengths prolongs further inwards. This suggests that the molecular gas in M87 could have been destroyed or excited by AGN activity, either by direct interaction with the radio plasma, or by the shock driven by the lobe into the X-ray emitting atmosphere. Finally, the CO(2-1) molecular filament in M87 located at south-east of the nucleus is extending along $\sim$0.6~kpc with a median width of $\sim$0.3~kpc.\\

\subsection{Velocity Structures of the Cold Molecular Gas}
\label{appendix:velo}

\noindent{\bf Centaurus} In Centaurus the molecular gas velocity is found in projection within $-100$ to $+250$~km~s$^{-1}$ relative to the galaxy's systemic velocity and shows two different velocity components, see Fig~\ref{fig:moments} (third and fourth panels). The extended southern filament (so-called ``the plume'') is characterised by smooth velocity gradient going from $+150$ to $+250$~km~s$^{-1}$ across its $\sim$4~kpc, increasing in velocity at larger radii. The filament\'s velocity is slower near the nucleus than at its larger distance. The filament have narrow velocity dispersion lying around 30~km~s$^{-1}$. However, at the northern position, the filament broadens in velocity dispersion, 90--150~km~s$^{-1}$. This could be a visual effect due a superposition of another perpendicular filament that is located behind the BCG, making it impossible to detect. A central gas component is present with a velocity dispersion below 100~km~s$^{-1}$ and moving with a roughly constant velocity of $+30$~km~s$^{-1}$. Finally a small blue-shifted filament is found WE at the north of the central component showing a shallow gradient of $-$160--$-$70~km~s$^{-1}$ at decreasing radii. The smooth velocity gradients and narrow velocity dispersion reveal a steady, ordered flow of molecular gas.\\

\noindent{\bf RXJ1539.5} RXJ1539.5 molecular nebula have a complex velocity structure across its spatial extension and appears to be very disturbed, see Fig~\ref{fig:moments} (third and fourth panels). The projected gas velocities that spans $\pm250$~km~s$^{-1}$ are spread almost symetrically around the systemic velocity of the galaxy. The central component is characterised by a broad velocity dispersion of 120~km~s$^{-1}$. The eastern clumpy filaments show largely quiescent structures with no-monotonic or coherent gradient along its structure. The nebula has roughly constant velocity centroids from $+50$--$+100$~km~s$^{-1}$ and with narrow velocity dispersion, 50--80~km~s$^{-1}$. On the other hand, the clumpy western filament shows a shallow gradient of projected velocities that span from $-200$ to $+200$~km~s$^{-1}$ with an almost constant velocity dispersion of 30--50~km~s$^{-1}$. However at the NW terminus of the western filament the velocity dispersion broadens to $>$180~km~s$^{-1}$. Finally the blue-shifted northern filament shows a projected smooth velocity gradient that spans over $-150$ to $-250$~km~s$^{-1}$ from SN at increasing radii.\\

\noindent{\bf Abell\,S1101} The cold molecular gas velocities in Abell\,S1101 lie within $+30$ to $-250$~km~s$^{-1}$ around the systemic velocity of the galaxy and shows two different filamentary structures, SN and SW filaments, both coming from the nuclear region of the BCG, see Fig~\ref{fig:moments} (third and fourth panels). These structures present sign of a shallow velocity gradient across its kilo-parsec length, decreasing in velocity while increasing galactocentric radius. This is consistent with either an inflow or an outflow of cold gas, since the inclination are not known. The SN filament goes from $+$20 through $-$100~km~s$^{-1}$ relative to the galaxy's systemic velocity while increasing radii with a low velocity dispersion that lies below 40~km~s$^{-1}$. Instead the SW filament goes from $+30$ to $-250$~km~s$^{-1}$ as we move outside of the nucleus, with a narrow velocity dispersion lying 40--70~km~s$^{-1}$.\\ 

\noindent{\bf Abell\,2597} The molecular nebula velocities in Abell\,2597 \citep{tremblay16, tremblay18} lie within -150 and +150~km~s$^{-1}$ relative of the systemic velocity of galaxy. The velocity structures are complex across its spatial extent, see Fig~\ref{fig:moments} (third and fourth panels). The central components show a roughly symetrical velocity structures lying within $-100$ to $100$~km~s$^{-1}$, suggesting some rotation. While the southern filament shows largely quiescent structure along its extent, going from +30 to +100~km~s$^{-1}$ and with a velocity dispersion of 30 to 60~km~s$^{-1}$. \\

\noindent{\bf PKS\,0745-191} The cold molecular projected velocities of the CO(1-0) line emission of PKS\,0745-191 \citep{russell16} lie in a velocity gradient within $-60$ to $+40$~km~s$^{-1}$ from the systemic velocity, see Fig~\ref{fig:moments} (third and fourth panels) and showing low velocity dispersions that lie below 70~km~s$^{-1}$, with the broadest component close to the nucleus of the galaxy cluster. The projected velocities of the molecular gas in the CO(3-2) line emission show more complex motions within the nebula. They are found in a narrow range of projected velocities, $-60$ to $+40$~km~s$^{-1}$ and have a low velocity dispersion as well, below 100~km~s$^{-1}$ across the most extended structure.\\

\noindent{\bf Abell\,1664} In Abell\,1664 \citep{russell14} the molecular gas projected velocities of the CO(1-0) emission line is blue-shifted from $+20$ to $-650$~km~s$^{-1}$ to the SN, see Fig~\ref{fig:moments} (third and fourth panels). Furthermore, the velocity dispersion of the gas lies below 250~km~s$^{-1}$.
The projected velocities from the CO(3-2) emission lie within $-600$ to $+150$~km~s$^{-1}$ from the systemic velocity. The velocity dispersions lie below 350~km~s$^{-1}$. The SE component appears to be moving in a quiescent structure going from $+90$ to $+120$~km~s$^{-1}$ with a velocity dispersion of 70~km~s$^{-1}$, while the NW component has a more disturbed velocity structure blue-shifting from $-130$ to $-600$~km~s$^{-1}$ with a velocity dispersion below 350~km~s$^{-1}$. \citet{russell14} suggest that those high-velocity gas structures are consistent with a massive outflow projected in from of the BCG and moving towards us along the line of sight. This is supported by the low optical emission also found in this region, that is consistent with a highly dusty region (the cold gas then behing in the front side).\\

\noindent{\bf Abell\,1835} In Abell\,1835 \citep{mcnamara14} the projected velocities of the molecular gas of the CO(1-0) emission line lie in a smooth gradient that goes from $-150$ to $+120$~km~s$^{-1}$ along its extent, see Fig~\ref{fig:moments} (third and fourth panels). Furthermore, it shows low velocity dispersion, lying well below 130~km~s$^{-1}$. Similarly, the projected velocity structures of CO(3-2) emission line lie between $-120$ and $+120$~km~s$^{-1}$, and the velocity dispersion below 110~km~s$^{-1}$.\\

\noindent{\bf Abell\,2597} In Abell\,2597 \citep{tremblay16, tremblay18} the velocities of the molecular gas of the CO(2-1) emission line of the central ``V'' shaped clumps lie in a complex and turbulent velocity distribution, suggesting a interaction with the radio jet from the AGN, that goes from $-100$ to $+200$~km~s$^{-1}$, see Fig~\ref{fig:moments} (third and fourth panels). While, the SW filaments shows a velocity gradient from $+5$ to $+40$~km~s$^{-1}$ (from north to south). This suggests that the filament is could be lifted gas from the centre of the galaxy by a shock pressure triggered by one of the radio bubbles. The velocity dispersions of the filament are low (30--60~km~s$^{-1}$) in comparison to the central component (30--150~km~s$^{-1}$).\\

\noindent{\bf M87} The small clump detected in M87 \citep{simionescu18} shows a blue-shifted velocity gradient that goes from $-150$ to $-110$~km~s$^{-1}$, see Fig~\ref{fig:moments} (third and fourth panels). The measured that velocity dispersion appears remarkably quiescent $\sim$30~km~s$^{-1}$.\\ 

\noindent{\bf Abell\,1795} In Abell\,1795 \citet{russell17b} the projected molecular velocities of the CO(2-1) emission line of the curved filaments shows a smooth gradients that goes from $-170$ to $+190$~km~s$^{-1}$ relative to the system velocity of the BCG, see Fig~\ref{fig:moments} (third and fourth panels). While the SW filaments goes from $-60$ to 0~km~s$^{-1}$ from the north-south direction consistent with an outflow of gas by southern radio lobes from the AGN. The velocity dispersion of Abell\,1795 are very quiescent lying well below 60~km~s$^{-1}$.\\

\noindent{\bf Phoenix-A} The projected velocities of molecular gas traced with the CO(2-1) emission line in Phoenix-A \citet{russell17} show a very complex velocity structure, see Fig.~\ref{fig:moments} (third and fourth panels). All three extended filaments have ordered velocity gradients along their lengths and show low velocity dispersions lying below 70~km~s$^{-1}$. The velocities at the furthest extent of each filament increase towards to the galaxy nucleus with the highest velocity at the smallest radii, going from to $+420$ to $+180$~km~s$^{-1}$ for the NE and SE filaments, while the velocity gradient along the NW filament goes from $+200$ to 0~km~s$^{-1}$ decreasing at the nucleus. The SW filament shows a shallow gradient where the velocity increases with decreasing radius from $+160$ to $+220$~km~s$^{-1}$. Finally, the central compact component of the emission at the center of the galaxy shows a separate structure with a much higher velocity dispersion of 200~km~s$^{-1}$ and a velocity gradient that goes from $-90$ to $+250$~km~s$^{-1}$ across the nucleus and perpendicular to the filaments. This is consistent with ordered motions around the nucleus.\\

\noindent{\bf RXJ0820.9+0752} The cold molecular gas in RXJ0820.9+0752 \citet{vantyghem17} from CO(1-0) and CO(3-2) emission lines are lying in two main components surrounded by a smooth emission outside of the nuclear region of the galaxy, see Fig.~\ref{fig:moments} (third and fourth panels). The velocity maps show that two components are moving with low regular velocities that go from $+70$ to $+40$~km~s$^{-1}$ across the two main components. While the SE filament in the CO(3-2) emission line shows a shallow gradient where the velocity decreases with increasing radius from $-10$ to $-70$~km~s$^{-1}$, the NE filament is moving with a constant velocity of a few $\sim$10~km~s$^{-1}$. In general the smooth emission that surrounds the two component shows low velocity dispersions lying below 30~km~s$^{-1}$, while the two components present higher velocity dispersions around 60~km~s$^{-1}$.\\

\noindent{\bf 2A 0335+096} In 2A 0335+096 \citet{vantyghem16} the cold molecular gas projected velocities of the CO(1-0) and CO(3-2) emission lines are co-moving, see Fig.~\ref{fig:moments} (third and fourth panels). A large velocity gradient is present within the core of the BCG going from $-140$ to $+200$~km~s$^{-1}$ increasing from the SN across the nucleus. The blue-shifted emission is relatively broad, with a velocity dispersion around 100~km~s$^{-1}$, while the redshifted emission is considerably narrower with a velocity dispersion of 50~km~s$^{-1}$. The NE filament seen in the CO(1-0) emission maps shows a smooth gradient that goes from $+10$ to $-110$~km~s$^{-1}$ decreasing with radii with as weel as a relatively narrow velocity dispersion of 50--60~km~s$^{-1}$ across the filament.\\

\noindent{\bf Abell\,3581} The velocities of the cold molecular gas in Abell\,3581 are embedded in two main filaments, both showing ordered velocity gradients along their lengths and decreasing in velocity at large radii, see Fig.~\ref{fig:moments} (third and fourth panels). The SN filaments show that the molecular gas is moving from $+120$ to $-190$~km~s$^{-1}$ from the centre of the system towards the outer regions, with a broader velocity dispersion component in the central region of 130~km~s$^{-1}$ and decreasing at 30~km~s$^{-1}$ outside. The EW filament shows a clear velocity gradient that goes from $+140$ to $-10$~km~s$^{-1}$ and increasing with radii with a narrow velocity dispersion of around 30~km~s$^{-1}$. A SW orientated high-velocity filament is found at $\sim$1.5~kpc (3$\arcsec$) north of the galaxy cluster center, moving with almost constant projected line-of-sigh velocity, $-510$--$-580$~km~s$^{-1}$ across its extent.\\

\noindent{\bf Abell\,262 and Hydra-A} Abell\,262 and and Hydra-A show velocity maps and double peak spectral profiles consistent with a molecular rotating disk, see Fig.~\ref{fig:moments} (third and fourth panels). In Abell\,262 the molecular gas gradient goes from $-240$ to $+240$~km~s$^{-1}$ increasing from the SN and showing a central broader component with a velocity dispersion of around 120~km~s$^{-1}$, decreasing towards large radii. Similarly, the molecular gas in Hydra-A \citep{rose19} is moving in a smooth gradient that goes from $+300$ to $-300$~km~s$^{-1}$, with a broader velocity dispersion component at the center of the disk of 100~km~s$^{-1}$ and a narrow velocity dispersion lying below 70~km~s$^{-1}$. A narrow continuum red-shifted absorption feature is found at $\sim-$40~km~s$^{-1}$ from the systemic velocity located near the galaxy\'s nucleus. This shows that those absorption features are cast in inflowing cold molecular clouds, eclipsing our line of sight to the black hole continuum, similarly to Abell\,2597 \citet{tremblay16}.\\

\end{appendix}
\end{document}